\documentclass[aps,prl,reprint,superscriptaddress]{revtex4-1}
\usepackage{blindtext}
\usepackage{lipsum}
\usepackage{graphics}
\usepackage{amsmath}
\usepackage{graphicx}
\usepackage{graphics}
\usepackage{amssymb}
\usepackage{verbatim}
\usepackage{physics}
\usepackage{float}
\usepackage{hyperref}
\usepackage[dvipsnames]{xcolor}

\makeatletter
\DeclareFontFamily{OMX}{MnSymbolE}{}
\DeclareSymbolFont{MnLargeSymbols}{OMX}{MnSymbolE}{m}{n}
\SetSymbolFont{MnLargeSymbols}{bold}{OMX}{MnSymbolE}{b}{n}
\DeclareFontShape{OMX}{MnSymbolE}{m}{n}{
    <-6>  MnSymbolE5
   <6-7>  MnSymbolE6
   <7-8>  MnSymbolE7
   <8-9>  MnSymbolE8
   <9-10> MnSymbolE9
  <10-12> MnSymbolE10
  <12->   MnSymbolE12
}{}
\DeclareFontShape{OMX}{MnSymbolE}{b}{n}{
    <-6>  MnSymbolE-Bold5
   <6-7>  MnSymbolE-Bold6
   <7-8>  MnSymbolE-Bold7
   <8-9>  MnSymbolE-Bold8
   <9-10> MnSymbolE-Bold9
  <10-12> MnSymbolE-Bold10
  <12->   MnSymbolE-Bold12
}{}

\let\llangle\@undefined
\let\rrangle\@undefined
\DeclareMathDelimiter{\llangle}{\mathopen}%
                     {MnLargeSymbols}{'164}{MnLargeSymbols}{'164}
\DeclareMathDelimiter{\rrangle}{\mathclose}%
                     {MnLargeSymbols}{'171}{MnLargeSymbols}{'171}
\makeatother

\newcommand{\eff}{\mathrm{eff}}

\begin{document}

\title{Emergent ergodicity at the transition between many-body localized phases}
\author{Rahul Sahay}
\thanks{These authors contributed equally to this work.}
\affiliation{Department of Physics, University of California, Berkeley, California 94720 USA}

\author{Francisco Machado}
\thanks{These authors contributed equally to this work.}
\affiliation{Department of Physics, University of California, Berkeley, California 94720 USA}

\author{Bingtian Ye}
\thanks{These authors contributed equally to this work.}
\affiliation{Department of Physics, University of California, Berkeley, California 94720 USA}

\author{Chris R. Laumann}
\affiliation{Department of Physics, Boston University, Boston, MA, 02215, USA}

\author{Norman Y. Yao}
\affiliation{Department of Physics, University of California, Berkeley, California 94720 USA}
\affiliation{Materials Science Division, Lawrence Berkeley National Laboratory, Berkeley, CA 94720, USA}

\begin{abstract}
Strongly disordered systems in the many-body localized (MBL) phase can exhibit ground state order in highly excited eigenstates.
The interplay between localization, symmetry, and topology has led to the characterization of a broad landscape of MBL phases ranging from spin glasses and time crystals to symmetry protected topological phases. 
Understanding the nature of phase transitions between these different forms of eigenstate order remains an essential open question.  
Here, we conjecture that no direct transition between distinct MBL orders can occur; rather, a thermal phase always intervenes.
Motivated by recent advances in Rydberg-atom-based quantum simulation, we propose an experimental protocol where the intervening thermal phase can be diagnosed via the dynamics of local observables.
\end{abstract}

\maketitle

Traditionally, the classification of phases of matter has focused on systems at or near thermal equilibrium. 
Many-body localization (MBL) offers an alternative to this paradigm \cite{Gornyi_Mirlin_Polyakov_2005, Basko2006_BAA, Basko2006_2, Nandkishore2015_Rev, vosk2015theory, Abanin2019_Rev}.
In particular, owing to the presence of strong disorder, MBL phases are characterized by their failure to thermalize \cite{Deutsch_1991,Srednicki_1994,canovi2011quantum,eisert2015quantum}.
This dynamical property imposes strong constraints on the structure of eigenstates; namely, that they exhibit area-law entanglement and can be described as the ground state of quasi-local Hamiltonians \cite{Serbyn2013_LBit, Bauer2013_AreaLawOrder}.
Perhaps the most striking consequence is that such systems can exhibit order -- previously restricted to the ground state -- throughout their entire many-body spectrum \cite{Huse2013_LPQO,Bauer2013_AreaLawOrder,Parameswaran2018_MBLOrder_Review,Chandran2014_MBLSPT, Kjall2014_MBLOrder,Pekker_2014}. 
This offers a particularly tantalizing prospect for near-term quantum simulators: The ability to observe phenomena, such as coherent topological edge modes, without the need to cool to the many-body ground state
\cite{Bahri2015_MBLSPT,yao2015many,Yao2016_DTC,Else2016_DTC,Potirniche_2017}.

The presence of eigenstate order in the many-body localized phase also raises a more fundamental question: What is the nature of phase transitions between different types of MBL order?
This question highlights a delicate balance between the properties of localization and phase transitions.
On the one hand, the stability of MBL is contingent upon the existence of an extensive number of quasi-\textit{local} conserved quantities (``$\ell$-bits'') \cite{Serbyn2013_LBit,Huse2014_LBit}. 
On the other hand, the correlation length at a second-order phase transition \emph{diverges} \cite{Subir_QPT}. 
%
%This divergent correlation length is incompatible with the ``$\ell$-bit'' picture and leads to  delocalization.
%Already in non-interacting systems, Anderson localization is known to host a critical point which is not \textit{strictly} localized \cite{RSRGX,?,?}.
%
Understanding and characterizing this interplay remains an outstanding challenge.
Indeed, while certain studies suggest the presence of a direct transition between distinct MBL phases \cite{Kjall2014_MBLOrder, Altman2014_RSRGX, Venderley2016_EmerErg, Friedman2018_EmerErg,Vasseur2016_PHS, Yao2016_DTC}, others have found signatures of delocalization at the transition \cite{Khemani2016_EmerErg, Chan2020_EmerErg}.

\begin{figure}
    \centering
    \hspace{-4 mm}\includegraphics[width =3.5in]{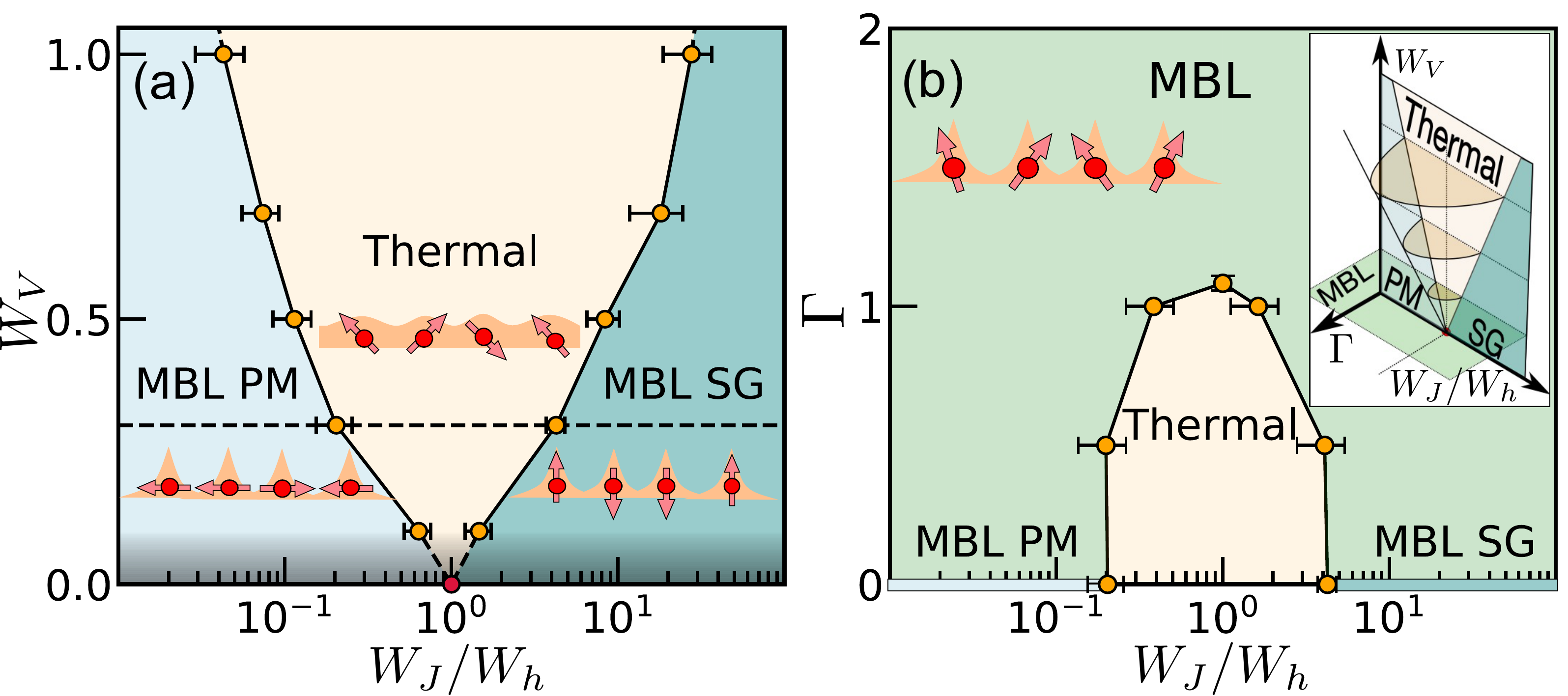}
    \caption{
    (a)
    Phase diagram of the symmetry breaking model, Eqn.~\ref{eq-SGModel}, as a function of $W_J/W_h$ and interaction strength $W_V$. 
    For all numerically accessible $W_V$, we observe a finite width thermal region between the two different MBL phases (PM and SG).
    %\footnote{Phase boundaries are obtained through finite-size scaling analysis of $\langle r \rangle$-ratio data for systems of length $L = 8, 10, 12, 14, 16$ with a minimum of $10^3$ disorder averages \cite{SM}.}.
    At $W_V=0$, the system is non-interacting and exhibits a critical point at $W_J/W_h=1$ (red point).
    % The red point indicates the critical point of the non-interacting disordered transverse field Ising model. 
    % Owing to finite size limitations, we are unable to access the low-interaction (shaded) strength region (see SM for details).
    (b)
    Phase diagram as a function of a symmetry breaking field $\Gamma$ and $W_J/W_h$ for $W_V = 0.3$.
    With increasing $\Gamma$, the size of the thermal region decreases until the system remain localized for all $W_J/W_h$.
    (inset) Schematic of the full phase diagram as a function of $W_J/W_h$, $W_V$ and $\Gamma$.
    }
    \label{fig:fig1}
\end{figure}

%First, focusing on symmetry breaking order, we construct the phase diagram, in which we observe the presence of a finite-width intervening thermal phase. 
%Second, we generalize our results to symmetry-protected topological (SPT) and discrete time-crystalline (DTC) order and demonstrate that such an intervening thermal phase is inevitable.

\begin{figure*}
    \centering
    \includegraphics[width = 7in]{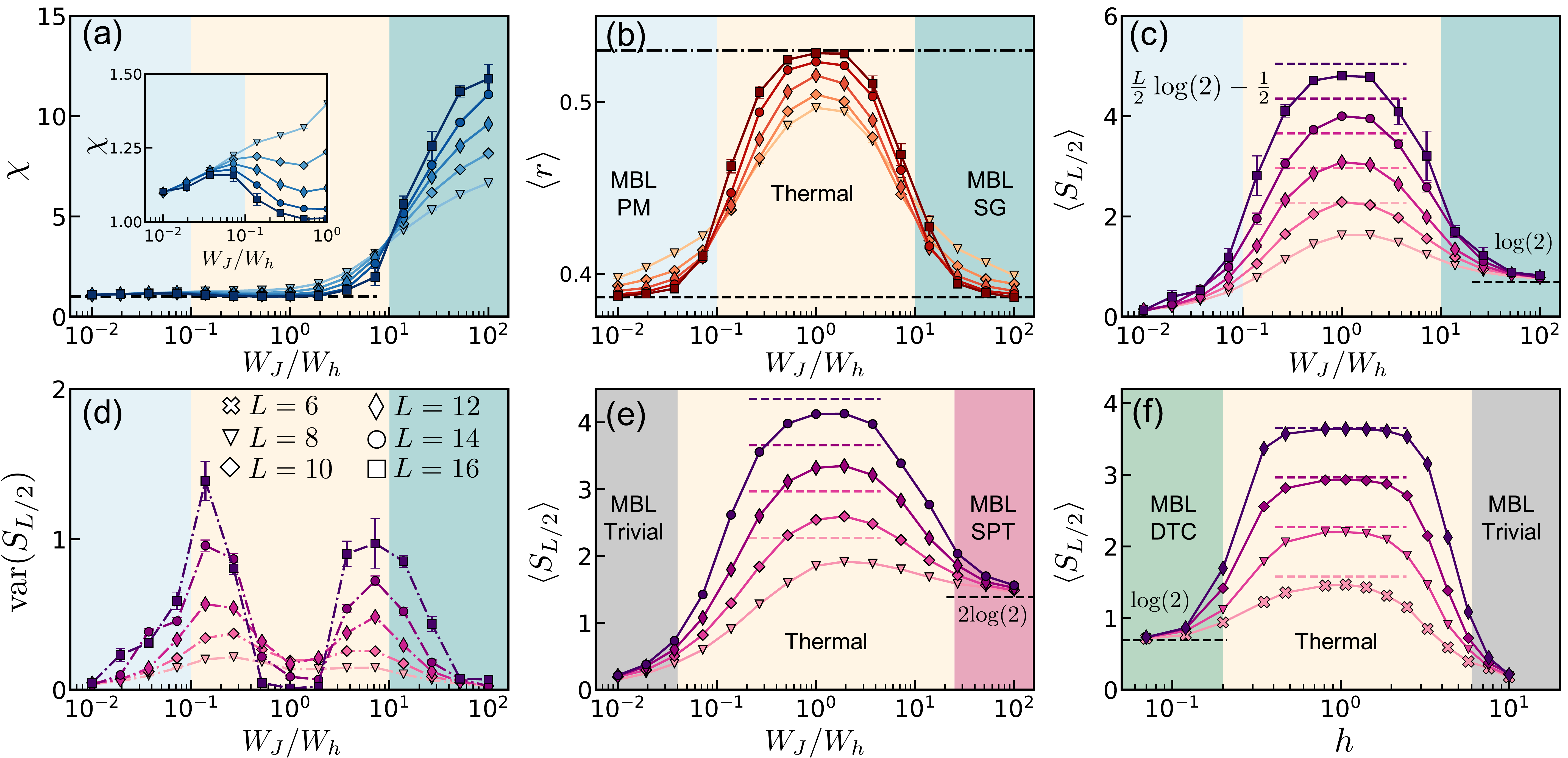}\vspace{-4 mm}
    \caption{
    %Spin-glass order parameter of Eq.~\ref{eq-SGOP} for $W_V = 0.7$. 
    % We perform $\sim 10^4$ disorder realizations for $L = 8, 10, 12,$ and $14$ and $\sim 10^3$ disorder realizations for $L = 16$. 
    % Inset: zoom-in of the regime for $W_J/W_h < 1$.
    % We find the separation of MBL PM and thermal behavior \by{Not sure what do you mean exactly}.
    % (b) Half-chain entanglement entropy, $S_{L/2}$ for $W_V = 0.7$, computed for an eigenstate at the center of the band. The curves show the a thermal phase (with $S_{L/2}$ approaching $S_{th}(L)$ (dashed lines in center)) separates an area law entangled ($S_{1/2} \sim 0$ dashed line on left) MBL PM and an area law entangled ( $S_{1/2} \sim \log(2)$ dashed line on right) MBL SG. The existence of two transitions can be resolved via two peaks in the variance of the entanglement (inset). (c) The ratio of adjacent level spacings at $W_V = 0.7$. The approach of $\langle r \rangle$ to the thermal (GOE) prediction (top dashed line) that separates two localized (Poisson) regions (bottom dashed line) once again demonstrates the existence of the intervening thermal phase. However, in the presence of a strong longitudinal field (d), with $W_V = 0.3, W_\Gamma = 4$, we see the existence of the intervening thermal phase disappear.  (e-f) $S_{1/2}$ for the MBL SPT transition and the MBL DTC transition which shows analogous behavior to the MBL SG transition. Dashed line in (e) indicates $2\log(2)$ entanglement and dashed line in (f) indicates $\log(2)$ entanglement. 
    (a-d) Characterization of the symmetry breaking model, Eqn.~\ref{eq-SGModel}, for $W_V = 0.7$.  
    % \footnote{We perform $\sim 10^4$ disorder realizations for $L = 8, 10, 12,$ and $14$ and $\sim 10^3$ disorder realizations for $L = 16$}. 
    (a) For $W_J/W_h \gtrsim 10$, $\chi$ increases with system size evincing the SG nature of the phase. 
    In the PM phase, $\chi$ approaches a finite constant, albeit exhibiting two distinct behaviors (inset).
    (b) 
    $\langle r \rangle$-ratio as a function of $W_J/W_h$ reveals an intervening thermal phase surrounded by two localized phases. 
    The dash-dotted [dashed] line corresponds to the GOE [Poisson] expectation.
    (c) The half-chain entanglement entropy $S_{L/2}$ increases with system size for intermediate $W_J/W_h$, in agreement with the expected thermal volume-law.
    In the two localized phases, $S_{L/2}$ saturates to different values, highlighting the distinct nature of the underlying eigenstate order.
    (d) The variance of $S_{L/2}$ exhibits two distinct peaks in agreement with  the presence of two distinct transitions.
    (e)[(f)] $S_{L/2}$ for the SPT [DTC] model of Eqn.~\ref{eq-SPTModel}~[Eqn.~\ref{eqn-DTC}] also demonstrates the presence of an intervening thermal phase.
    Each data point corresponds to averaging over  at least $10^3$ disorder realizations.
    }
    \label{fig-Data}
\end{figure*}

In this Letter, we conjecture that any transition between distinct MBL phases is invariably forbidden and that an intervening thermal phase always emerges (Fig.~\ref{fig:fig1}a). 
This conjecture is motivated by an extensive numerical study of three classes of MBL transitions: 
(i) a symmetry-breaking transition, %paramagnet to spin glass, 
(ii) a symmetry-protected topological (SPT) transition, and % to topologically trivial, 
(iii) a discrete time crystalline transition (in a Floquet system).  
By systematically constructing the various phase diagrams, we show that an intervening ergodic region emerges for all numerically-accessible interaction strengths.
%
%thermal phase down to very small values of the interactions, where finite-size effects are expected to "hide" the behavior \cite{Dima, Challenges in Finite Size}. 
%Indeed this regime agrees with previous signatures of a direct transition.
%
Moreover, we demonstrate that this emergent ergodicity is intimately tied to the presence of a phase transition; a  \emph{disorder-less},  symmetry-breaking  field  suppresses the intervening ergodic phase.
In addition to numerics, we analyze two instabilities which could induce thermalization near the putative transition:
(i) the proliferation of two-body resonances \cite{Basko2006_BAA, Altshuler_1997, Nandkishore2014_drewrahul} and 
(ii) the run-away of avalanches \cite{DeRoeck2017_Avalanche,Crowley2019_Avalanche}. 
We find that the latter is marginal.
%
%We relate our observations to the phase transition by considering the effect of an explicitly symmetry breaking symmetry field....
Finally, we propose and analyze an experimental platform capable of directly exploring the emergence of ergodicity at the transition between MBL phases. 
Our proposal is motivated by recent advances in Rydberg-dressed, neutral-atom quantum simulators \cite{ Balewski2014_Dressing,Bloch2016_2DMBL, Bloch2017_Dressing, Lukin2017_tweezer, Cooper2018_AlkalineEarth, Leseleuc_2019, Thompson2019_AlkalineEarth, Madjarov_2020}; we demonstrate that the phase diagram depicted in Fig.~\ref{fig:fig1} can be directly probed via quench dynamics of local observables within experimental decoherence time-scales.

%To probe the stability of localization between two distinct MBL phases, we start by considering a paradigmatic model capable of exhibiting two such phases -- the interacting, disordered spin-1/2 transverse field Ising chain \cite{RSRGX, HuseLPQO, KjallMBLSG}.
Let us start by considering the paradigmatic example of a disordered one dimensional spin chain, which hosts two distinct MBL phases:
\begin{equation}\label{eq-SGModel}
H =  \sum_{i} J_i \sigma_i^z \sigma_{i+1}^z +\sum_{i} h_i \sigma_i^x + \sum_{i} V_i (\sigma_i^x\sigma_{i+1}^x + \sigma_i^z\sigma_{i+2}^z),
\end{equation}
where $\vec{\sigma}$ are Pauli operators and all coupling strengths are disordered, with $J_i \in [ -W_J, W_J]$, $h_i \in [-W_h, W_h]$, and $V_i \in [-W_V, W_V]$ \cite{ft2}. % \footnote{zz nnn and xx nn basically the same, self dual allows us to keep track of the critical point but we break this and the analysis is more complicated but all conclusions remain unchanged.}.
We choose to work with the normalization $\sqrt{W_J W_h} = 1$ and perform extensive exact
diagonalization studies up to system size $L=16$ \cite{SM}.
In the absence of $V_i$, the system reduces to the non-interacting, Anderson localized limit and for sufficiently strong disorder (in $J_i$ and $h_i$), this localization persists in the presence of interactions.
% \by{Should comment on the duality} \footnote{dual on average, we know where the critical point even when we turn on interactions}

%At large disorder (where MBL is stable), the two phases are defined with respect to the $\mathbb{Z}_2$ symmetry generated by a global spin-flip $G = \prod_i \sigma_i^x $---

%these correspond to the MBL paramagnet (PM) and the MBL spin-glass (SG).
%The properties of these two phases can be easily understood in the $\ell$-bit picture of MBL (Fig.~\ref{fig:Phase_Diagram}a).
% In order to gain some intuition as to the nature. of these phases and in what limits they arise, we examine the structure of the model's ``$\ell$-bits'' in some illuminating limits (See the cartoon above Fig.~\ref{fig:Phase_Diagram}). 

The Hamiltonian (Eqn.~1) exhibits a $\mathbb{Z}_2$ symmetry corresponding to a global spin-flip, $G = \prod_i \sigma_i^x$.
In the many-body localized regime, two distinct forms of eigenstate order emerge with respect to the breaking of this symmetry. 
For $W_h \gg W_J, W_V$, the transverse field dominates and the system is in the MBL paramagnetic (PM) phase.
The conserved $\ell$-bits simply correspond to dressed versions of the physical $\sigma_i^x$ operators. 
%
%There are $L$ such $\tau_i$ which are quasi-local and commute with each other and their expectation values index each eigenstate.
%
For $W_J \gg W_h, W_V$, the Ising interaction dominates and the eigenstates correspond to ``cat states'' of  spin configurations in the $\hat{z}$ direction.
Physical states break the associated $\mathbb{Z}_2$ symmetry, the $\ell$-bits are dressed versions of $\sigma^z_i\sigma^z_{i+1}$, and the system is in the so-called MBL spin-glass (SG) phase \cite{Huse2013_LPQO, Kjall2014_MBLOrder}.

%and the system is in the SG phase.
%Analogous to the PM phase, the $\tau_i$ are dressed versions of $\sigma^z_i\sigma^z_{i+1}$.
%Unlike the PM phase, there are only $L-1$ such $\tau_i$; the missing degree of freedom gives rise to a two-fold degeneracy (in the thermodynamic limit) of the above mentioned cat states. 

These two types of eigenstate order can be distinguished via the Edwards-Anderson order parameter which probes the presence of long-range Ising correlations in eigenstates $\ket{n}$, 
$\chi =  \left \llangle L^{-1} \sum_{i, j} \bra{n} \sigma_i^z \sigma_j^z \ket{n}^2  \right \rrangle$, where $\llangle \cdots \rrangle$ denotes averaging over disorder realizations \cite{Kjall2014_MBLOrder, Vasseur2016_PHS}.
In the SG phase, this order parameter scales extensively with system size, $\chi \propto L$, while in the PM phase, it  approaches a constant $\mathcal{O}(1)$ value.
Fixing $W_V = 0.7$, $\chi$ exhibits a clear transition from PM to SG as one tunes the ratio of $W_J / W_h$ (Fig.~2a). 
The finite-size flow of $\chi$ is consistent with the presence of a \emph{single} critical point at $W_J = 3.2, W_h =0.32$ ($W_J/W_h \approx 10$).

However, thermalization diagnostics tell a different story. 
In particular, we compute the $\langle r \rangle$-ratio, a measure of the rigidity of the many-body spectrum: $\langle r \rangle = \left\llangle \text{min}\{\delta_n, \delta_{n+1}\}/\text{max}\{\delta_n, \delta_{n+1}\}\right\rrangle$, where $\delta_n = E_{n+1} - E_{n}$, $E_n$ is the $n^\textrm{th}$ eigenenergy and averaging is also done across the entire many-body spectrum \cite{Oganesyan_2007, Pal_2010}.
In the MBL phase, energy levels exhibit Poisson statistics with $\langle r \rangle \approx 0.39$, while in the ergodic phase, level repulsion leads to the  GOE expectation $\langle r \rangle  \approx 0.53$ \cite{Abanin2019_Rev, Nandkishore2015_Rev, Srednicki1994_ChaosMBL}. 
Unlike $\chi$, which exhibits a single transition, the $\langle r \rangle$-ratio exhibits two distinct critical points, each characterized by its own finite-size flow (Fig.~\ref{fig-Data}b).
This demarcates three distinct phases: two many-body localized phases (for $W_J/W_h \lesssim 0.1$ and $W_J/W_h \gtrsim 10$) separated by an intervening thermal phase. 
Interestingly, the location of the thermal-MBL transition at $W_J/W_h \approx 10$ matches the location of the spin-glass transition observed via $\chi$.
The fact that an additional thermal-MBL transition is observed in the $\langle r \rangle$-ratio, but not in  $\chi$, suggests that the PM regime has slightly more structure. 

In order to further probe this structure, we turn to the half-chain entanglement entropy, $S_{L/2} = -\text{Tr}[ \rho_{\mathrm{s}} \log(\rho_{\mathrm{s}})]$, where $\rho_{\mathrm{s}} = \text{Tr}_{i\le L/2}[ \ket{n}\bra{n}]$.
The behavior of $S_{L/2}$, illustrated in Fig.~\ref{fig-Data}c, allows us to clearly distinguish three phases: the MBL paramagnet, the thermal paramagnet, and the MBL spin-glass. 
%
%\crl{Removed asymptotic $W_J/W_h$ values of $S_{L/2}$}
For $W_J/W_h \ll 0.1$, the eigenstates are close to product states and the entanglement entropy $S_{L/2}$ is independent of $L$, consistent with a localized paramagnet. %[1overL correction]
Near $W_J/W_h \approx 1$, $S_{L/2}$ increases with system size, approaching $(L\log 2-1)/2$, consistent with a thermal paramagnet. %
Finally, for $W_J/W_h \gg 10$, the half-chain entanglement again becomes independent of $L$ and, for very large $W_J/W_h$, approaches $\log 2$, consistent with the cat-state-nature of  eigenstates in the MBL SG phase.

A few remarks are in order. 
First, the variance of $S_{L/2}$ provides a complementary diagnostic to confirm the presence of two distinct thermal-MBL transitions (Fig.~\ref{fig-Data}d). 
Indeed, one observes two well-separated peaks in $\textrm{var}(S_{L/2})$, whose locations are consistent with the transitions found in the $\langle r \rangle$-ratio.
Second, although $\chi$ only scales with system size in the SG phase, one expects its behavior to be qualitatively different in the MBL versus thermal paramagnet. 
In particular, in the MBL paramagnet, the $\ell$-bits have a small overlap with $\sigma^z_i\sigma^z_j$ and one expects $\chi > 1$; meanwhile, in the thermal paramagnet (at infinite temperature) one expects $\chi \to 1$ rapidly with increasing system size. 
This is indeed borne out by the numerics, as shown in the inset of Fig.~\ref{fig-Data}a.
%
%Moreover, this cross-over behavior in $\chi$

\begin{figure}
    \centering
    \hspace{-5mm}\includegraphics[width = 3.5in]{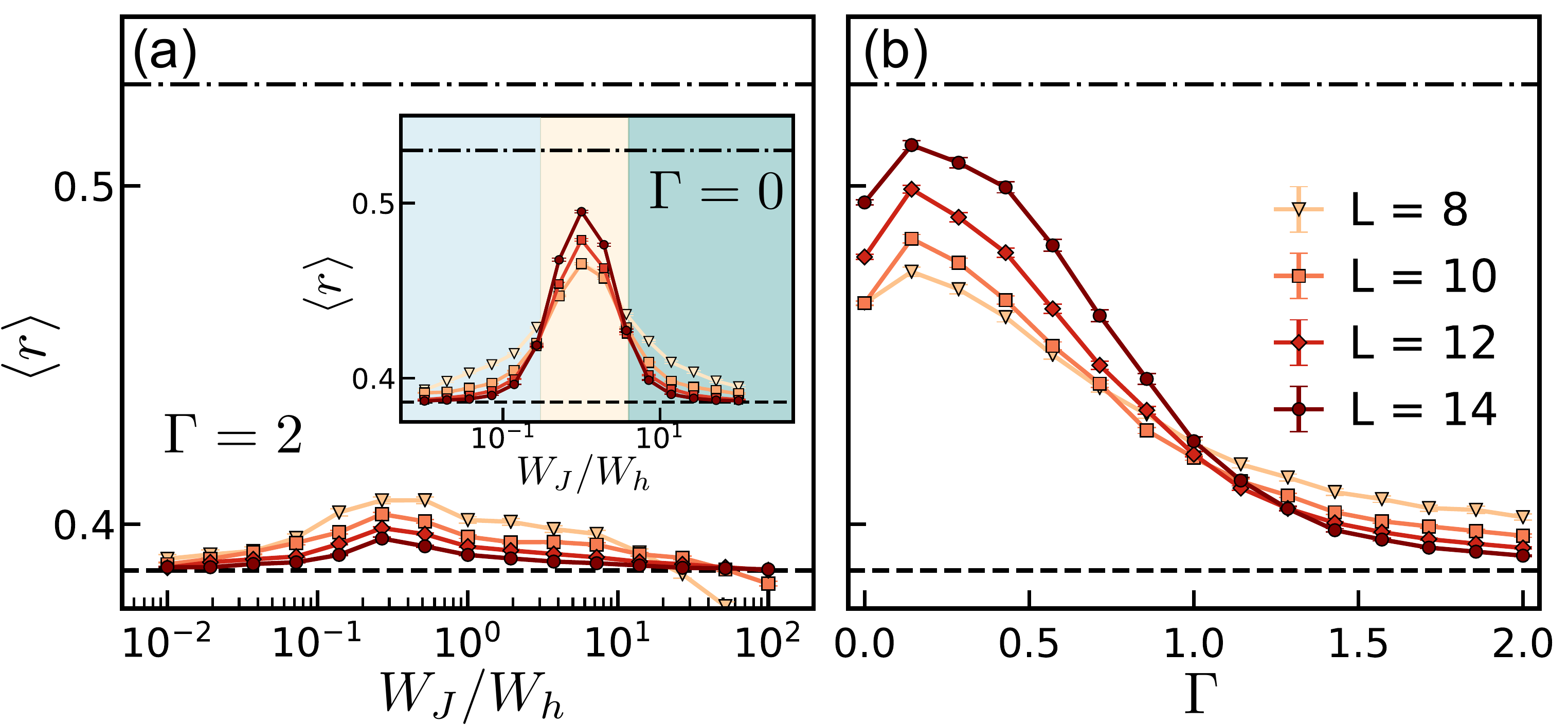}
    \caption{
        (a) $\langle r \rangle$-ratio as a function of $W_J/W_h$ at $W_V = 0.3$ in the presence of an explicit symmetry breaking field $\Gamma=2$.
        The dash-dotted [dashed] line corresponds to the GOE [Poisson] expectation.
        Unlike the symmetry respecting case ($\Gamma=0$, inset), the system remains localized for all values of $W_J/W_h$.
        (b) Within the thermal region (here with $W_J/W_h=1$), an increasing symmetry-breaking field drives the system towards localization. Each data point corresponds to averaging over at least $3 \cdot 10^2$ disorder realizations.
    }
    \label{fig:my_label}
\end{figure}

%Interestingly, this cross-over matches with the transition observed in $\langle r\rangle$-ration and $S_{L/2}$.

Diagnostics in hand, we now construct the full phase diagram as a function of $W_V$ and $W_J/W_h$ (Fig.~\ref{fig:fig1}a). 
Even for the smallest interaction strengths accessible, $W_V\sim 0.07$, one observes a finite width region where the $\langle r \rangle$-ratio increases with system size \cite{SM}. 
Although clearly indicative of a thermal intrusion, it is possible that our analysis underestimates the size of the intervening ergodic phase \cite{Abanin2019_FiniteSize, Panda2020_FiniteSize, Papic2019_Miniband}.
%we note that it is difficult to use the $\langle r \rangle$-ratio to quantitatively bound the region of this intrusion;
%
Extrapolating toward $W_V = 0$, our phase diagram suggests the presence of a finite-width thermal region between the two MBL phases, which terminates at the non-interacting critical point (Fig.~\ref{fig:fig1}a).  
%\ny{approach to Wv0 and how it should scale?}

In order to verify that the presence of a phase transition is indeed responsible for the intervening ergodic region, one can explicitly break the $\mathbb{Z}_2$ symmetry in Eqn.~\ref{eq-SGModel}. 
We do so by adding a \emph{disorder-less}, on-site longitudinal field, $ \Gamma \sum_i \sigma_i^z$.
Despite the fact that the field is uniform, it causes the $\langle r \rangle$-ratio to systematically decrease (Figs.~\ref{fig:my_label}a,b), and for a sufficiently large symmetry breaking field, all finite-size flow tends toward localization.
This allows us to construct the  phase diagram in the presence of finite $\Gamma$, as depicted in  Fig.~\ref{fig:fig1}b.

To understand the generality of an emergent ergodic region between many-body localized phases, we now consider two additional types of MBL transitions: a symmetry-protected topological (SPT) transition and
a discrete time-crystalline (DTC) transition. 
% \fm{From discussion with Chris, shouldn't we state that these correspond to representative examples of all possible continuous phase transitions?}
%
The Hamiltonian of the SPT model is given by \cite{ft3}:
\begin{align}\label{eq-SPTModel}
\begin{split}
  H_{\mathrm{SPT}} = \sum_i J_i \sigma_{i-1}^z \sigma_i^x \sigma_{i+1}^z + \sum_i h_i \sigma_i^x\\
  + \sum_i V_i(\sigma_i^x \sigma_{i+1}^x + \sigma_{i-1}^{z} \sigma_i^y \sigma_{i+1}^y \sigma_{i+2}^z)~,
\end{split}
\end{align}
with $J_i \in [ -W_J, W_J]$, $h_i \in [-W_h, W_h]$, and $V_i \in [-W_V, W_V]$.
$H_{\mathrm{SPT}}$ exhibits a $\mathbb{Z}_2 \times \mathbb{Z}_2$ symmetry, which gives rise to an MBL SPT (Haldane) phase for $W_J \gg W_h, W_V$ and a topologically-trivial MBL phase for $W_J \ll W_h, W_V$ \cite{Bahri2015_MBLSPT, Chandran2014_MBLSPT}.
For the DTC model, we consider a stroboscopic Floquet system:
\begin{equation}\label{eqn-DTC}
    H_F(t) = \begin{cases}\sum_i J_i \sigma_i^z \sigma_{i+1}^z + h_i \sigma_i^x + V_i \sigma_i^z & t \in[0, T/2) \\-\frac{\pi}{T}\sum_i \sigma_i^x & t \in [T/2,T)  \end{cases}
\end{equation}
where $J_i \in [0.5, 1.5]$, $T=2$, $h_i \in [0, h]$ and $V_i \in [0, 2V]$. 
%
%This Hamiltonian exhibits a discrete time-translation symmetry with a period of $T$; we set $T=2$ without loss of generality. 
When $h \ll 1$, the Floquet system spontaneously breaks time-translation symmetry and is in the so-called DTC phase, while for $h \gg 1$, the system is in a Floquet paramagnetic phase \cite{Else2016_DTC, Yao2016_DTC, Khemani2016_EmerErg}.
We analyze each of these models using the four diagnostics previously described: (i) the order parameter, (ii) the $\langle r \rangle$-ratio, (iii) the half-chain entanglement, and (iv) the variance, $\textrm{var}(S_{L/2})$.
We observe the same qualitative behavior for both transitions across all diagnostics: An intervening ergodic phase emerges which terminates at the non-interacting critical point.
This is illustrated in Figs.~\ref{fig-Data}e,f for both the SPT model and the DTC model using $S_{L/2}$; all additional data for the different diagnostics can be found in the supplemental material \cite{SM}.
We further analyze the finite-size effects arising from small couplings \cite{SM}, which we believe underlie previous numerical observations of apparent direct transitions \cite{Yao2016_DTC, Friedman2018_EmerErg, Vasseur2016_PHS, Venderley2016_EmerErg}.

%uch systems have recently emerged as powerful platforms for building up many-body quantum systems atom-by-atom. 

%Here, we envision the effective spin degree of freedom in Eq. (1) to be formed by two hyperfine atomic ground states. The most natural Hamiltonian available in such a system is a long-range transverse field Ising model:

%The experimental observation of the phenomena described above is most naturally realized in trapped neutral alkali or alkaline-earth atoms \cite{}.
%
%In such systems, strong interactions arise from the manipulation of associated Rydberg states, enabling the simulation of the coherent dynamics of strongly interacting many-body systems in the context of optical lattices or optical tweezer arrays \cite{}.
%
%Indeed, by rapidly toggling different local optical fields, one can generate an effective Hamiltonian that closely resembles Eq.~\eqref{eq-SGModel}.

\begin{figure}
    \centering
    \includegraphics[width =3.2in]{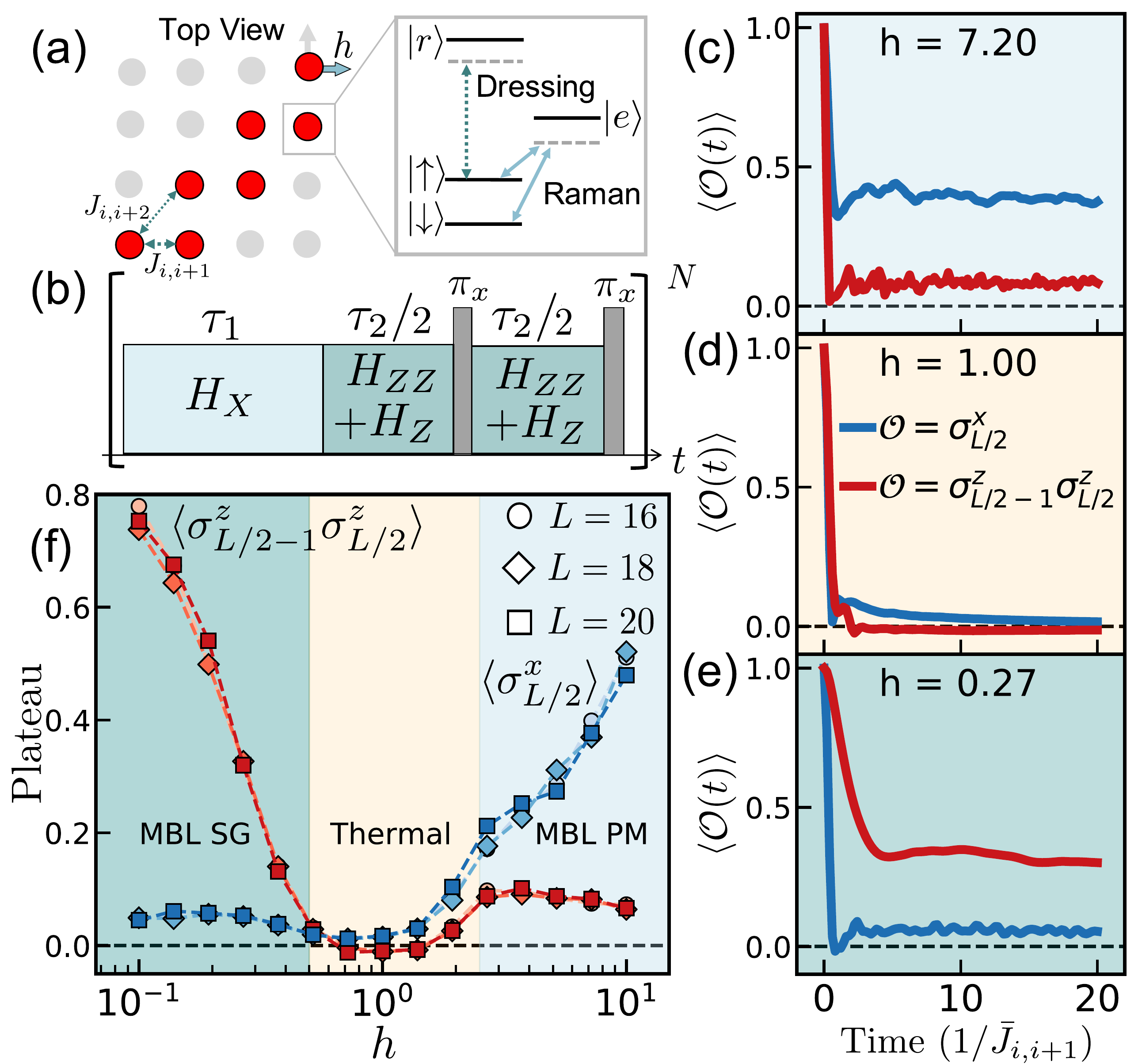} 
    \caption{
    (a) Schematic of the proposed experimental protocol. 
    Within an optical lattice, neutral atoms are prepared along two adjacent diagonals (i.e.~with a gas microscope), defining a zig-zag spin chain configuration.
    Dressing with a Rydberg state $\ket{r}$ leads to $H_{ZZ} + H_Z$, while a two-photon Raman transition mediated by an excited state $\ket{e}$ leads to $H_{X}$.
    (b) By combining rapid spin echo pulses with Floquet evolution under $H_{X}$ and $H_{ZZ}+H_Z$, one can  engineer $H_{\eff}$ (Eqn.~\ref{eq:Exp}).
    (c-e) Dynamics of $\sigma^x_{L/2}$ (blue) and $\sigma^z_{L/2-1}\sigma^z_{L/2}$ (red) under $H_\eff$ starting with initial states $\ket{\psi_x}$ and $\ket{\psi_{zz}}$, respectively.
    Different panels correspond to representative behaviors for the three distinct phases (tuned via $h$).
    (f) The height of the late-time plateau  distinguishes between the three phases. Each data point corresponds to averaging over at least $10^2$ disorder realizations.
    }
    \label{fig:Experiment}
\end{figure}

\textit{Experimental Realization}.---Motivated by recent advances in the characterization and control of Rydberg states, we propose an experimental protocol to directly explore the emergence of ergodicity between MBL phases.
Our protocol is most naturally implemented in one dimensional chains of either alkali or alkaline-earth atoms \cite{ Balewski2014_Dressing,Bloch2016_2DMBL, Bloch2017_Dressing, Lukin2017_tweezer, Cooper2018_AlkalineEarth, Leseleuc_2019, Thompson2019_AlkalineEarth, Madjarov_2020}. 
To be specific, we consider $^{87}\rm Rb$ with an effective spin-1/2 encoded in hyperfine states:~$\ket{\downarrow} = \ket{F=1,m_F=-1}$ and $\ket{\uparrow} = \ket{F=2,m_F=-2}$.
Recent experiments have demonstrated the ability to generate strong interactions via either Rydberg-dressing in an optical lattice (where atoms are typically spaced by $\sim 0.5~\mathrm{\mu m}$) or via Rydberg-blockade in a tweezer array (where atoms are typically spaced by $\sim 3~ \mathrm{\mu m}$) \cite{Balewski2014_Dressing,Bloch2016_2DMBL, Bloch2017_Dressing, Lukin2017_tweezer, Cooper2018_AlkalineEarth, Leseleuc_2019, Thompson2019_AlkalineEarth, Madjarov_2020}.
%Yao2017_DressingTheory, Zoller2015_DressingTheory
%Lukin2017_tweezer, , Balewski2014_Dressing
%
Focusing on the optical lattice setup, dressing enables the generation of tunable, long-range soft-core Ising interactions, $H_{ZZ} = \sum_{i,j} J_{ij}\sigma_i^z \sigma_j^z$, with a spatial profile that interpolates between a constant at short distances (determined by the blockade radius) and a $1/r^6$ van der Waals tail.
A particularly simple implementation of a PM-SG Hamiltonian (closely related to Eqn.~1)  is to alternate time evolution under  $H_{ZZ}$ and $H_X = \sum_i h_i \sigma^x_i$, with the latter being implemented via a two-photon Raman transition (Fig.~\ref{fig:Experiment}a). 
In the high frequency limit, the dynamics are governed by an effective Hamiltonian:
\begin{align} \label{eq:Exp}
    H_{\eff} = \frac{\tau_1}{\tau_1 + \tau_2}\sum_i h_i \sigma_i^x  + \frac{\tau_2}{\tau_1 + \tau_2}\sum_{ij} J_{ij} \sigma_i^z \sigma_{j}^z ~,
\end{align}
where $H_X$ is applied for time $\tau_1$, $H_{ZZ}$ is applied for time $\tau_2$, and the Floquet frequency $\omega = 2\pi/(\tau_1 + \tau_2) \gg h_i, J_{ij}$.
This latter inequality ensures that both Floquet heating and higher-order corrections to $H_\eff$ can be safely neglected on experimentally relevant time-scales \cite{Abanin2015_Heating, Machado2019_Heating, SM}.
Note that unlike the DTC model (Eqn.~3), here  Floquet engineering is being used to emulate a static MBL PM-SG Hamiltonian \cite{Yao2017_DressingTheory}.

A few remarks are in order. 
First, by applying the Rydberg dressing to only one of the two hyperfine  states (Fig.~\ref{fig:Experiment}a), an additional longitudinal field $H_Z \propto \sigma^z_i$ is naturally generated. 
To restore the $\mathbb{Z}_2$ symmetry, one can exactly cancel this field by embedding a spin echo  into the Floquet sequence (Fig.~\ref{fig:Experiment}b).
In addition, varying the spacing between the echo $\pi$-pulses (Fig.~\ref{fig:Experiment}b) directly controls the degree of cancellation, enabling one to   experimentally  probe the effect of an explicit symmetry breaking field. 
%
%Second, there are a number of experimentally demonstrated methods to implement disorder in $h_i$ and $J_{ij}$.
%
%\ny{Immanuel: hx need Raman disorder. Jij need disorder in the dressing, can be achieved by spatially varying Rabi or detuning? DMD?? Currently no DMD for UV light? Is there a better way?}
%
Second, although our prior analysis has focused on eigenstate properties, these are inaccessible to experiment. 
Fortunately, as we will demonstrate, the phase diagram can also be characterized via the dynamics of local observables. 
The intuition behind this is simple: observables that overlap with an $\ell$-bit exhibit a plateau at late times. 

To investigate this behavior, we use Krylov subspace methods \cite{dynamite, SLEPc1, SLEPc2, Parallel} to numerically simulate the dynamics of $H_{\textrm{eff}}$ with $\tau_1=\tau_2=1$, $J_{i,i+1} \in [-1, -3]$, $J_{i,i+2} = 0.6J_{i,i+1}$ and $h_i \in [h, 3h]$.
We note that the ratio of the nearest- to next-nearest-neighbor coupling strength is chosen based upon the experimentally measured Rydberg-dressing-interaction profile and a 1D zig-zag chain geometry (Fig.~\ref{fig:Experiment}a)  \cite{Bloch2017_Dressing,Gross2017_GasMicroscope,ft1}. 
%

% \fm{supp info: and such that the local observable take a maximal initial value. }

% Norm's sentence
% For system sizes up to $L=20$, we compute the dynamics of two local observables, $\sigma^x_{L/2}$ and $\sigma^z_{L/2}\sigma^z_{L/2+1}$, starting from two different initial product states $\ket{\psi_x}$ and $\ket{\psi_{zz}}$; both states are easily preparable in experiment, are close to zero energy density, and $\ket{\psi_x}$ is chosen to have expectation value $\bra{\psi_x} \sigma^x_{L/2} \ket{\psi_x} = 1$, while $\ket{\psi_{zz}}$ is chosen to expectation value $\bra{\psi_{zz}} \sigma^z_{L/2}\sigma^z_{L/2+1} \ket{\psi_{zz}} = 1$.

For system sizes up to $L=20$, we compute the dynamics of initial states $\ket{\psi_x}$ and $\ket{\psi_{zz}}$ \cite{ft4};
both states are easily preparable in experiment, close to zero energy density, and chosen such that $\bra{\psi_x}\sigma^x_{L/2}\ket{\psi_x} = 1$ and $\bra{\psi_{zz}}\sigma^z_{L/2-1}\sigma^z_{L/2}\ket{\psi_{zz}} = 1$.
Starting with $\ket{\psi_x}$ as our initial state and large $h$, we observe that  $\langle \sigma^x_{L/2} (t) \rangle$ plateaus to a finite value at late-times, indicating the system is in the MBL PM phase (Fig.~\ref{fig:Experiment}c).
Analogously, for $\ket{\psi_{zz}}$  and small $h$, we observe that  $\langle \sigma^z_{L/2-1}(t) \sigma^z_{L/2} (t) \rangle$ plateaus to a finite value at late-times, indicating the system is in the MBL SG phase (Fig.~\ref{fig:Experiment}e).
For $h\sim 1$, \emph{both} observables decay to zero, indicating the system is the thermal phase (Fig.~\ref{fig:Experiment}d).
%
% \fm{To quantify these observations, we consider a phenomenological decay functional form $(1-A) e^{-t/T} + A$ and extract value of the plateau $A$ for both operators as a function of $h$.}
%As depicted in Fig.~\ref{fig:Experiment}d, by plotting the plateau value of the two observables as a function of $h$, one can clearly identify the intervening ergodic region. 
The plateau value of the two observables as a function of $h$ clearly identifies the intervening ergodic region (Fig.~\ref{fig:Experiment}f).

%
%Practically, to extract the late-time offset from the dynamics of the chosen local observables by fitting the data to $(1-A) e^{-t/T} + A$ and extract value of the plateau $A$ for both operators as a function of $h$.

To ensure that one can observe the intervening thermal phase within experimental coherence times, we now estimate the  time-scales necessary to carry out our protocol. 
Previous experiments using Rydberg dressing have demonstrated coherence times $T_2 \sim 1~\mathrm{ms}$, with nearest neighbor couplings $J_{i,i+1} \sim (2\pi) \times 13~\mathrm{kHz}$ %transverse fields $h_i \sim (2\pi) \times 20~\mathrm{kHz}$, 
and a microwave-induced $\pi$-pulse duration $\sim 25~\mathrm{\mu s}$ \cite{Bloch2017_Dressing}.
Taken together, this leads to an estimate of  $\sim 55~\mathrm{\mu s}$ for the Floquet period (Fig.~4b).
Crucially, within $T_2$ (i.e.$\sim 20$ Floquet cycles), all observables approach their late-time plateaus~\cite{SM}.
\textit{Discussion and outlook}.---We conclude by discussing previous analytical results and how they may shed light on the origins of the intervening thermal phase.
In the absence of interactions, the Hamiltonian transitions we consider all fall into  infinite-randomness universality classes characterized by both a divergent single-particle density of states (DOS, $D(\varepsilon) \sim |\varepsilon \log^3 \varepsilon |^{-1}$ near zero single-particle energy $\varepsilon$) and single-particle orbitals with diverging mean and typical localization lengths ($\xi_{\rm mean}\sim |\log^2\varepsilon|$ and $\xi_{\rm typ}\sim |\log \varepsilon |$ respectively) \cite{Evers2008_AndersonTransitions,Fisher1995_RTFIM,Balents1997_SUSY,McCoy1968_TFIM}.
These divergences suggest that two-body resonances might directly destabilize MBL upon the introduction of interactions; however, a simple counting of resonances in typical blocks does not produce such an instability: In a block of length $l$, there are $l N(\varepsilon)$ ``active'' single particle orbitals with $\xi_{\rm typ}(\varepsilon) \ge l$, where $N(\varepsilon) = \int^\varepsilon d\varepsilon'\ D(\varepsilon')$ is the integrated DOS \cite{Nandkishore2014_drewrahul,YaoLau2014_PowerLaw,SM}. 
These orbitals overlap in real space and are thus susceptible to participating in perturbative two-body resonances. 
A perturbative instability of the localized state arises if $l N$ diverges as $\varepsilon \to 0$; even for arbitrarily small interactions, a large network of resonant pairs can be found at low enough energy.
Using the DOS and localization lengths of the infinite-randomness transition, we find $l N \sim 1/|\log \varepsilon|$ which vanishes slowly as $\varepsilon \to 0$.

Alternatively, one might consider the susceptibility to `avalanches'  due to rare thermal bubbles induced by the interactions \cite{Luitz2017_Avalanche,Thiery2018_Avalanche, DeRoeck2017_Avalanche}.
For a system with a distribution of localization lengths, it has recently been shown that the \emph{average} localization length controls this instability \cite{Crowley2019_Avalanche}: for $\overline{\xi} > 2/\log 2 $, thermal bubbles avalanche.
However, this is within a model where the orbitals have a single localization center. 
Near the infinite-randomness transition, the orbitals have two centers whose separation is controlled by $\xi_{\textrm{mean}}$ but whose overlap onto a putative thermal bubble is controlled by $\xi_{\textrm{typ}}$. 
Thus, while $\overline{\xi_{\mathrm{mean}}}$ diverges logarithmically, the more appropriate $\overline{\xi_{\mathrm{typ}}}$ remains finite and this criterion does not produce an absolute instability \cite{SM}.
We highlight that it is only a logarithmic correction which causes the convergence of the average localization length; 
unaccounted channels might provide an additional logarithm leading to an absolute avalanche instability. We leave this to future work.

Finally, let us note that the direct numerical observation of avalanche instabilities remains extremely challenging \cite{DeRoeck2017_Avalanche, Potirniche_2019}; 
the presence of a robust intervening thermal region in our study suggests that an alternate mechanism might be at the heart of our observations.
%
% %An interesting future direction could be to verify this by performing the same numerical study in the presence of quasi-periodic disorder where avalanche effects are not believed to exist.

% \rs{Professor Laumann: Would it be useful to include a sentence about how there are no disordered first-order phase transition? Also perhaps we should mention 'truly' interacting MBL that does not smoothly connect to any AL phases. Additionally, I found a paper regarding PHS localization https://arxiv.org/pdf/1510.04282.pdf. Is this something we want to include? They try to argue that interactions are irrelevant at high disorder but there claims are contingent on employing a model with a U(1) symmetry.}

\emph{Note added:} During the completion of this work, we became aware of complementary work on the presence of intervening thermal phases between MBL transitions \cite{sanjay2020} which will appear in the same arXiv posting.

\begin{acknowledgements}
\textit{Acknowledgements}---We gratefully acknowledge discussions with Ehud Altman, Anushya Chandran, Soonwon Choi, Phillip J. D. Crowley, Simon Hollerith, David Huse, Gregory D. Kahanamoku-Meyer  and Antonio Rubio-Abadal. 
We thank Immanuel Bloch for detailed comments on a draft. 
Krylov subspace numerics are performed using the “dynamite” package \cite{dynamite}, a \texttt{PYTHON} wrapper for Krylov subscape methods based upon the SLEPc/PETSC library.
This work was supported by the NSF (QII-TAQS program and grant PHY-1654740) and the DOE (DE-SC0019241).
C.R.L. acknowledges support from the NSF through grant PHY-1752727.
R.S. acknowledges support from the Barry M. Goldwater Scholarship, the Berkeley Physics Undergraduate Research Scholarship, and UC Berkeley's Summer Undergraduate Research Fellowship.
\end{acknowledgements}

\bibliographystyle{apsrev4-1} % Tell bibtex which bibliography style to use
\bibliography{refs} % Tell bibtex which .bib file to use (this one is some example file in TexLive's file tree)

%merlin.mbs apsrev4-1.bst 2010-07-25 4.21a (PWD, AO, DPC) hacked
%Control: key (0)
%Control: author (72) initials jnrlst
%Control: editor formatted (1) identically to author
%Control: production of article title (-1) disabled
%Control: page (0) single
%Control: year (1) truncated
%Control: production of eprint (0) enabled
\begin{thebibliography}{70}%
\makeatletter
\providecommand \@ifxundefined [1]{%
 \@ifx{#1\undefined}
}%
\providecommand \@ifnum [1]{%
 \ifnum #1\expandafter \@firstoftwo
 \else \expandafter \@secondoftwo
 \fi
}%
\providecommand \@ifx [1]{%
 \ifx #1\expandafter \@firstoftwo
 \else \expandafter \@secondoftwo
 \fi
}%
\providecommand \natexlab [1]{#1}%
\providecommand \enquote  [1]{``#1''}%
\providecommand \bibnamefont  [1]{#1}%
\providecommand \bibfnamefont [1]{#1}%
\providecommand \citenamefont [1]{#1}%
\providecommand \href@noop [0]{\@secondoftwo}%
\providecommand \href [0]{\begingroup \@sanitize@url \@href}%
\providecommand \@href[1]{\@@startlink{#1}\@@href}%
\providecommand \@@href[1]{\endgroup#1\@@endlink}%
\providecommand \@sanitize@url [0]{\catcode `\\12\catcode `\$12\catcode
  `\&12\catcode `\#12\catcode `\^12\catcode `\_12\catcode `\%12\relax}%
\providecommand \@@startlink[1]{}%
\providecommand \@@endlink[0]{}%
\providecommand \url  [0]{\begingroup\@sanitize@url \@url }%
\providecommand \@url [1]{\endgroup\@href {#1}{\urlprefix }}%
\providecommand \urlprefix  [0]{URL }%
\providecommand \Eprint [0]{\href }%
\providecommand \doibase [0]{http://dx.doi.org/}%
\providecommand \selectlanguage [0]{\@gobble}%
\providecommand \bibinfo  [0]{\@secondoftwo}%
\providecommand \bibfield  [0]{\@secondoftwo}%
\providecommand \translation [1]{[#1]}%
\providecommand \BibitemOpen [0]{}%
\providecommand \bibitemStop [0]{}%
\providecommand \bibitemNoStop [0]{.\EOS\space}%
\providecommand \EOS [0]{\spacefactor3000\relax}%
\providecommand \BibitemShut  [1]{\csname bibitem#1\endcsname}%
\let\auto@bib@innerbib\@empty
%</preamble>
\bibitem [{\citenamefont {Gornyi}\ \emph {et~al.}(2005)\citenamefont {Gornyi},
  \citenamefont {Mirlin},\ and\ \citenamefont
  {Polyakov}}]{Gornyi_Mirlin_Polyakov_2005}%
  \BibitemOpen
  \bibfield  {author} {\bibinfo {author} {\bibfnamefont {I.~V.}\ \bibnamefont
  {Gornyi}}, \bibinfo {author} {\bibfnamefont {A.~D.}\ \bibnamefont {Mirlin}},
  \ and\ \bibinfo {author} {\bibfnamefont {D.~G.}\ \bibnamefont {Polyakov}},\
  }\href {\doibase 10.1103/PhysRevLett.95.206603} {\bibfield  {journal}
  {\bibinfo  {journal} {Physical Review Letters}\ }\textbf {\bibinfo {volume}
  {95}},\ \bibinfo {pages} {206603} (\bibinfo {year} {2005})}\BibitemShut
  {NoStop}%
\bibitem [{\citenamefont {Basko}\ \emph
  {et~al.}(2006{\natexlab{a}})\citenamefont {Basko}, \citenamefont {Aleiner},\
  and\ \citenamefont {Altshuler}}]{Basko2006_BAA}%
  \BibitemOpen
  \bibfield  {author} {\bibinfo {author} {\bibfnamefont {D.}~\bibnamefont
  {Basko}}, \bibinfo {author} {\bibfnamefont {I.}~\bibnamefont {Aleiner}}, \
  and\ \bibinfo {author} {\bibfnamefont {B.}~\bibnamefont {Altshuler}},\ }\href
  {\doibase https://doi.org/10.1016/j.aop.2005.11.014} {\bibfield  {journal}
  {\bibinfo  {journal} {Annals of Physics}\ }\textbf {\bibinfo {volume}
  {321}},\ \bibinfo {pages} {1126 } (\bibinfo {year}
  {2006}{\natexlab{a}})}\BibitemShut {NoStop}%
\bibitem [{\citenamefont {Basko}\ \emph
  {et~al.}(2006{\natexlab{b}})\citenamefont {Basko}, \citenamefont {Aleiner},\
  and\ \citenamefont {Altshuler}}]{Basko2006_2}%
  \BibitemOpen
  \bibfield  {author} {\bibinfo {author} {\bibfnamefont {D.~M.}\ \bibnamefont
  {Basko}}, \bibinfo {author} {\bibfnamefont {I.~L.}\ \bibnamefont {Aleiner}},
  \ and\ \bibinfo {author} {\bibfnamefont {B.~L.}\ \bibnamefont {Altshuler}},\
  }\href {http://arxiv.org/abs/cond-mat/0602510} {\bibfield  {journal}
  {\bibinfo  {journal} {arXiv:cond-mat/0602510}\ } (\bibinfo {year}
  {2006}{\natexlab{b}})},\ \bibinfo {note} {arXiv:
  cond-mat/0602510}\BibitemShut {NoStop}%
\bibitem [{\citenamefont {Nandkishore}\ and\ \citenamefont
  {Huse}(2015)}]{Nandkishore2015_Rev}%
  \BibitemOpen
  \bibfield  {author} {\bibinfo {author} {\bibfnamefont {R.}~\bibnamefont
  {Nandkishore}}\ and\ \bibinfo {author} {\bibfnamefont {D.~A.}\ \bibnamefont
  {Huse}},\ }\href {\doibase 10.1146/annurev-conmatphys-031214-014726}
  {\bibfield  {journal} {\bibinfo  {journal} {Annual Review of Condensed Matter
  Physics}\ }\textbf {\bibinfo {volume} {6}},\ \bibinfo {pages} {15} (\bibinfo
  {year} {2015})}\BibitemShut {NoStop}%
\bibitem [{\citenamefont {Vosk}\ \emph {et~al.}(2015)\citenamefont {Vosk},
  \citenamefont {Huse},\ and\ \citenamefont {Altman}}]{vosk2015theory}%
  \BibitemOpen
  \bibfield  {author} {\bibinfo {author} {\bibfnamefont {R.}~\bibnamefont
  {Vosk}}, \bibinfo {author} {\bibfnamefont {D.~A.}\ \bibnamefont {Huse}}, \
  and\ \bibinfo {author} {\bibfnamefont {E.}~\bibnamefont {Altman}},\
  }\href@noop {} {\bibfield  {journal} {\bibinfo  {journal} {Physical Review
  X}\ }\textbf {\bibinfo {volume} {5}},\ \bibinfo {pages} {031032} (\bibinfo
  {year} {2015})}\BibitemShut {NoStop}%
\bibitem [{\citenamefont {Abanin}\ \emph
  {et~al.}(2019{\natexlab{a}})\citenamefont {Abanin}, \citenamefont {Altman},
  \citenamefont {Bloch},\ and\ \citenamefont {Serbyn}}]{Abanin2019_Rev}%
  \BibitemOpen
  \bibfield  {author} {\bibinfo {author} {\bibfnamefont {D.~A.}\ \bibnamefont
  {Abanin}}, \bibinfo {author} {\bibfnamefont {E.}~\bibnamefont {Altman}},
  \bibinfo {author} {\bibfnamefont {I.}~\bibnamefont {Bloch}}, \ and\ \bibinfo
  {author} {\bibfnamefont {M.}~\bibnamefont {Serbyn}},\ }\href {\doibase
  10.1103/RevModPhys.91.021001} {\bibfield  {journal} {\bibinfo  {journal}
  {Rev. Mod. Phys.}\ }\textbf {\bibinfo {volume} {91}},\ \bibinfo {pages}
  {021001} (\bibinfo {year} {2019}{\natexlab{a}})}\BibitemShut {NoStop}%
\bibitem [{\citenamefont {Deutsch}(1991)}]{Deutsch_1991}%
  \BibitemOpen
  \bibfield  {author} {\bibinfo {author} {\bibfnamefont {J.~M.}\ \bibnamefont
  {Deutsch}},\ }\href {\doibase 10.1103/PhysRevA.43.2046} {\bibfield  {journal}
  {\bibinfo  {journal} {Physical Review A}\ }\textbf {\bibinfo {volume} {43}},\
  \bibinfo {pages} {2046–2049} (\bibinfo {year} {1991})}\BibitemShut
  {NoStop}%
\bibitem [{\citenamefont {Srednicki}(1994{\natexlab{a}})}]{Srednicki_1994}%
  \BibitemOpen
  \bibfield  {author} {\bibinfo {author} {\bibfnamefont {M.}~\bibnamefont
  {Srednicki}},\ }\href {\doibase 10.1103/PhysRevE.50.888} {\bibfield
  {journal} {\bibinfo  {journal} {Physical Review E}\ }\textbf {\bibinfo
  {volume} {50}},\ \bibinfo {pages} {888–901} (\bibinfo {year}
  {1994}{\natexlab{a}})}\BibitemShut {NoStop}%
\bibitem [{\citenamefont {Canovi}\ \emph {et~al.}(2011)\citenamefont {Canovi},
  \citenamefont {Rossini}, \citenamefont {Fazio}, \citenamefont {Santoro},\
  and\ \citenamefont {Silva}}]{canovi2011quantum}%
  \BibitemOpen
  \bibfield  {author} {\bibinfo {author} {\bibfnamefont {E.}~\bibnamefont
  {Canovi}}, \bibinfo {author} {\bibfnamefont {D.}~\bibnamefont {Rossini}},
  \bibinfo {author} {\bibfnamefont {R.}~\bibnamefont {Fazio}}, \bibinfo
  {author} {\bibfnamefont {G.~E.}\ \bibnamefont {Santoro}}, \ and\ \bibinfo
  {author} {\bibfnamefont {A.}~\bibnamefont {Silva}},\ }\href@noop {}
  {\bibfield  {journal} {\bibinfo  {journal} {Physical Review B}\ }\textbf
  {\bibinfo {volume} {83}},\ \bibinfo {pages} {094431} (\bibinfo {year}
  {2011})}\BibitemShut {NoStop}%
\bibitem [{\citenamefont {Eisert}\ \emph {et~al.}(2015)\citenamefont {Eisert},
  \citenamefont {Friesdorf},\ and\ \citenamefont
  {Gogolin}}]{eisert2015quantum}%
  \BibitemOpen
  \bibfield  {author} {\bibinfo {author} {\bibfnamefont {J.}~\bibnamefont
  {Eisert}}, \bibinfo {author} {\bibfnamefont {M.}~\bibnamefont {Friesdorf}}, \
  and\ \bibinfo {author} {\bibfnamefont {C.}~\bibnamefont {Gogolin}},\
  }\href@noop {} {\bibfield  {journal} {\bibinfo  {journal} {Nature Physics}\
  }\textbf {\bibinfo {volume} {11}},\ \bibinfo {pages} {124} (\bibinfo {year}
  {2015})}\BibitemShut {NoStop}%
\bibitem [{\citenamefont {Serbyn}\ \emph {et~al.}(2013)\citenamefont {Serbyn},
  \citenamefont {Papi\ifmmode~\acute{c}\else \'{c}\fi{}},\ and\ \citenamefont
  {Abanin}}]{Serbyn2013_LBit}%
  \BibitemOpen
  \bibfield  {author} {\bibinfo {author} {\bibfnamefont {M.}~\bibnamefont
  {Serbyn}}, \bibinfo {author} {\bibfnamefont {Z.}~\bibnamefont
  {Papi\ifmmode~\acute{c}\else \'{c}\fi{}}}, \ and\ \bibinfo {author}
  {\bibfnamefont {D.~A.}\ \bibnamefont {Abanin}},\ }\href {\doibase
  10.1103/PhysRevLett.111.127201} {\bibfield  {journal} {\bibinfo  {journal}
  {Phys. Rev. Lett.}\ }\textbf {\bibinfo {volume} {111}},\ \bibinfo {pages}
  {127201} (\bibinfo {year} {2013})}\BibitemShut {NoStop}%
\bibitem [{\citenamefont {Bauer}\ and\ \citenamefont
  {Nayak}(2013)}]{Bauer2013_AreaLawOrder}%
  \BibitemOpen
  \bibfield  {author} {\bibinfo {author} {\bibfnamefont {B.}~\bibnamefont
  {Bauer}}\ and\ \bibinfo {author} {\bibfnamefont {C.}~\bibnamefont {Nayak}},\
  }\href {\doibase 10.1088/1742-5468/2013/09/p09005} {\bibfield  {journal}
  {\bibinfo  {journal} {Journal of Statistical Mechanics: Theory and
  Experiment}\ }\textbf {\bibinfo {volume} {2013}},\ \bibinfo {pages} {P09005}
  (\bibinfo {year} {2013})}\BibitemShut {NoStop}%
\bibitem [{\citenamefont {Huse}\ \emph {et~al.}(2013)\citenamefont {Huse},
  \citenamefont {Nandkishore}, \citenamefont {Oganesyan}, \citenamefont {Pal},\
  and\ \citenamefont {Sondhi}}]{Huse2013_LPQO}%
  \BibitemOpen
  \bibfield  {author} {\bibinfo {author} {\bibfnamefont {D.~A.}\ \bibnamefont
  {Huse}}, \bibinfo {author} {\bibfnamefont {R.}~\bibnamefont {Nandkishore}},
  \bibinfo {author} {\bibfnamefont {V.}~\bibnamefont {Oganesyan}}, \bibinfo
  {author} {\bibfnamefont {A.}~\bibnamefont {Pal}}, \ and\ \bibinfo {author}
  {\bibfnamefont {S.~L.}\ \bibnamefont {Sondhi}},\ }\href {\doibase
  10.1103/PhysRevB.88.014206} {\bibfield  {journal} {\bibinfo  {journal} {Phys.
  Rev. B}\ }\textbf {\bibinfo {volume} {88}},\ \bibinfo {pages} {014206}
  (\bibinfo {year} {2013})}\BibitemShut {NoStop}%
\bibitem [{\citenamefont {Parameswaran}\ and\ \citenamefont
  {Vasseur}(2018)}]{Parameswaran2018_MBLOrder_Review}%
  \BibitemOpen
  \bibfield  {author} {\bibinfo {author} {\bibfnamefont {S.~A.}\ \bibnamefont
  {Parameswaran}}\ and\ \bibinfo {author} {\bibfnamefont {R.}~\bibnamefont
  {Vasseur}},\ }\href {\doibase 10.1088/1361-6633/aac9ed} {\bibfield  {journal}
  {\bibinfo  {journal} {Reports on Progress in Physics}\ }\textbf {\bibinfo
  {volume} {81}},\ \bibinfo {pages} {082501} (\bibinfo {year}
  {2018})}\BibitemShut {NoStop}%
\bibitem [{\citenamefont {Chandran}\ \emph {et~al.}(2014)\citenamefont
  {Chandran}, \citenamefont {Khemani}, \citenamefont {Laumann},\ and\
  \citenamefont {Sondhi}}]{Chandran2014_MBLSPT}%
  \BibitemOpen
  \bibfield  {author} {\bibinfo {author} {\bibfnamefont {A.}~\bibnamefont
  {Chandran}}, \bibinfo {author} {\bibfnamefont {V.}~\bibnamefont {Khemani}},
  \bibinfo {author} {\bibfnamefont {C.~R.}\ \bibnamefont {Laumann}}, \ and\
  \bibinfo {author} {\bibfnamefont {S.~L.}\ \bibnamefont {Sondhi}},\ }\href
  {\doibase 10.1103/PhysRevB.89.144201} {\bibfield  {journal} {\bibinfo
  {journal} {Phys. Rev. B}\ }\textbf {\bibinfo {volume} {89}},\ \bibinfo
  {pages} {144201} (\bibinfo {year} {2014})}\BibitemShut {NoStop}%
\bibitem [{\citenamefont {Kj\"all}\ \emph {et~al.}(2014)\citenamefont
  {Kj\"all}, \citenamefont {Bardarson},\ and\ \citenamefont
  {Pollmann}}]{Kjall2014_MBLOrder}%
  \BibitemOpen
  \bibfield  {author} {\bibinfo {author} {\bibfnamefont {J.~A.}\ \bibnamefont
  {Kj\"all}}, \bibinfo {author} {\bibfnamefont {J.~H.}\ \bibnamefont
  {Bardarson}}, \ and\ \bibinfo {author} {\bibfnamefont {F.}~\bibnamefont
  {Pollmann}},\ }\href {\doibase 10.1103/PhysRevLett.113.107204} {\bibfield
  {journal} {\bibinfo  {journal} {Phys. Rev. Lett.}\ }\textbf {\bibinfo
  {volume} {113}},\ \bibinfo {pages} {107204} (\bibinfo {year}
  {2014})}\BibitemShut {NoStop}%
\bibitem [{\citenamefont {Pekker}\ \emph
  {et~al.}(2014{\natexlab{a}})\citenamefont {Pekker}, \citenamefont {Refael},
  \citenamefont {Altman}, \citenamefont {Demler},\ and\ \citenamefont
  {Oganesyan}}]{Pekker_2014}%
  \BibitemOpen
  \bibfield  {author} {\bibinfo {author} {\bibfnamefont {D.}~\bibnamefont
  {Pekker}}, \bibinfo {author} {\bibfnamefont {G.}~\bibnamefont {Refael}},
  \bibinfo {author} {\bibfnamefont {E.}~\bibnamefont {Altman}}, \bibinfo
  {author} {\bibfnamefont {E.}~\bibnamefont {Demler}}, \ and\ \bibinfo {author}
  {\bibfnamefont {V.}~\bibnamefont {Oganesyan}},\ }\href {\doibase
  10.1103/PhysRevX.4.011052} {\bibfield  {journal} {\bibinfo  {journal}
  {Physical Review X}\ }\textbf {\bibinfo {volume} {4}},\ \bibinfo {pages}
  {011052} (\bibinfo {year} {2014}{\natexlab{a}})}\BibitemShut {NoStop}%
\bibitem [{\citenamefont {Bahri}\ \emph {et~al.}(2015)\citenamefont {Bahri},
  \citenamefont {Vosk}, \citenamefont {Altman},\ and\ \citenamefont
  {Vishwanath}}]{Bahri2015_MBLSPT}%
  \BibitemOpen
  \bibfield  {author} {\bibinfo {author} {\bibfnamefont {Y.}~\bibnamefont
  {Bahri}}, \bibinfo {author} {\bibfnamefont {R.}~\bibnamefont {Vosk}},
  \bibinfo {author} {\bibfnamefont {E.}~\bibnamefont {Altman}}, \ and\ \bibinfo
  {author} {\bibfnamefont {A.}~\bibnamefont {Vishwanath}},\ }\href {\doibase
  10.1038/ncomms8341} {\bibfield  {journal} {\bibinfo  {journal} {Nature
  Communications}\ }\textbf {\bibinfo {volume} {6}},\ \bibinfo {pages} {7341}
  (\bibinfo {year} {2015})}\BibitemShut {NoStop}%
\bibitem [{\citenamefont {Yao}\ \emph {et~al.}(2015)\citenamefont {Yao},
  \citenamefont {Laumann},\ and\ \citenamefont {Vishwanath}}]{yao2015many}%
  \BibitemOpen
  \bibfield  {author} {\bibinfo {author} {\bibfnamefont {N.~Y.}\ \bibnamefont
  {Yao}}, \bibinfo {author} {\bibfnamefont {C.~R.}\ \bibnamefont {Laumann}}, \
  and\ \bibinfo {author} {\bibfnamefont {A.}~\bibnamefont {Vishwanath}},\
  }\href@noop {} {\bibfield  {journal} {\bibinfo  {journal} {arXiv preprint
  arXiv:1508.06995}\ } (\bibinfo {year} {2015})}\BibitemShut {NoStop}%
\bibitem [{\citenamefont {Yao}\ \emph {et~al.}(2017)\citenamefont {Yao},
  \citenamefont {Potter}, \citenamefont {Potirniche},\ and\ \citenamefont
  {Vishwanath}}]{Yao2016_DTC}%
  \BibitemOpen
  \bibfield  {author} {\bibinfo {author} {\bibfnamefont {N.~Y.}\ \bibnamefont
  {Yao}}, \bibinfo {author} {\bibfnamefont {A.~C.}\ \bibnamefont {Potter}},
  \bibinfo {author} {\bibfnamefont {I.-D.}\ \bibnamefont {Potirniche}}, \ and\
  \bibinfo {author} {\bibfnamefont {A.}~\bibnamefont {Vishwanath}},\ }\href
  {\doibase 10.1103/PhysRevLett.118.030401} {\bibfield  {journal} {\bibinfo
  {journal} {Phys. Rev. Lett.}\ }\textbf {\bibinfo {volume} {118}},\ \bibinfo
  {pages} {030401} (\bibinfo {year} {2017})}\BibitemShut {NoStop}%
\bibitem [{\citenamefont {Else}\ \emph {et~al.}(2016)\citenamefont {Else},
  \citenamefont {Bauer},\ and\ \citenamefont {Nayak}}]{Else2016_DTC}%
  \BibitemOpen
  \bibfield  {author} {\bibinfo {author} {\bibfnamefont {D.~V.}\ \bibnamefont
  {Else}}, \bibinfo {author} {\bibfnamefont {B.}~\bibnamefont {Bauer}}, \ and\
  \bibinfo {author} {\bibfnamefont {C.}~\bibnamefont {Nayak}},\ }\href
  {\doibase 10.1103/PhysRevLett.117.090402} {\bibfield  {journal} {\bibinfo
  {journal} {Phys. Rev. Lett.}\ }\textbf {\bibinfo {volume} {117}},\ \bibinfo
  {pages} {090402} (\bibinfo {year} {2016})}\BibitemShut {NoStop}%
\bibitem [{\citenamefont {Potirniche}\ \emph
  {et~al.}(2017{\natexlab{a}})\citenamefont {Potirniche}, \citenamefont
  {Potter}, \citenamefont {Schleier-Smith}, \citenamefont {Vishwanath},\ and\
  \citenamefont {Yao}}]{Potirniche_2017}%
  \BibitemOpen
  \bibfield  {author} {\bibinfo {author} {\bibfnamefont {I.-D.}\ \bibnamefont
  {Potirniche}}, \bibinfo {author} {\bibfnamefont {A.~C.}\ \bibnamefont
  {Potter}}, \bibinfo {author} {\bibfnamefont {M.}~\bibnamefont
  {Schleier-Smith}}, \bibinfo {author} {\bibfnamefont {A.}~\bibnamefont
  {Vishwanath}}, \ and\ \bibinfo {author} {\bibfnamefont {N.~Y.}\ \bibnamefont
  {Yao}},\ }\href {\doibase 10.1103/PhysRevLett.119.123601} {\bibfield
  {journal} {\bibinfo  {journal} {Physical Review Letters}\ }\textbf {\bibinfo
  {volume} {119}},\ \bibinfo {pages} {123601} (\bibinfo {year}
  {2017}{\natexlab{a}})},\ \bibinfo {note} {arXiv: 1610.07611}\BibitemShut
  {NoStop}%
\bibitem [{\citenamefont {Huse}\ \emph {et~al.}(2014)\citenamefont {Huse},
  \citenamefont {Nandkishore},\ and\ \citenamefont
  {Oganesyan}}]{Huse2014_LBit}%
  \BibitemOpen
  \bibfield  {author} {\bibinfo {author} {\bibfnamefont {D.~A.}\ \bibnamefont
  {Huse}}, \bibinfo {author} {\bibfnamefont {R.}~\bibnamefont {Nandkishore}}, \
  and\ \bibinfo {author} {\bibfnamefont {V.}~\bibnamefont {Oganesyan}},\ }\href
  {\doibase 10.1103/PhysRevB.90.174202} {\bibfield  {journal} {\bibinfo
  {journal} {Phys. Rev. B}\ }\textbf {\bibinfo {volume} {90}},\ \bibinfo
  {pages} {174202} (\bibinfo {year} {2014})}\BibitemShut {NoStop}%
\bibitem [{\citenamefont {Sachdev}(2001)}]{Subir_QPT}%
  \BibitemOpen
  \bibfield  {author} {\bibinfo {author} {\bibfnamefont {S.}~\bibnamefont
  {Sachdev}},\ }\href {https://books.google.com/books?id=Ih\_E05N5TZQC} {\emph
  {\bibinfo {title} {Quantum Phase Transitions}}}\ (\bibinfo  {publisher}
  {Cambridge University Press},\ \bibinfo {year} {2001})\BibitemShut {NoStop}%
\bibitem [{\citenamefont {Pekker}\ \emph
  {et~al.}(2014{\natexlab{b}})\citenamefont {Pekker}, \citenamefont {Refael},
  \citenamefont {Altman}, \citenamefont {Demler},\ and\ \citenamefont
  {Oganesyan}}]{Altman2014_RSRGX}%
  \BibitemOpen
  \bibfield  {author} {\bibinfo {author} {\bibfnamefont {D.}~\bibnamefont
  {Pekker}}, \bibinfo {author} {\bibfnamefont {G.}~\bibnamefont {Refael}},
  \bibinfo {author} {\bibfnamefont {E.}~\bibnamefont {Altman}}, \bibinfo
  {author} {\bibfnamefont {E.}~\bibnamefont {Demler}}, \ and\ \bibinfo {author}
  {\bibfnamefont {V.}~\bibnamefont {Oganesyan}},\ }\href {\doibase
  10.1103/PhysRevX.4.011052} {\bibfield  {journal} {\bibinfo  {journal} {Phys.
  Rev. X}\ }\textbf {\bibinfo {volume} {4}},\ \bibinfo {pages} {011052}
  (\bibinfo {year} {2014}{\natexlab{b}})}\BibitemShut {NoStop}%
\bibitem [{\citenamefont {Venderley}\ \emph {et~al.}(2018)\citenamefont
  {Venderley}, \citenamefont {Khemani},\ and\ \citenamefont
  {Kim}}]{Venderley2016_EmerErg}%
  \BibitemOpen
  \bibfield  {author} {\bibinfo {author} {\bibfnamefont {J.}~\bibnamefont
  {Venderley}}, \bibinfo {author} {\bibfnamefont {V.}~\bibnamefont {Khemani}},
  \ and\ \bibinfo {author} {\bibfnamefont {E.-A.}\ \bibnamefont {Kim}},\ }\href
  {\doibase 10.1103/PhysRevLett.120.257204} {\bibfield  {journal} {\bibinfo
  {journal} {Phys. Rev. Lett.}\ }\textbf {\bibinfo {volume} {120}},\ \bibinfo
  {pages} {257204} (\bibinfo {year} {2018})}\BibitemShut {NoStop}%
\bibitem [{\citenamefont {Friedman}\ \emph {et~al.}(2018)\citenamefont
  {Friedman}, \citenamefont {Vasseur}, \citenamefont {Potter},\ and\
  \citenamefont {Parameswaran}}]{Friedman2018_EmerErg}%
  \BibitemOpen
  \bibfield  {author} {\bibinfo {author} {\bibfnamefont {A.~J.}\ \bibnamefont
  {Friedman}}, \bibinfo {author} {\bibfnamefont {R.}~\bibnamefont {Vasseur}},
  \bibinfo {author} {\bibfnamefont {A.~C.}\ \bibnamefont {Potter}}, \ and\
  \bibinfo {author} {\bibfnamefont {S.~A.}\ \bibnamefont {Parameswaran}},\
  }\href {\doibase 10.1103/PhysRevB.98.064203} {\bibfield  {journal} {\bibinfo
  {journal} {Phys. Rev. B}\ }\textbf {\bibinfo {volume} {98}},\ \bibinfo
  {pages} {064203} (\bibinfo {year} {2018})}\BibitemShut {NoStop}%
\bibitem [{\citenamefont {Vasseur}\ \emph {et~al.}(2016)\citenamefont
  {Vasseur}, \citenamefont {Friedman}, \citenamefont {Parameswaran},\ and\
  \citenamefont {Potter}}]{Vasseur2016_PHS}%
  \BibitemOpen
  \bibfield  {author} {\bibinfo {author} {\bibfnamefont {R.}~\bibnamefont
  {Vasseur}}, \bibinfo {author} {\bibfnamefont {A.~J.}\ \bibnamefont
  {Friedman}}, \bibinfo {author} {\bibfnamefont {S.~A.}\ \bibnamefont
  {Parameswaran}}, \ and\ \bibinfo {author} {\bibfnamefont {A.~C.}\
  \bibnamefont {Potter}},\ }\href {\doibase 10.1103/PhysRevB.93.134207}
  {\bibfield  {journal} {\bibinfo  {journal} {Phys. Rev. B}\ }\textbf {\bibinfo
  {volume} {93}},\ \bibinfo {pages} {134207} (\bibinfo {year}
  {2016})}\BibitemShut {NoStop}%
\bibitem [{\citenamefont {Khemani}\ \emph {et~al.}(2016)\citenamefont
  {Khemani}, \citenamefont {Lazarides}, \citenamefont {Moessner},\ and\
  \citenamefont {Sondhi}}]{Khemani2016_EmerErg}%
  \BibitemOpen
  \bibfield  {author} {\bibinfo {author} {\bibfnamefont {V.}~\bibnamefont
  {Khemani}}, \bibinfo {author} {\bibfnamefont {A.}~\bibnamefont {Lazarides}},
  \bibinfo {author} {\bibfnamefont {R.}~\bibnamefont {Moessner}}, \ and\
  \bibinfo {author} {\bibfnamefont {S.~L.}\ \bibnamefont {Sondhi}},\ }\href
  {\doibase 10.1103/PhysRevLett.116.250401} {\bibfield  {journal} {\bibinfo
  {journal} {Phys. Rev. Lett.}\ }\textbf {\bibinfo {volume} {116}},\ \bibinfo
  {pages} {250401} (\bibinfo {year} {2016})}\BibitemShut {NoStop}%
\bibitem [{\citenamefont {Chan}\ and\ \citenamefont
  {Wahl}(2020)}]{Chan2020_EmerErg}%
  \BibitemOpen
  \bibfield  {author} {\bibinfo {author} {\bibfnamefont {A.}~\bibnamefont
  {Chan}}\ and\ \bibinfo {author} {\bibfnamefont {T.~B.}\ \bibnamefont
  {Wahl}},\ }\href {\doibase 10.1088/1361-648x/ab7f01} {\bibfield  {journal}
  {\bibinfo  {journal} {Journal of Physics: Condensed Matter}\ }\textbf
  {\bibinfo {volume} {32}},\ \bibinfo {pages} {305601} (\bibinfo {year}
  {2020})}\BibitemShut {NoStop}%
\bibitem [{\citenamefont {Altshuler}\ \emph {et~al.}(1997)\citenamefont
  {Altshuler}, \citenamefont {Gefen}, \citenamefont {Kamenev},\ and\
  \citenamefont {Levitov}}]{Altshuler_1997}%
  \BibitemOpen
  \bibfield  {author} {\bibinfo {author} {\bibfnamefont {B.~L.}\ \bibnamefont
  {Altshuler}}, \bibinfo {author} {\bibfnamefont {Y.}~\bibnamefont {Gefen}},
  \bibinfo {author} {\bibfnamefont {A.}~\bibnamefont {Kamenev}}, \ and\
  \bibinfo {author} {\bibfnamefont {L.~S.}\ \bibnamefont {Levitov}},\ }\href
  {\doibase 10.1103/PhysRevLett.78.2803} {\bibfield  {journal} {\bibinfo
  {journal} {Physical Review Letters}\ }\textbf {\bibinfo {volume} {78}},\
  \bibinfo {pages} {2803–2806} (\bibinfo {year} {1997})},\ \bibinfo {note}
  {arXiv: cond-mat/9609132}\BibitemShut {NoStop}%
\bibitem [{\citenamefont {Nandkishore}\ and\ \citenamefont
  {Potter}(2014)}]{Nandkishore2014_drewrahul}%
  \BibitemOpen
  \bibfield  {author} {\bibinfo {author} {\bibfnamefont {R.}~\bibnamefont
  {Nandkishore}}\ and\ \bibinfo {author} {\bibfnamefont {A.~C.}\ \bibnamefont
  {Potter}},\ }\href {\doibase 10.1103/PhysRevB.90.195115} {\bibfield
  {journal} {\bibinfo  {journal} {Phys. Rev. B}\ }\textbf {\bibinfo {volume}
  {90}},\ \bibinfo {pages} {195115} (\bibinfo {year} {2014})}\BibitemShut
  {NoStop}%
\bibitem [{\citenamefont {De~Roeck}\ and\ \citenamefont
  {Huveneers}(2017)}]{DeRoeck2017_Avalanche}%
  \BibitemOpen
  \bibfield  {author} {\bibinfo {author} {\bibfnamefont {W.}~\bibnamefont
  {De~Roeck}}\ and\ \bibinfo {author} {\bibfnamefont {F.~m.~c.}\ \bibnamefont
  {Huveneers}},\ }\href {\doibase 10.1103/PhysRevB.95.155129} {\bibfield
  {journal} {\bibinfo  {journal} {Phys. Rev. B}\ }\textbf {\bibinfo {volume}
  {95}},\ \bibinfo {pages} {155129} (\bibinfo {year} {2017})}\BibitemShut
  {NoStop}%
\bibitem [{\citenamefont {Crowley}\ and\ \citenamefont
  {Chandran}(2019)}]{Crowley2019_Avalanche}%
  \BibitemOpen
  \bibfield  {author} {\bibinfo {author} {\bibfnamefont {P.~J.~D.}\
  \bibnamefont {Crowley}}\ and\ \bibinfo {author} {\bibfnamefont
  {A.}~\bibnamefont {Chandran}},\ }\href@noop {} {\  (\bibinfo {year}
  {2019})},\ \Eprint {http://arxiv.org/abs/arXiv:1910.10812} {arXiv:1910.10812}
  \BibitemShut {NoStop}%
\bibitem [{\citenamefont {Balewski}\ \emph {et~al.}(2014)\citenamefont
  {Balewski}, \citenamefont {Krupp}, \citenamefont {Gaj}, \citenamefont
  {Hofferberth}, \citenamefont {Löw},\ and\ \citenamefont
  {Pfau}}]{Balewski2014_Dressing}%
  \BibitemOpen
  \bibfield  {author} {\bibinfo {author} {\bibfnamefont {J.~B.}\ \bibnamefont
  {Balewski}}, \bibinfo {author} {\bibfnamefont {A.~T.}\ \bibnamefont {Krupp}},
  \bibinfo {author} {\bibfnamefont {A.}~\bibnamefont {Gaj}}, \bibinfo {author}
  {\bibfnamefont {S.}~\bibnamefont {Hofferberth}}, \bibinfo {author}
  {\bibfnamefont {R.}~\bibnamefont {Löw}}, \ and\ \bibinfo {author}
  {\bibfnamefont {T.}~\bibnamefont {Pfau}},\ }\href {\doibase
  10.1088/1367-2630/16/6/063012} {\bibfield  {journal} {\bibinfo  {journal}
  {New Journal of Physics}\ }\textbf {\bibinfo {volume} {16}},\ \bibinfo
  {pages} {063012} (\bibinfo {year} {2014})}\BibitemShut {NoStop}%
\bibitem [{\citenamefont {Choi}\ \emph {et~al.}(2016)\citenamefont {Choi},
  \citenamefont {Hild}, \citenamefont {Zeiher}, \citenamefont {Schau{\ss}},
  \citenamefont {Rubio-Abadal}, \citenamefont {Yefsah}, \citenamefont
  {Khemani}, \citenamefont {Huse}, \citenamefont {Bloch},\ and\ \citenamefont
  {Gross}}]{Bloch2016_2DMBL}%
  \BibitemOpen
  \bibfield  {author} {\bibinfo {author} {\bibfnamefont {J.-y.}\ \bibnamefont
  {Choi}}, \bibinfo {author} {\bibfnamefont {S.}~\bibnamefont {Hild}}, \bibinfo
  {author} {\bibfnamefont {J.}~\bibnamefont {Zeiher}}, \bibinfo {author}
  {\bibfnamefont {P.}~\bibnamefont {Schau{\ss}}}, \bibinfo {author}
  {\bibfnamefont {A.}~\bibnamefont {Rubio-Abadal}}, \bibinfo {author}
  {\bibfnamefont {T.}~\bibnamefont {Yefsah}}, \bibinfo {author} {\bibfnamefont
  {V.}~\bibnamefont {Khemani}}, \bibinfo {author} {\bibfnamefont {D.~A.}\
  \bibnamefont {Huse}}, \bibinfo {author} {\bibfnamefont {I.}~\bibnamefont
  {Bloch}}, \ and\ \bibinfo {author} {\bibfnamefont {C.}~\bibnamefont
  {Gross}},\ }\href {\doibase 10.1126/science.aaf8834} {\bibfield  {journal}
  {\bibinfo  {journal} {Science}\ }\textbf {\bibinfo {volume} {352}},\ \bibinfo
  {pages} {1547} (\bibinfo {year} {2016})}\BibitemShut {NoStop}%
\bibitem [{\citenamefont {Zeiher}\ \emph {et~al.}(2017)\citenamefont {Zeiher},
  \citenamefont {Choi}, \citenamefont {Rubio-Abadal}, \citenamefont {Pohl},
  \citenamefont {van Bijnen}, \citenamefont {Bloch},\ and\ \citenamefont
  {Gross}}]{Bloch2017_Dressing}%
  \BibitemOpen
  \bibfield  {author} {\bibinfo {author} {\bibfnamefont {J.}~\bibnamefont
  {Zeiher}}, \bibinfo {author} {\bibfnamefont {J.-y.}\ \bibnamefont {Choi}},
  \bibinfo {author} {\bibfnamefont {A.}~\bibnamefont {Rubio-Abadal}}, \bibinfo
  {author} {\bibfnamefont {T.}~\bibnamefont {Pohl}}, \bibinfo {author}
  {\bibfnamefont {R.}~\bibnamefont {van Bijnen}}, \bibinfo {author}
  {\bibfnamefont {I.}~\bibnamefont {Bloch}}, \ and\ \bibinfo {author}
  {\bibfnamefont {C.}~\bibnamefont {Gross}},\ }\href {\doibase
  10.1103/PhysRevX.7.041063} {\bibfield  {journal} {\bibinfo  {journal} {Phys.
  Rev. X}\ }\textbf {\bibinfo {volume} {7}},\ \bibinfo {pages} {041063}
  (\bibinfo {year} {2017})}\BibitemShut {NoStop}%
\bibitem [{\citenamefont {Bernien}\ \emph {et~al.}(2017)\citenamefont
  {Bernien}, \citenamefont {Schwartz}, \citenamefont {Keesling}, \citenamefont
  {Levine}, \citenamefont {Omran}, \citenamefont {Pichler}, \citenamefont
  {Choi}, \citenamefont {Zibrov}, \citenamefont {Endres}, \citenamefont
  {Greiner}, \citenamefont {Vuleti{\'c}},\ and\ \citenamefont
  {Lukin}}]{Lukin2017_tweezer}%
  \BibitemOpen
  \bibfield  {author} {\bibinfo {author} {\bibfnamefont {H.}~\bibnamefont
  {Bernien}}, \bibinfo {author} {\bibfnamefont {S.}~\bibnamefont {Schwartz}},
  \bibinfo {author} {\bibfnamefont {A.}~\bibnamefont {Keesling}}, \bibinfo
  {author} {\bibfnamefont {H.}~\bibnamefont {Levine}}, \bibinfo {author}
  {\bibfnamefont {A.}~\bibnamefont {Omran}}, \bibinfo {author} {\bibfnamefont
  {H.}~\bibnamefont {Pichler}}, \bibinfo {author} {\bibfnamefont
  {S.}~\bibnamefont {Choi}}, \bibinfo {author} {\bibfnamefont {A.~S.}\
  \bibnamefont {Zibrov}}, \bibinfo {author} {\bibfnamefont {M.}~\bibnamefont
  {Endres}}, \bibinfo {author} {\bibfnamefont {M.}~\bibnamefont {Greiner}},
  \bibinfo {author} {\bibfnamefont {V.}~\bibnamefont {Vuleti{\'c}}}, \ and\
  \bibinfo {author} {\bibfnamefont {M.~D.}\ \bibnamefont {Lukin}},\ }\href
  {\doibase 10.1038/nature24622} {\bibfield  {journal} {\bibinfo  {journal}
  {Nature}\ }\textbf {\bibinfo {volume} {551}},\ \bibinfo {pages} {579}
  (\bibinfo {year} {2017})}\BibitemShut {NoStop}%
\bibitem [{\citenamefont {Cooper}\ \emph {et~al.}(2018)\citenamefont {Cooper},
  \citenamefont {Covey}, \citenamefont {Madjarov}, \citenamefont {Porsev},
  \citenamefont {Safronova},\ and\ \citenamefont
  {Endres}}]{Cooper2018_AlkalineEarth}%
  \BibitemOpen
  \bibfield  {author} {\bibinfo {author} {\bibfnamefont {A.}~\bibnamefont
  {Cooper}}, \bibinfo {author} {\bibfnamefont {J.~P.}\ \bibnamefont {Covey}},
  \bibinfo {author} {\bibfnamefont {I.~S.}\ \bibnamefont {Madjarov}}, \bibinfo
  {author} {\bibfnamefont {S.~G.}\ \bibnamefont {Porsev}}, \bibinfo {author}
  {\bibfnamefont {M.~S.}\ \bibnamefont {Safronova}}, \ and\ \bibinfo {author}
  {\bibfnamefont {M.}~\bibnamefont {Endres}},\ }\href {\doibase
  10.1103/PhysRevX.8.041055} {\bibfield  {journal} {\bibinfo  {journal} {Phys.
  Rev. X}\ }\textbf {\bibinfo {volume} {8}},\ \bibinfo {pages} {041055}
  (\bibinfo {year} {2018})}\BibitemShut {NoStop}%
\bibitem [{\citenamefont {Léséleuc}\ \emph {et~al.}(2019)\citenamefont
  {Léséleuc}, \citenamefont {Lienhard}, \citenamefont {Scholl}, \citenamefont
  {Barredo}, \citenamefont {Weber}, \citenamefont {Lang}, \citenamefont
  {Büchler}, \citenamefont {Lahaye},\ and\ \citenamefont
  {Browaeys}}]{Leseleuc_2019}%
  \BibitemOpen
  \bibfield  {author} {\bibinfo {author} {\bibfnamefont {S.~d.}\ \bibnamefont
  {Léséleuc}}, \bibinfo {author} {\bibfnamefont {V.}~\bibnamefont
  {Lienhard}}, \bibinfo {author} {\bibfnamefont {P.}~\bibnamefont {Scholl}},
  \bibinfo {author} {\bibfnamefont {D.}~\bibnamefont {Barredo}}, \bibinfo
  {author} {\bibfnamefont {S.}~\bibnamefont {Weber}}, \bibinfo {author}
  {\bibfnamefont {N.}~\bibnamefont {Lang}}, \bibinfo {author} {\bibfnamefont
  {H.~P.}\ \bibnamefont {Büchler}}, \bibinfo {author} {\bibfnamefont
  {T.}~\bibnamefont {Lahaye}}, \ and\ \bibinfo {author} {\bibfnamefont
  {A.}~\bibnamefont {Browaeys}},\ }\href {\doibase 10.1126/science.aav9105}
  {\bibfield  {journal} {\bibinfo  {journal} {Science}\ }\textbf {\bibinfo
  {volume} {365}},\ \bibinfo {pages} {775–780} (\bibinfo {year}
  {2019})}\BibitemShut {NoStop}%
\bibitem [{\citenamefont {Wilson}\ \emph {et~al.}(2019)\citenamefont {Wilson},
  \citenamefont {Saskin}, \citenamefont {Meng}, \citenamefont {Ma},
  \citenamefont {Dilip}, \citenamefont {Burgers},\ and\ \citenamefont
  {Thompson}}]{Thompson2019_AlkalineEarth}%
  \BibitemOpen
  \bibfield  {author} {\bibinfo {author} {\bibfnamefont {J.}~\bibnamefont
  {Wilson}}, \bibinfo {author} {\bibfnamefont {S.}~\bibnamefont {Saskin}},
  \bibinfo {author} {\bibfnamefont {Y.}~\bibnamefont {Meng}}, \bibinfo {author}
  {\bibfnamefont {S.}~\bibnamefont {Ma}}, \bibinfo {author} {\bibfnamefont
  {R.}~\bibnamefont {Dilip}}, \bibinfo {author} {\bibfnamefont
  {A.}~\bibnamefont {Burgers}}, \ and\ \bibinfo {author} {\bibfnamefont
  {J.}~\bibnamefont {Thompson}},\ }\href {http://arxiv.org/abs/1912.08754}
  {\bibfield  {journal} {\bibinfo  {journal} {arXiv:1912.08754 [physics,
  physics:quant-ph]}\ } (\bibinfo {year} {2019})},\ \bibinfo {note} {arXiv:
  1912.08754}\BibitemShut {NoStop}%
\bibitem [{\citenamefont {Madjarov}\ \emph {et~al.}(2020)\citenamefont
  {Madjarov}, \citenamefont {Covey}, \citenamefont {Shaw}, \citenamefont
  {Choi}, \citenamefont {Kale}, \citenamefont {Cooper}, \citenamefont
  {Pichler}, \citenamefont {Schkolnik}, \citenamefont {Williams},\ and\
  \citenamefont {Endres}}]{Madjarov_2020}%
  \BibitemOpen
  \bibfield  {author} {\bibinfo {author} {\bibfnamefont {I.~S.}\ \bibnamefont
  {Madjarov}}, \bibinfo {author} {\bibfnamefont {J.~P.}\ \bibnamefont {Covey}},
  \bibinfo {author} {\bibfnamefont {A.~L.}\ \bibnamefont {Shaw}}, \bibinfo
  {author} {\bibfnamefont {J.}~\bibnamefont {Choi}}, \bibinfo {author}
  {\bibfnamefont {A.}~\bibnamefont {Kale}}, \bibinfo {author} {\bibfnamefont
  {A.}~\bibnamefont {Cooper}}, \bibinfo {author} {\bibfnamefont
  {H.}~\bibnamefont {Pichler}}, \bibinfo {author} {\bibfnamefont
  {V.}~\bibnamefont {Schkolnik}}, \bibinfo {author} {\bibfnamefont {J.~R.}\
  \bibnamefont {Williams}}, \ and\ \bibinfo {author} {\bibfnamefont
  {M.}~\bibnamefont {Endres}},\ }\href {\doibase 10.1038/s41567-020-0903-z}
  {\bibfield  {journal} {\bibinfo  {journal} {Nature Physics}\ ,\ \bibinfo
  {pages} {1–5}} (\bibinfo {year} {2020})}\BibitemShut {NoStop}%
\bibitem [{ft2()}]{ft2}%
  \BibitemOpen
  \href@noop {} {}\bibinfo {note} {We remark that up to edge effects, the model
  is dual under the Kramers-Wannier map ensuring that any direct transition
  between the MBL SG and MBL PM phases must occur at $W_J/W_h = 1$
  \cite{SM}.}\BibitemShut {Stop}%
\bibitem [{SM()}]{SM}%
  \BibitemOpen
  \href@noop {} {\bibinfo  {journal} {For details see supplemental material}\
  }\BibitemShut {NoStop}%
\bibitem [{\citenamefont {Oganesyan}\ and\ \citenamefont
  {Huse}(2007)}]{Oganesyan_2007}%
  \BibitemOpen
\bibfield  {journal} {  }\bibfield  {author} {\bibinfo {author} {\bibfnamefont
  {V.}~\bibnamefont {Oganesyan}}\ and\ \bibinfo {author} {\bibfnamefont
  {D.~A.}\ \bibnamefont {Huse}},\ }\href {\doibase 10.1103/PhysRevB.75.155111}
  {\bibfield  {journal} {\bibinfo  {journal} {Physical Review B}\ }\textbf
  {\bibinfo {volume} {75}},\ \bibinfo {pages} {155111} (\bibinfo {year}
  {2007})}\BibitemShut {NoStop}%
\bibitem [{\citenamefont {Pal}\ and\ \citenamefont {Huse}(2010)}]{Pal_2010}%
  \BibitemOpen
  \bibfield  {author} {\bibinfo {author} {\bibfnamefont {A.}~\bibnamefont
  {Pal}}\ and\ \bibinfo {author} {\bibfnamefont {D.~A.}\ \bibnamefont {Huse}},\
  }\href {\doibase 10.1103/PhysRevB.82.174411} {\bibfield  {journal} {\bibinfo
  {journal} {Physical Review B}\ }\textbf {\bibinfo {volume} {82}},\ \bibinfo
  {pages} {174411} (\bibinfo {year} {2010})}\BibitemShut {NoStop}%
\bibitem [{\citenamefont
  {Srednicki}(1994{\natexlab{b}})}]{Srednicki1994_ChaosMBL}%
  \BibitemOpen
  \bibfield  {author} {\bibinfo {author} {\bibfnamefont {M.}~\bibnamefont
  {Srednicki}},\ }\href {\doibase 10.1103/PhysRevE.50.888} {\bibfield
  {journal} {\bibinfo  {journal} {Phys. Rev. E}\ }\textbf {\bibinfo {volume}
  {50}},\ \bibinfo {pages} {888} (\bibinfo {year}
  {1994}{\natexlab{b}})}\BibitemShut {NoStop}%
\bibitem [{\citenamefont {Abanin}\ \emph
  {et~al.}(2019{\natexlab{b}})\citenamefont {Abanin}, \citenamefont
  {Bardarson}, \citenamefont {Tomasi}, \citenamefont {Gopalakrishnan},
  \citenamefont {Khemani}, \citenamefont {Parameswaran}, \citenamefont
  {Pollmann}, \citenamefont {Potter}, \citenamefont {Serbyn},\ and\
  \citenamefont {Vasseur}}]{Abanin2019_FiniteSize}%
  \BibitemOpen
  \bibfield  {author} {\bibinfo {author} {\bibfnamefont {D.~A.}\ \bibnamefont
  {Abanin}}, \bibinfo {author} {\bibfnamefont {J.~H.}\ \bibnamefont
  {Bardarson}}, \bibinfo {author} {\bibfnamefont {G.~D.}\ \bibnamefont
  {Tomasi}}, \bibinfo {author} {\bibfnamefont {S.}~\bibnamefont
  {Gopalakrishnan}}, \bibinfo {author} {\bibfnamefont {V.}~\bibnamefont
  {Khemani}}, \bibinfo {author} {\bibfnamefont {S.~A.}\ \bibnamefont
  {Parameswaran}}, \bibinfo {author} {\bibfnamefont {F.}~\bibnamefont
  {Pollmann}}, \bibinfo {author} {\bibfnamefont {A.~C.}\ \bibnamefont
  {Potter}}, \bibinfo {author} {\bibfnamefont {M.}~\bibnamefont {Serbyn}}, \
  and\ \bibinfo {author} {\bibfnamefont {R.}~\bibnamefont {Vasseur}},\
  }\href@noop {} {\  (\bibinfo {year} {2019}{\natexlab{b}})},\ \Eprint
  {http://arxiv.org/abs/arXiv:1911.04501} {arXiv:1911.04501} \BibitemShut
  {NoStop}%
\bibitem [{\citenamefont {Panda}\ \emph {et~al.}(2020)\citenamefont {Panda},
  \citenamefont {Scardicchio}, \citenamefont {Schulz}, \citenamefont {Taylor},\
  and\ \citenamefont {{\v{Z}}nidari{\v{c}}}}]{Panda2020_FiniteSize}%
  \BibitemOpen
  \bibfield  {author} {\bibinfo {author} {\bibfnamefont {R.~K.}\ \bibnamefont
  {Panda}}, \bibinfo {author} {\bibfnamefont {A.}~\bibnamefont {Scardicchio}},
  \bibinfo {author} {\bibfnamefont {M.}~\bibnamefont {Schulz}}, \bibinfo
  {author} {\bibfnamefont {S.~R.}\ \bibnamefont {Taylor}}, \ and\ \bibinfo
  {author} {\bibfnamefont {M.}~\bibnamefont {{\v{Z}}nidari{\v{c}}}},\ }\href
  {\doibase 10.1209/0295-5075/128/67003} {\bibfield  {journal} {\bibinfo
  {journal} {{EPL} (Europhysics Letters)}\ }\textbf {\bibinfo {volume} {128}},\
  \bibinfo {pages} {67003} (\bibinfo {year} {2020})}\BibitemShut {NoStop}%
\bibitem [{\citenamefont {Papić}\ \emph {et~al.}(2015)\citenamefont {Papić},
  \citenamefont {Stoudenmire},\ and\ \citenamefont
  {Abanin}}]{Papic2019_Miniband}%
  \BibitemOpen
  \bibfield  {author} {\bibinfo {author} {\bibfnamefont {Z.}~\bibnamefont
  {Papić}}, \bibinfo {author} {\bibfnamefont {E.~M.}\ \bibnamefont
  {Stoudenmire}}, \ and\ \bibinfo {author} {\bibfnamefont {D.~A.}\ \bibnamefont
  {Abanin}},\ }\href {\doibase https://doi.org/10.1016/j.aop.2015.08.024}
  {\bibfield  {journal} {\bibinfo  {journal} {Annals of Physics}\ }\textbf
  {\bibinfo {volume} {362}},\ \bibinfo {pages} {714 } (\bibinfo {year}
  {2015})}\BibitemShut {NoStop}%
\bibitem [{ft3()}]{ft3}%
  \BibitemOpen
  \href@noop {} {}\bibinfo {note} {We chose the form of our interaction such
  that, in the thermodynamic limit, the MBL SPT and MBL PM are dual to one
  another under the duality transformation $\sigma^z_i \to \sigma_i^z
  \sigma_{i+1}^x$ and $\sigma_i^x \to \sigma_{i-1}^z \sigma_i^x
  \sigma_{i+1}^z$.}\BibitemShut {Stop}%
\bibitem [{\citenamefont {Abanin}\ \emph {et~al.}(2015)\citenamefont {Abanin},
  \citenamefont {De~Roeck},\ and\ \citenamefont
  {Huveneers}}]{Abanin2015_Heating}%
  \BibitemOpen
  \bibfield  {author} {\bibinfo {author} {\bibfnamefont {D.~A.}\ \bibnamefont
  {Abanin}}, \bibinfo {author} {\bibfnamefont {W.}~\bibnamefont {De~Roeck}}, \
  and\ \bibinfo {author} {\bibfnamefont {F.~m.~c.}\ \bibnamefont {Huveneers}},\
  }\href {\doibase 10.1103/PhysRevLett.115.256803} {\bibfield  {journal}
  {\bibinfo  {journal} {Phys. Rev. Lett.}\ }\textbf {\bibinfo {volume} {115}},\
  \bibinfo {pages} {256803} (\bibinfo {year} {2015})}\BibitemShut {NoStop}%
\bibitem [{\citenamefont {Machado}\ \emph {et~al.}(2019)\citenamefont
  {Machado}, \citenamefont {Kahanamoku-Meyer}, \citenamefont {Else},
  \citenamefont {Nayak},\ and\ \citenamefont {Yao}}]{Machado2019_Heating}%
  \BibitemOpen
  \bibfield  {author} {\bibinfo {author} {\bibfnamefont {F.}~\bibnamefont
  {Machado}}, \bibinfo {author} {\bibfnamefont {G.~D.}\ \bibnamefont
  {Kahanamoku-Meyer}}, \bibinfo {author} {\bibfnamefont {D.~V.}\ \bibnamefont
  {Else}}, \bibinfo {author} {\bibfnamefont {C.}~\bibnamefont {Nayak}}, \ and\
  \bibinfo {author} {\bibfnamefont {N.~Y.}\ \bibnamefont {Yao}},\ }\href
  {\doibase 10.1103/PhysRevResearch.1.033202} {\bibfield  {journal} {\bibinfo
  {journal} {Phys. Rev. Research}\ }\textbf {\bibinfo {volume} {1}},\ \bibinfo
  {pages} {033202} (\bibinfo {year} {2019})}\BibitemShut {NoStop}%
\bibitem [{\citenamefont {Potirniche}\ \emph
  {et~al.}(2017{\natexlab{b}})\citenamefont {Potirniche}, \citenamefont
  {Potter}, \citenamefont {Schleier-Smith}, \citenamefont {Vishwanath},\ and\
  \citenamefont {Yao}}]{Yao2017_DressingTheory}%
  \BibitemOpen
  \bibfield  {author} {\bibinfo {author} {\bibfnamefont {I.-D.}\ \bibnamefont
  {Potirniche}}, \bibinfo {author} {\bibfnamefont {A.~C.}\ \bibnamefont
  {Potter}}, \bibinfo {author} {\bibfnamefont {M.}~\bibnamefont
  {Schleier-Smith}}, \bibinfo {author} {\bibfnamefont {A.}~\bibnamefont
  {Vishwanath}}, \ and\ \bibinfo {author} {\bibfnamefont {N.~Y.}\ \bibnamefont
  {Yao}},\ }\href {\doibase 10.1103/PhysRevLett.119.123601} {\bibfield
  {journal} {\bibinfo  {journal} {Phys. Rev. Lett.}\ }\textbf {\bibinfo
  {volume} {119}},\ \bibinfo {pages} {123601} (\bibinfo {year}
  {2017}{\natexlab{b}})}\BibitemShut {NoStop}%
\bibitem [{dyn()}]{dynamite}%
  \BibitemOpen
  \href@noop {} {\ }\bibinfo {note} {\hspace{-1mm} For more information, see
  \url{https://zenodo.org/record/3606826}}\BibitemShut {NoStop}%
\bibitem [{\citenamefont {Hernandez}\ \emph {et~al.}(2005)\citenamefont
  {Hernandez}, \citenamefont {Roman},\ and\ \citenamefont {Vidal}}]{SLEPc1}%
  \BibitemOpen
  \bibfield  {author} {\bibinfo {author} {\bibfnamefont {V.}~\bibnamefont
  {Hernandez}}, \bibinfo {author} {\bibfnamefont {J.~E.}\ \bibnamefont
  {Roman}}, \ and\ \bibinfo {author} {\bibfnamefont {V.}~\bibnamefont
  {Vidal}},\ }\href {\doibase 10.1145/1089014.1089019} {\bibfield  {journal}
  {\bibinfo  {journal} {ACM Trans. Math. Softw.}\ }\textbf {\bibinfo {volume}
  {31}},\ \bibinfo {pages} {351–362} (\bibinfo {year} {2005})}\BibitemShut
  {NoStop}%
\bibitem [{\citenamefont {Hernandez}\ \emph {et~al.}(2003)\citenamefont
  {Hernandez}, \citenamefont {Roman},\ and\ \citenamefont {Vidal}}]{SLEPc2}%
  \BibitemOpen
  \bibfield  {author} {\bibinfo {author} {\bibfnamefont {V.}~\bibnamefont
  {Hernandez}}, \bibinfo {author} {\bibfnamefont {J.~E.}\ \bibnamefont
  {Roman}}, \ and\ \bibinfo {author} {\bibfnamefont {V.}~\bibnamefont
  {Vidal}},\ }\href@noop {} {\bibfield  {journal} {\bibinfo  {journal} {Lect.
  Notes Comput. Sci.}\ }\textbf {\bibinfo {volume} {2565}},\ \bibinfo {pages}
  {377} (\bibinfo {year} {2003})}\BibitemShut {NoStop}%
\bibitem [{\citenamefont {Balay}\ \emph {et~al.}(1997)\citenamefont {Balay},
  \citenamefont {Gropp}, \citenamefont {McInnes},\ and\ \citenamefont
  {Smith}}]{Parallel}%
  \BibitemOpen
  \bibfield  {author} {\bibinfo {author} {\bibfnamefont {S.}~\bibnamefont
  {Balay}}, \bibinfo {author} {\bibfnamefont {W.~D.}\ \bibnamefont {Gropp}},
  \bibinfo {author} {\bibfnamefont {L.~C.}\ \bibnamefont {McInnes}}, \ and\
  \bibinfo {author} {\bibfnamefont {B.~F.}\ \bibnamefont {Smith}},\ }\enquote
  {\bibinfo {title} {Efficient management of parallelism in object-oriented
  numerical software libraries},}\ in\ \href {\doibase
  10.1007/978-1-4612-1986-6_8} {\emph {\bibinfo {booktitle} {Modern Software
  Tools for Scientific Computing}}},\ \bibinfo {editor} {edited by\ \bibinfo
  {editor} {\bibfnamefont {E.}~\bibnamefont {Arge}}, \bibinfo {editor}
  {\bibfnamefont {A.~M.}\ \bibnamefont {Bruaset}}, \ and\ \bibinfo {editor}
  {\bibfnamefont {H.~P.}\ \bibnamefont {Langtangen}}}\ (\bibinfo  {publisher}
  {Birkh{\"a}user Boston},\ \bibinfo {address} {Boston, MA},\ \bibinfo {year}
  {1997})\ pp.\ \bibinfo {pages} {163--202}\BibitemShut {NoStop}%
\bibitem [{\citenamefont {Gross}\ and\ \citenamefont
  {Bloch}(2017)}]{Gross2017_GasMicroscope}%
  \BibitemOpen
  \bibfield  {author} {\bibinfo {author} {\bibfnamefont {C.}~\bibnamefont
  {Gross}}\ and\ \bibinfo {author} {\bibfnamefont {I.}~\bibnamefont {Bloch}},\
  }\href@noop {} {\bibfield  {journal} {\bibinfo  {journal} {Science}\ }\textbf
  {\bibinfo {volume} {357}},\ \bibinfo {pages} {995} (\bibinfo {year}
  {2017})}\BibitemShut {NoStop}%
\bibitem [{ft1()}]{ft1}%
  \BibitemOpen
  \href@noop {} {}\bibinfo {note} {An analogous behavior can be found in a
  linear geometry where the ration between nearest and next-nearest neighbor
  interactions is $J_{i,i+2} = 0.2J_{i,i+1}$ \cite{SM}.}\BibitemShut {Stop}%
\bibitem [{ft4()}]{ft4}%
  \BibitemOpen
  \href@noop {} {}\bibinfo {note} {$\ket{\psi_x}$ is polarized along $+\hat{y}$
  except at sites $L/2, L/2 + 1$ where it is polarized in the $+\hat{x}$ and
  $-\hat{x}$ direction respectively. Analogously $\ket{\psi_{zz}}$ is polarized
  along $+\hat{y}$ except at sites $L/2-1$ through $L/2+2$ where the spins are
  polarized in the $\hat{z}$ direction with the pattern
  $\uparrow\uparrow\downarrow\downarrow$.}\BibitemShut {Stop}%
\bibitem [{\citenamefont {Evers}\ and\ \citenamefont
  {Mirlin}(2008)}]{Evers2008_AndersonTransitions}%
  \BibitemOpen
  \bibfield  {author} {\bibinfo {author} {\bibfnamefont {F.}~\bibnamefont
  {Evers}}\ and\ \bibinfo {author} {\bibfnamefont {A.~D.}\ \bibnamefont
  {Mirlin}},\ }\href {\doibase 10.1103/RevModPhys.80.1355} {\bibfield
  {journal} {\bibinfo  {journal} {Rev. Mod. Phys.}\ }\textbf {\bibinfo {volume}
  {80}},\ \bibinfo {pages} {1355} (\bibinfo {year} {2008})}\BibitemShut
  {NoStop}%
\bibitem [{\citenamefont {Fisher}(1995)}]{Fisher1995_RTFIM}%
  \BibitemOpen
  \bibfield  {author} {\bibinfo {author} {\bibfnamefont {D.~S.}\ \bibnamefont
  {Fisher}},\ }\href {\doibase 10.1103/PhysRevB.51.6411} {\bibfield  {journal}
  {\bibinfo  {journal} {Phys. Rev. B}\ }\textbf {\bibinfo {volume} {51}},\
  \bibinfo {pages} {6411} (\bibinfo {year} {1995})}\BibitemShut {NoStop}%
\bibitem [{\citenamefont {Balents}\ and\ \citenamefont
  {Fisher}(1997)}]{Balents1997_SUSY}%
  \BibitemOpen
  \bibfield  {author} {\bibinfo {author} {\bibfnamefont {L.}~\bibnamefont
  {Balents}}\ and\ \bibinfo {author} {\bibfnamefont {M.~P.~A.}\ \bibnamefont
  {Fisher}},\ }\href {\doibase 10.1103/PhysRevB.56.12970} {\bibfield  {journal}
  {\bibinfo  {journal} {Phys. Rev. B}\ }\textbf {\bibinfo {volume} {56}},\
  \bibinfo {pages} {12970} (\bibinfo {year} {1997})}\BibitemShut {NoStop}%
\bibitem [{\citenamefont {McCoy}\ and\ \citenamefont
  {Wu}(1968)}]{McCoy1968_TFIM}%
  \BibitemOpen
  \bibfield  {author} {\bibinfo {author} {\bibfnamefont {B.~M.}\ \bibnamefont
  {McCoy}}\ and\ \bibinfo {author} {\bibfnamefont {T.~T.}\ \bibnamefont {Wu}},\
  }\href {\doibase 10.1103/PhysRev.176.631} {\bibfield  {journal} {\bibinfo
  {journal} {Phys. Rev.}\ }\textbf {\bibinfo {volume} {176}},\ \bibinfo {pages}
  {631} (\bibinfo {year} {1968})}\BibitemShut {NoStop}%
\bibitem [{\citenamefont {Yao}\ \emph {et~al.}(2014)\citenamefont {Yao},
  \citenamefont {Laumann}, \citenamefont {Gopalakrishnan}, \citenamefont
  {Knap}, \citenamefont {M\"uller}, \citenamefont {Demler},\ and\ \citenamefont
  {Lukin}}]{YaoLau2014_PowerLaw}%
  \BibitemOpen
  \bibfield  {author} {\bibinfo {author} {\bibfnamefont {N.~Y.}\ \bibnamefont
  {Yao}}, \bibinfo {author} {\bibfnamefont {C.~R.}\ \bibnamefont {Laumann}},
  \bibinfo {author} {\bibfnamefont {S.}~\bibnamefont {Gopalakrishnan}},
  \bibinfo {author} {\bibfnamefont {M.}~\bibnamefont {Knap}}, \bibinfo {author}
  {\bibfnamefont {M.}~\bibnamefont {M\"uller}}, \bibinfo {author}
  {\bibfnamefont {E.~A.}\ \bibnamefont {Demler}}, \ and\ \bibinfo {author}
  {\bibfnamefont {M.~D.}\ \bibnamefont {Lukin}},\ }\href {\doibase
  10.1103/PhysRevLett.113.243002} {\bibfield  {journal} {\bibinfo  {journal}
  {Phys. Rev. Lett.}\ }\textbf {\bibinfo {volume} {113}},\ \bibinfo {pages}
  {243002} (\bibinfo {year} {2014})}\BibitemShut {NoStop}%
\bibitem [{\citenamefont {Luitz}\ \emph {et~al.}(2017)\citenamefont {Luitz},
  \citenamefont {Huveneers},\ and\ \citenamefont
  {De~Roeck}}]{Luitz2017_Avalanche}%
  \BibitemOpen
  \bibfield  {author} {\bibinfo {author} {\bibfnamefont {D.~J.}\ \bibnamefont
  {Luitz}}, \bibinfo {author} {\bibfnamefont {F.~m.~c.}\ \bibnamefont
  {Huveneers}}, \ and\ \bibinfo {author} {\bibfnamefont {W.}~\bibnamefont
  {De~Roeck}},\ }\href {\doibase 10.1103/PhysRevLett.119.150602} {\bibfield
  {journal} {\bibinfo  {journal} {Phys. Rev. Lett.}\ }\textbf {\bibinfo
  {volume} {119}},\ \bibinfo {pages} {150602} (\bibinfo {year}
  {2017})}\BibitemShut {NoStop}%
\bibitem [{\citenamefont {Thiery}\ \emph {et~al.}(2018)\citenamefont {Thiery},
  \citenamefont {Huveneers}, \citenamefont {M\"uller},\ and\ \citenamefont
  {De~Roeck}}]{Thiery2018_Avalanche}%
  \BibitemOpen
  \bibfield  {author} {\bibinfo {author} {\bibfnamefont {T.}~\bibnamefont
  {Thiery}}, \bibinfo {author} {\bibfnamefont {F.~m.~c.}\ \bibnamefont
  {Huveneers}}, \bibinfo {author} {\bibfnamefont {M.}~\bibnamefont {M\"uller}},
  \ and\ \bibinfo {author} {\bibfnamefont {W.}~\bibnamefont {De~Roeck}},\
  }\href {\doibase 10.1103/PhysRevLett.121.140601} {\bibfield  {journal}
  {\bibinfo  {journal} {Phys. Rev. Lett.}\ }\textbf {\bibinfo {volume} {121}},\
  \bibinfo {pages} {140601} (\bibinfo {year} {2018})}\BibitemShut {NoStop}%
\bibitem [{\citenamefont {Potirniche}\ \emph {et~al.}(2019)\citenamefont
  {Potirniche}, \citenamefont {Banerjee},\ and\ \citenamefont
  {Altman}}]{Potirniche_2019}%
  \BibitemOpen
  \bibfield  {author} {\bibinfo {author} {\bibfnamefont {I.-D.}\ \bibnamefont
  {Potirniche}}, \bibinfo {author} {\bibfnamefont {S.}~\bibnamefont
  {Banerjee}}, \ and\ \bibinfo {author} {\bibfnamefont {E.}~\bibnamefont
  {Altman}},\ }\href {\doibase 10.1103/PhysRevB.99.205149} {\bibfield
  {journal} {\bibinfo  {journal} {Physical Review B}\ }\textbf {\bibinfo
  {volume} {99}},\ \bibinfo {pages} {205149} (\bibinfo {year} {2019})},\
  \bibinfo {note} {arXiv: 1805.01475}\BibitemShut {NoStop}%
\bibitem [{\citenamefont {Moudgalya}\ \emph {et~al.}(pear)\citenamefont
  {Moudgalya}, \citenamefont {Huse},\ and\ \citenamefont
  {Khemani}}]{sanjay2020}%
  \BibitemOpen
  \bibfield  {author} {\bibinfo {author} {\bibfnamefont {S.}~\bibnamefont
  {Moudgalya}}, \bibinfo {author} {\bibfnamefont {D.~A.}\ \bibnamefont {Huse}},
  \ and\ \bibinfo {author} {\bibfnamefont {V.}~\bibnamefont {Khemani}},\
  }\href@noop {} {\  (\bibinfo {year} {to appear})}\BibitemShut {NoStop}%
\end{thebibliography}%


%merlin.mbs apsrev4-1.bst 2010-07-25 4.21a (PWD, AO, DPC) hacked
%Control: key (0)
%Control: author (72) initials jnrlst
%Control: editor formatted (1) identically to author
%Control: production of article title (-1) disabled
%Control: page (0) single
%Control: year (1) truncated
%Control: production of eprint (0) enabled
\begin{thebibliography}{14}%
\makeatletter
\providecommand \@ifxundefined [1]{%
 \@ifx{#1\undefined}
}%
\providecommand \@ifnum [1]{%
 \ifnum #1\expandafter \@firstoftwo
 \else \expandafter \@secondoftwo
 \fi
}%
\providecommand \@ifx [1]{%
 \ifx #1\expandafter \@firstoftwo
 \else \expandafter \@secondoftwo
 \fi
}%
\providecommand \natexlab [1]{#1}%
\providecommand \enquote  [1]{``#1''}%
\providecommand \bibnamefont  [1]{#1}%
\providecommand \bibfnamefont [1]{#1}%
\providecommand \citenamefont [1]{#1}%
\providecommand \href@noop [0]{\@secondoftwo}%
\providecommand \href [0]{\begingroup \@sanitize@url \@href}%
\providecommand \@href[1]{\@@startlink{#1}\@@href}%
\providecommand \@@href[1]{\endgroup#1\@@endlink}%
\providecommand \@sanitize@url [0]{\catcode `\\12\catcode `\$12\catcode
  `\&12\catcode `\#12\catcode `\^12\catcode `\_12\catcode `\%12\relax}%
\providecommand \@@startlink[1]{}%
\providecommand \@@endlink[0]{}%
\providecommand \url  [0]{\begingroup\@sanitize@url \@url }%
\providecommand \@url [1]{\endgroup\@href {#1}{\urlprefix }}%
\providecommand \urlprefix  [0]{URL }%
\providecommand \Eprint [0]{\href }%
\providecommand \doibase [0]{http://dx.doi.org/}%
\providecommand \selectlanguage [0]{\@gobble}%
\providecommand \bibinfo  [0]{\@secondoftwo}%
\providecommand \bibfield  [0]{\@secondoftwo}%
\providecommand \translation [1]{[#1]}%
\providecommand \BibitemOpen [0]{}%
\providecommand \bibitemStop [0]{}%
\providecommand \bibitemNoStop [0]{.\EOS\space}%
\providecommand \EOS [0]{\spacefactor3000\relax}%
\providecommand \BibitemShut  [1]{\csname bibitem#1\endcsname}%
\let\auto@bib@innerbib\@empty
%</preamble>
\bibitem [{\citenamefont {Panda}\ \emph {et~al.}(2020)\citenamefont {Panda},
  \citenamefont {Scardicchio}, \citenamefont {Schulz}, \citenamefont {Taylor},\
  and\ \citenamefont {{\v{Z}}nidari{\v{c}}}}]{Panda2020_FiniteSize}%
  \BibitemOpen
  \bibfield  {author} {\bibinfo {author} {\bibfnamefont {R.~K.}\ \bibnamefont
  {Panda}}, \bibinfo {author} {\bibfnamefont {A.}~\bibnamefont {Scardicchio}},
  \bibinfo {author} {\bibfnamefont {M.}~\bibnamefont {Schulz}}, \bibinfo
  {author} {\bibfnamefont {S.~R.}\ \bibnamefont {Taylor}}, \ and\ \bibinfo
  {author} {\bibfnamefont {M.}~\bibnamefont {{\v{Z}}nidari{\v{c}}}},\ }\href
  {\doibase 10.1209/0295-5075/128/67003} {\bibfield  {journal} {\bibinfo
  {journal} {{EPL} (Europhysics Letters)}\ }\textbf {\bibinfo {volume} {128}},\
  \bibinfo {pages} {67003} (\bibinfo {year} {2020})}\BibitemShut {NoStop}%
\bibitem [{\citenamefont {Abanin}\ \emph {et~al.}(2019)\citenamefont {Abanin},
  \citenamefont {Bardarson}, \citenamefont {Tomasi}, \citenamefont
  {Gopalakrishnan}, \citenamefont {Khemani}, \citenamefont {Parameswaran},
  \citenamefont {Pollmann}, \citenamefont {Potter}, \citenamefont {Serbyn},\
  and\ \citenamefont {Vasseur}}]{Abanin2019_FiniteSize}%
  \BibitemOpen
  \bibfield  {author} {\bibinfo {author} {\bibfnamefont {D.~A.}\ \bibnamefont
  {Abanin}}, \bibinfo {author} {\bibfnamefont {J.~H.}\ \bibnamefont
  {Bardarson}}, \bibinfo {author} {\bibfnamefont {G.~D.}\ \bibnamefont
  {Tomasi}}, \bibinfo {author} {\bibfnamefont {S.}~\bibnamefont
  {Gopalakrishnan}}, \bibinfo {author} {\bibfnamefont {V.}~\bibnamefont
  {Khemani}}, \bibinfo {author} {\bibfnamefont {S.~A.}\ \bibnamefont
  {Parameswaran}}, \bibinfo {author} {\bibfnamefont {F.}~\bibnamefont
  {Pollmann}}, \bibinfo {author} {\bibfnamefont {A.~C.}\ \bibnamefont
  {Potter}}, \bibinfo {author} {\bibfnamefont {M.}~\bibnamefont {Serbyn}}, \
  and\ \bibinfo {author} {\bibfnamefont {R.}~\bibnamefont {Vasseur}},\
  }\href@noop {} {\  (\bibinfo {year} {2019})},\ \Eprint
  {http://arxiv.org/abs/arXiv:1911.04501} {arXiv:1911.04501} \BibitemShut
  {NoStop}%
\bibitem [{\citenamefont {Papić}\ \emph {et~al.}(2015)\citenamefont {Papić},
  \citenamefont {Stoudenmire},\ and\ \citenamefont
  {Abanin}}]{Papic2019_Miniband}%
  \BibitemOpen
  \bibfield  {author} {\bibinfo {author} {\bibfnamefont {Z.}~\bibnamefont
  {Papić}}, \bibinfo {author} {\bibfnamefont {E.~M.}\ \bibnamefont
  {Stoudenmire}}, \ and\ \bibinfo {author} {\bibfnamefont {D.~A.}\ \bibnamefont
  {Abanin}},\ }\href {\doibase https://doi.org/10.1016/j.aop.2015.08.024}
  {\bibfield  {journal} {\bibinfo  {journal} {Annals of Physics}\ }\textbf
  {\bibinfo {volume} {362}},\ \bibinfo {pages} {714 } (\bibinfo {year}
  {2015})}\BibitemShut {NoStop}%
\bibitem [{\citenamefont {Fisher}(1995)}]{Fisher_1995}%
  \BibitemOpen
  \bibfield  {author} {\bibinfo {author} {\bibfnamefont {D.~S.}\ \bibnamefont
  {Fisher}},\ }\href {\doibase 10.1103/PhysRevB.51.6411} {\bibfield  {journal}
  {\bibinfo  {journal} {Phys. Rev. B}\ }\textbf {\bibinfo {volume} {51}},\
  \bibinfo {pages} {6411} (\bibinfo {year} {1995})}\BibitemShut {NoStop}%
\bibitem [{\citenamefont {Evers}\ and\ \citenamefont
  {Mirlin}(2008)}]{Evers_2005}%
  \BibitemOpen
  \bibfield  {author} {\bibinfo {author} {\bibfnamefont {F.}~\bibnamefont
  {Evers}}\ and\ \bibinfo {author} {\bibfnamefont {A.~D.}\ \bibnamefont
  {Mirlin}},\ }\href {\doibase 10.1103/RevModPhys.80.1355} {\bibfield
  {journal} {\bibinfo  {journal} {Rev. Mod. Phys.}\ }\textbf {\bibinfo {volume}
  {80}},\ \bibinfo {pages} {1355} (\bibinfo {year} {2008})}\BibitemShut
  {NoStop}%
\bibitem [{\citenamefont {Bahri}\ and\ \citenamefont
  {Vishwanath}(2014)}]{Bahri_SPTOP}%
  \BibitemOpen
  \bibfield  {author} {\bibinfo {author} {\bibfnamefont {Y.}~\bibnamefont
  {Bahri}}\ and\ \bibinfo {author} {\bibfnamefont {A.}~\bibnamefont
  {Vishwanath}},\ }\href {\doibase 10.1103/PhysRevB.89.155135} {\bibfield
  {journal} {\bibinfo  {journal} {Phys. Rev. B}\ }\textbf {\bibinfo {volume}
  {89}},\ \bibinfo {pages} {155135} (\bibinfo {year} {2014})}\BibitemShut
  {NoStop}%
\bibitem [{\citenamefont {Else}\ \emph {et~al.}(2016)\citenamefont {Else},
  \citenamefont {Bauer},\ and\ \citenamefont {Nayak}}]{Else_2016}%
  \BibitemOpen
  \bibfield  {author} {\bibinfo {author} {\bibfnamefont {D.~V.}\ \bibnamefont
  {Else}}, \bibinfo {author} {\bibfnamefont {B.}~\bibnamefont {Bauer}}, \ and\
  \bibinfo {author} {\bibfnamefont {C.}~\bibnamefont {Nayak}},\ }\href
  {\doibase 10.1103/PhysRevLett.117.090402} {\bibfield  {journal} {\bibinfo
  {journal} {Phys. Rev. Lett.}\ }\textbf {\bibinfo {volume} {117}},\ \bibinfo
  {pages} {090402} (\bibinfo {year} {2016})}\BibitemShut {NoStop}%
\bibitem [{\citenamefont {Luitz}\ \emph {et~al.}(2015)\citenamefont {Luitz},
  \citenamefont {Laflorencie},\ and\ \citenamefont {Alet}}]{Luitz}%
  \BibitemOpen
  \bibfield  {author} {\bibinfo {author} {\bibfnamefont {D.~J.}\ \bibnamefont
  {Luitz}}, \bibinfo {author} {\bibfnamefont {N.}~\bibnamefont {Laflorencie}},
  \ and\ \bibinfo {author} {\bibfnamefont {F.}~\bibnamefont {Alet}},\ }\href
  {\doibase 10.1103/PhysRevB.91.081103} {\bibfield  {journal} {\bibinfo
  {journal} {Phys. Rev. B}\ }\textbf {\bibinfo {volume} {91}},\ \bibinfo
  {pages} {081103} (\bibinfo {year} {2015})}\BibitemShut {NoStop}%
\bibitem [{\citenamefont {Harris}(1974)}]{Harris_1974}%
  \BibitemOpen
  \bibfield  {author} {\bibinfo {author} {\bibfnamefont {A.~B.}\ \bibnamefont
  {Harris}},\ }\href {\doibase 10.1088/0022-3719/7/9/009} {\bibfield  {journal}
  {\bibinfo  {journal} {Journal of Physics C: Solid State Physics}\ }\textbf
  {\bibinfo {volume} {7}},\ \bibinfo {pages} {1671} (\bibinfo {year}
  {1974})}\BibitemShut {NoStop}%
\bibitem [{\citenamefont {Chandran}\ \emph {et~al.}(2015)\citenamefont
  {Chandran}, \citenamefont {Laumann},\ and\ \citenamefont
  {Oganesyan}}]{Chandran2015_Harris}%
  \BibitemOpen
  \bibfield  {author} {\bibinfo {author} {\bibfnamefont {A.}~\bibnamefont
  {Chandran}}, \bibinfo {author} {\bibfnamefont {C.~R.}\ \bibnamefont
  {Laumann}}, \ and\ \bibinfo {author} {\bibfnamefont {V.}~\bibnamefont
  {Oganesyan}},\ }\href@noop {} {\enquote {\bibinfo {title} {Finite size
  scaling bounds on many-body localized phase transitions},}\ } (\bibinfo
  {year} {2015}),\ \Eprint {http://arxiv.org/abs/arXiv:1509.04285}
  {arXiv:1509.04285} \BibitemShut {NoStop}%
\bibitem [{\citenamefont {Kj\"all}\ \emph {et~al.}(2014)\citenamefont
  {Kj\"all}, \citenamefont {Bardarson},\ and\ \citenamefont
  {Pollmann}}]{Kjall2014_MBLOrder}%
  \BibitemOpen
  \bibfield  {author} {\bibinfo {author} {\bibfnamefont {J.~A.}\ \bibnamefont
  {Kj\"all}}, \bibinfo {author} {\bibfnamefont {J.~H.}\ \bibnamefont
  {Bardarson}}, \ and\ \bibinfo {author} {\bibfnamefont {F.}~\bibnamefont
  {Pollmann}},\ }\href {\doibase 10.1103/PhysRevLett.113.107204} {\bibfield
  {journal} {\bibinfo  {journal} {Phys. Rev. Lett.}\ }\textbf {\bibinfo
  {volume} {113}},\ \bibinfo {pages} {107204} (\bibinfo {year}
  {2014})}\BibitemShut {NoStop}%
\bibitem [{ft1()}]{ft1}%
  \BibitemOpen
  \href@noop {} {}\bibinfo {note} {Here, $\ket{\psi_x}$ is a state that is spin
  polarized in the $+y$-direction except for sites $L/2-1$ and $L/2$ which are
  prepared in the $+x$ and $-x$ directions respectively. Similarly,
  $\ket{\psi_z}$ is a state that is similarly spin polarized along $+y$ except
  for sites $L/2-1$ through $L/2+2$ where the spins are polarized in the
  $\hat{z}$ direction with the pattern
  $\uparrow\uparrow\downarrow\downarrow$.}\BibitemShut {Stop}%
\bibitem [{\citenamefont {Choi}\ \emph {et~al.}(2020)\citenamefont {Choi},
  \citenamefont {Zhou}, \citenamefont {Knowles}, \citenamefont {Landig},
  \citenamefont {Choi},\ and\ \citenamefont {Lukin}}]{Soonwon_Pulse}%
  \BibitemOpen
  \bibfield  {author} {\bibinfo {author} {\bibfnamefont {J.}~\bibnamefont
  {Choi}}, \bibinfo {author} {\bibfnamefont {H.}~\bibnamefont {Zhou}}, \bibinfo
  {author} {\bibfnamefont {H.~S.}\ \bibnamefont {Knowles}}, \bibinfo {author}
  {\bibfnamefont {R.}~\bibnamefont {Landig}}, \bibinfo {author} {\bibfnamefont
  {S.}~\bibnamefont {Choi}}, \ and\ \bibinfo {author} {\bibfnamefont {M.~D.}\
  \bibnamefont {Lukin}},\ }\href {\doibase 10.1103/PhysRevX.10.031002}
  {\bibfield  {journal} {\bibinfo  {journal} {Phys. Rev. X}\ }\textbf {\bibinfo
  {volume} {10}},\ \bibinfo {pages} {031002} (\bibinfo {year}
  {2020})}\BibitemShut {NoStop}%
\bibitem [{\citenamefont {Nandkishore}\ and\ \citenamefont
  {Potter}(2014)}]{Potter_2014}%
  \BibitemOpen
  \bibfield  {author} {\bibinfo {author} {\bibfnamefont {R.}~\bibnamefont
  {Nandkishore}}\ and\ \bibinfo {author} {\bibfnamefont {A.~C.}\ \bibnamefont
  {Potter}},\ }\href {\doibase 10.1103/PhysRevB.90.195115} {\bibfield
  {journal} {\bibinfo  {journal} {Phys. Rev. B}\ }\textbf {\bibinfo {volume}
  {90}},\ \bibinfo {pages} {195115} (\bibinfo {year} {2014})}\BibitemShut
  {NoStop}%
\end{thebibliography}%

\end{document}

\title{Supplemental Material:\\Emergent ergodicity at the transition between many-body localized phases}
\author{Rahul Sahay}
\thanks{These authors contributed equally to this work.}
\affiliation{Department of Physics, University of California, Berkeley, California 94720 USA}

\author{Francisco Machado}
\thanks{These authors contributed equally to this work.}
\affiliation{Department of Physics, University of California, Berkeley, California 94720 USA}

\author{Bingtian Ye}
\thanks{These authors contributed equally to this work.}
\affiliation{Department of Physics, University of California, Berkeley, California 94720 USA}

\author{Chris R. Laumann}
\affiliation{Department of Physics, Boston University, Boston, MA, 02215, USA}

\author{Norman Y. Yao}
\affiliation{Department of Physics, University of California, Berkeley, California 94720 USA}
\affiliation{Materials Science Division, Lawrence Berkeley National Laboratory, Berkeley, CA 94720, USA}

\maketitle

\section{Additional Numerical Data for Order Parameters, Level Statistics, and Entanglement}

\subsection{Symmetry-Breaking Model}

\begin{figure}
    \centering
    \includegraphics[width = \textwidth]{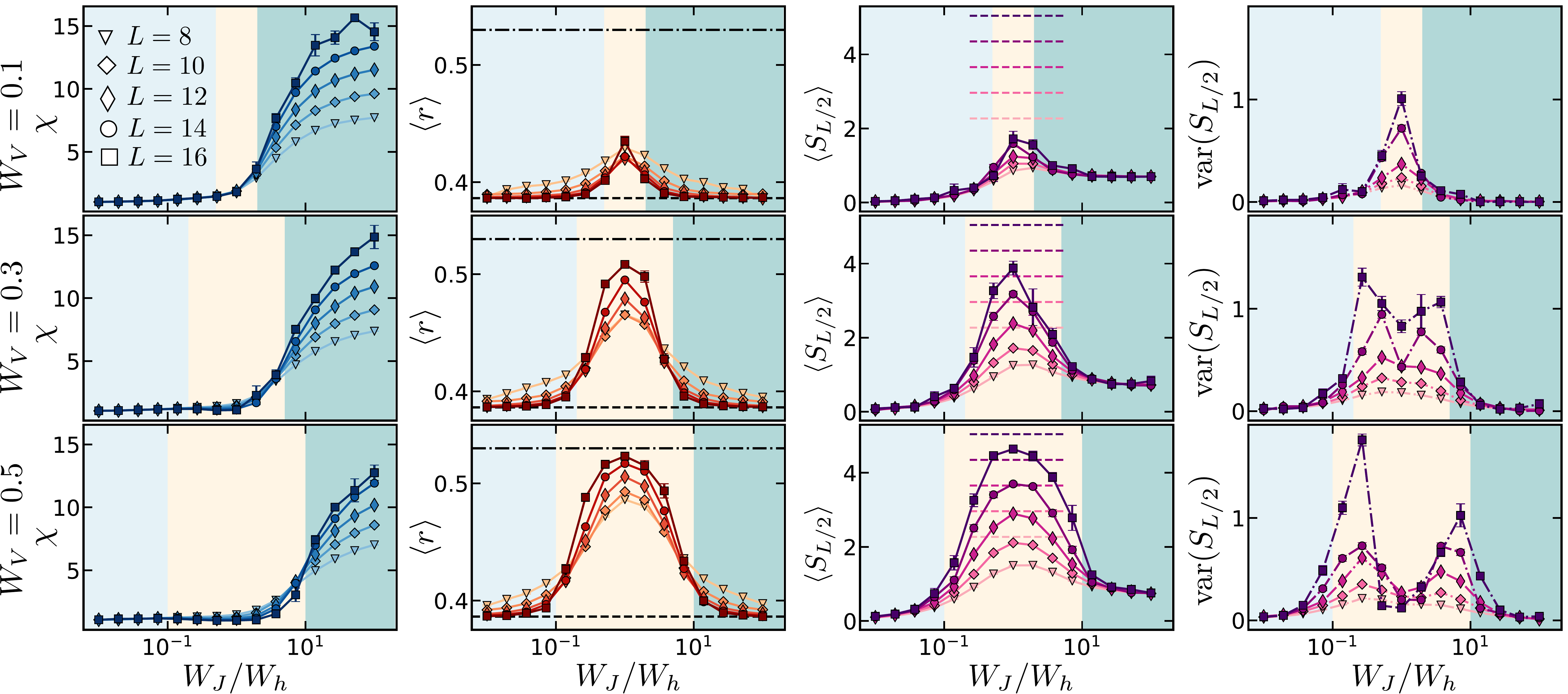}
    \caption{The order parameter $\chi$, the $\langle r \rangle$-ratio, the half-chain entanglement $S_{L/2}$, and the variance of the entanglement $\mathrm{var}(S_{L/2})$ as a function of $W_J/W_h$ for the model of Eqn.~1 of the main text. 
    From the top row to the bottom, the interaction strengths are chosen to bess $W_V\in \{0.1, 0.3, 0.5\}$ (from left to right).  For panels depicting the $\langle r \rangle$-ratio, the dash-dotted [dashed] line corresponds to the GOE [Poisson] expectation.
    A minimum of $3 \cdot 10^2$ disorder averages are performed for each quantity.}
    \label{fig:1}
\end{figure}

In the main text, we employed the order parameter $\chi$, the $\langle r \rangle$-ratio, the half-chain entanglement $S_{L/2}$, and the variance of the entanglement $\mathrm{var}(S_{L/2})$ to diagnose the phase diagram of the symmetry breaking model (Eqn.~1) and reveal the existence of an intervening thermal phase.
In this section, we provide additional data for these quantities at a variety of interaction strengths $W_V \in \{0.07, 0.1, 0.3, 0.5\}$ (Fig.~\ref{fig:1}) to complement the data shown at $W_V = 0.7$ in Fig.~2(a-d) of the main text. 
%Here, we provide additional data of the order parameter $\chi$, $\langle r \rangle$-ratio, half-chain entanglement $S_{L/2}$, and variance of the entanglement $\mathrm{var}(S_{L/2})$ for the symmetry breaking model (Eqn.~1). 
%
%In particular, we focus on different interaction strengths $W_V\in \{0.1, 0.3, 0.5\}$, which complement the $W_V = 0.7$ considered in the main text, Fig.~\ref{fig:1}.
%
%
%\begin{figure}
%    \centering
%    \includegraphics[width = \textwidth]{Supp_Figure_2_Schematic.pdf}
%    \caption{Schematic Layout of the Final Figure. Top row is $V = 0.1$, Middle Row is $V = 0.3$, and Bottom Row is $V = 0.5$. }
%    \label{fig:2}
%\end{figure}

\begin{figure}
    \centering
    \includegraphics[width = 0.4\textwidth]{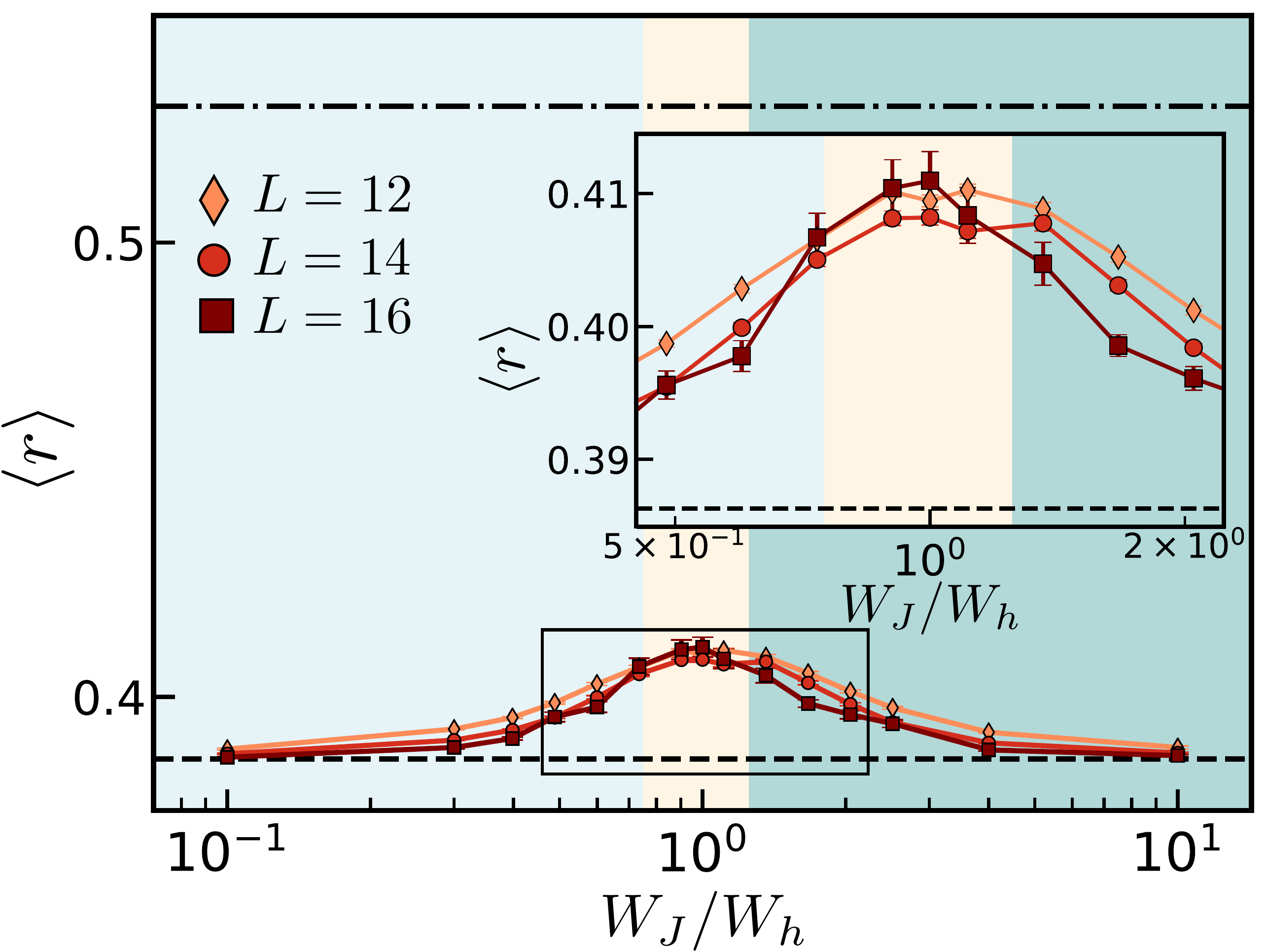}
    \caption{$\langle r \rangle$-ratio as a function of the tuning parameter $W_J/W_h$ at low-interaction $W_V = 0.07$. The dash-dotted [dashed] line corresponds to GOE [Poisson] value. Zooming into the region around criticality (inset), one can see evidence for an intervening thermal phase---at the largest system size considered ($L = 16$) and around $W_J / W_h = 1$, the $\langle r \rangle$ begins increasing toward the thermal value, suggesting a flow toward the thermal value. A minimum of $10^3$ disorder realizations were averaged over for the $L = 12, 14$ curves and $10^2$ disorder realizations were averaged over for the $L = 16$ curve.}  
    \label{fig:1.5}
\end{figure}

%We note that across all $W_V$, signatures of the intervening thermal phase remain present; however, due to finite size effects, these become less prominent at low interactions \cite{} \fm{what do you want to cite? Shouldn't we just say that it becomes thinner at smaller interaction strength?}. %I say interactions twice here
We note that across all $W_V$, signatures of the intervening thermal phase remain present. 
In the case of the $\langle r \rangle$-ratio and $S_{L/2}$ (shown in the middle two columns of Fig.~\ref{fig:1}), this signature manifests itself in a distinct change in finite size flow around $W_J/W_h = 1$. 
However, we comment that, at low interactions, due to both the small width of the intervening thermal phase and finite-size limitations \cite{Panda2020_FiniteSize, Abanin2019_FiniteSize, Papic2019_Miniband}, the thermal phase cannot be resolved in certain numerical probes.
For example, at $W_V = 0.1$, the double peak structure of the variance of entanglement disappears; this is due to the fact that the width of these peaks (at the system sizes accessible) exceeds the width of the intervening thermal phase.
We remark that we see evidence of the intervening thermal phase to interaction strengths down to $W_V = 0.07$ (See Fig.~\ref{fig:1.5}).
However, in this parameter regime, we are limited by strong finite size effects and as such conventional methods for extracting critical points (e.g. finite-size scaling) are not feasible.

%yet from the scaling  hypothesis of $\langle r \rangle$-ratio (Sec.~\ref{sec:scaling}) one extracts a finite thermal region.

\begin{figure}
    \centering
    \includegraphics[width = 0.9\textwidth]{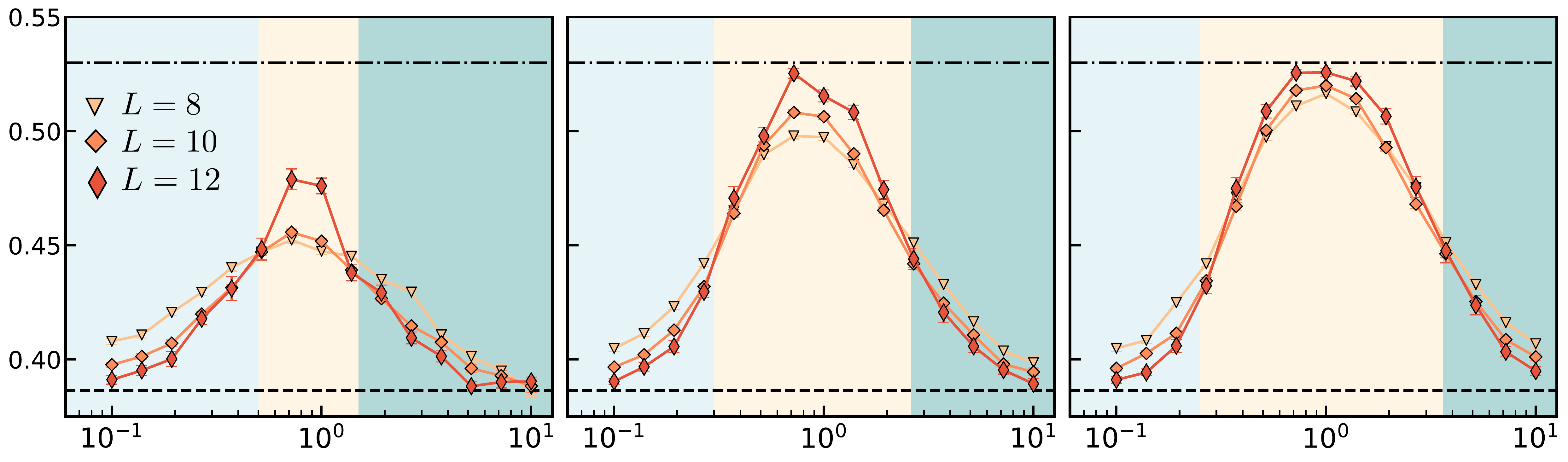}
    \caption{$\langle r\rangle$-ratio as a function of $h$ for the model of Eqn.~\ref{eq-AltSBModel}. The dash-dotted [dashed] line corresponds to the GOE [Poisson] expectation. From left to right, the panels show the $\langle r \rangle$-ratio at $V = 0.1, 0.3,$ and $0.5$. Observe that the behavior of the $\langle r \rangle$-ratio is qualitatively the same for this parameterization as it is for the parameterization in the main text. A minimum of $5 \cdot 10^2$ disorder realizations were averaged over for the $L = 8, 10$ curves and 20 disorder averages were performed for the $L = 12$ curve.}
    \label{fig:2}
\end{figure}

We conclude this section by demonstrating that all results for this model are independent of the parameterization and interactions chosen.
Focusing on the behavior of the  $\langle r \rangle$-ratio, we consider the model: 
\begin{equation}\label{eq-AltSBModel}
H = \sum_i J_i \sigma_i^z \sigma_{i+1}^z + \sum_i h_i \sigma_i^x + \sum_i V_i \sigma_{i}^z \sigma_{i+2}^z
\end{equation}
with fixed strength of the Ising coupling $J_i \in [1/2, 3/2]$, a variable transverse field $h_i \in [W_h/2, 3W_h/2]$ and interaction strength $V_i \in [W_V/2, 3W_V/2]$. 
Indeed, the observed behavior (Fig.~\ref{fig:2}) exhibits the same finite width thermal region as the model in main text (Fig.~1, Fig.~2b, and Fig.~\ref{fig:1}).
Unlike the model considered in the main text, Eqn.~(\ref{eq-AltSBModel}) is not self-dual under the Kramers-Wannier map ($\sigma_i^x \to \sigma_{i}^z \sigma_{i+1}^z$, $\sigma_i^z \to \prod_{j < i} \sigma_i^x$) which takes $W_J/W_h \to W_h/W_J$ and maps between the MBL SG and the MBL PM phases.
As such, the intervening thermal phase need not be centered at $W_J/W_h = 1$.

% Namely, we find that under such a map, $\sigma_i^z \sigma_{i+2}^z \to \sigma_i^x \sigma_{i+1}^x$ which will change the underlying Hamiltonian. As such, the phase diagram is not constrained to be symmetric about $W_h = 1$.

\subsection{Characterizing Finite-Size Effects}

In the main text, we commented on the severity of finite-size effects in exact diagonalization numerics performed with either weak interaction strength or disorder distributions strongly peaked at zero.
In this section, we expand this discussion by focusing on a single-parameter family of disorder distributions that emerge in the context of strong disorder RG \cite{Fisher_1995, Evers_2005}. 

Heuristically, if there are links in the chain whose coupling is smaller than the many-body level spacing, we expect the system to be 'cut' across those links, producing apparent localization in any finite-size diagnostics.  
As the many-body level spacing decays as $ \delta \sim L / 2^L$ (possibly with additional sub-exponential corrections due to symmetries), this does not reflect the thermodynamic limit for any system with extensively many couplings sampled from an $O(1)$ distribution. 
On the other hand, for system sizes accessible to exact diagonalization $\delta$ can remain larger than the sampled coupling strengths.
This is particularly important when attempting to characterize the phase diagram of the model at weak interaction strength or when the disorder distribution is strongly peaked at zero.
Consider the symmetry breaking model of Eqn.~1 of the main text with a generalized disorder distribution: $J_i = J X^J_i, h_i = h X^h_i$, and $V_i = V X^V_i$ where $X^{J}_i$, $X^{J}_i$, and $X^{V}_i$ are distributed according to 
\begin{equation}\label{eq-WDistro}
P_W(X) = \frac{1}{W} \left(\frac{1}{X}\right)^{1 - 1/W}
\end{equation}
with $X \in (0, 1]$. 
Here, $W$ characterizes the strength of the disorder: 
the numerics reported in the main-text are performed with uniform distributions ($W=1$),  $W > 1$  produces a power law divergence at small coupling strength and $W \to \infty$ corresponds to the infinite-randomness fixed point.
For a chain of length $L$, the typical minimum coupling is of order $J_{\textrm{min}} \sim L^{-W}$. 
Thus, at any given system size $L$, there is a large enough $W$ such that $J_{\textrm{min}}$ is typically smaller than the level spacing $\delta$, which is only weakly dependent on $W$. 

\begin{figure}
    \centering
    \includegraphics[width = 0.7\textwidth]{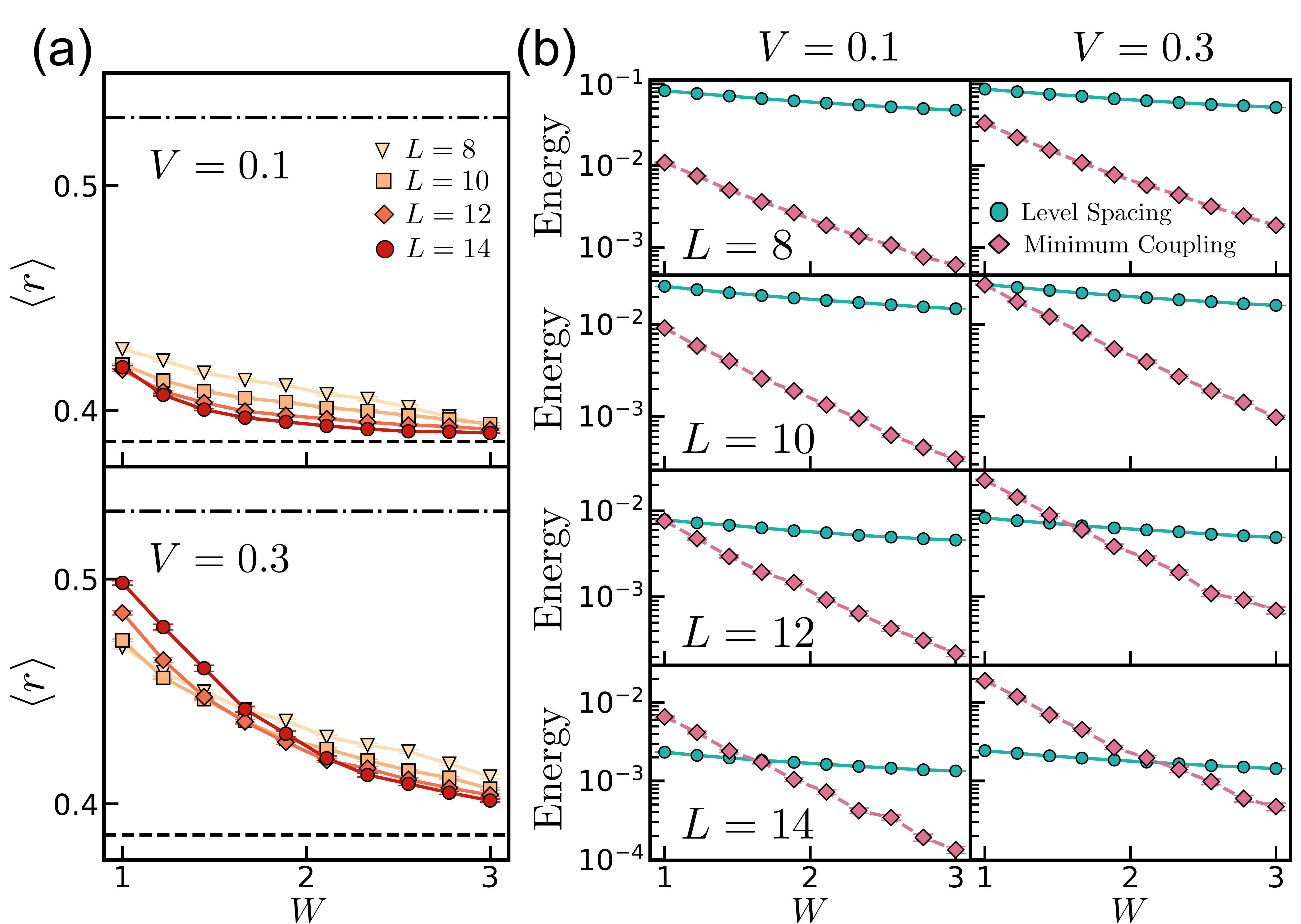}
    \caption{(a) The $\langle r \rangle$-ratio as a function of $W$ for $V = 0.1$ and $V = 0.3$ (top and bottom row respectively). The dash-dotted [dashed] line corresponds to the GOE [Poisson] expectation of the $\langle r \rangle$-ratio.
    (b) The disorder averaged mean level spacing and minimum coupling as a function of $W$ for $L = 8, 10, 12, 14$ (top to bottom). The middle column shows the $V = 0.1$ case and the right column shows the $V = 0.3$ case. At least $10^3$ disorder realization were averaged over for each data point. }
    \label{fig:Wplot}
\end{figure}

In Fig.~\ref{fig:Wplot} we compare the disorder averaged mean level spacing and minimum interaction coupling as a function of $W$ for $J = h = 1$ and $V = 0.1$ and $0.3$. Furthermore, we report the behavior of the $\langle r \rangle$-ratio as a function of $W$ for these same parameters.

For $V = 0.1$ and $L = 8, 10$, the mean level spacing $\delta$ is much greater than the average minimum coupling for all $W$ suggesting strong finite-size limitations.
By contrast, at larger system size ($L = 12, 14$) and near the uniform distribution case ($W \sim 1$) the average minimum interaction coupling becomes larger than the mean level spacing.
In parallel, the $\langle r \rangle$-ratio appears to flow localized between system sizes $L = 8, 10, 12$ but starts to flow thermal for system sizes between $L = 12, 14$ for $W \sim 1$ indicating that apparent localized behavior is an artifact of the chain cutting finite-size effect, Fig.~\ref{fig:Wplot}a inset.

For $V = 0.3$ and the smallest size considered ($L=8$) the mean level spacing is larger than the average minimum coupling, but already for $L\gtrsim 10$ and $W\lesssim 2$ this trend reverses.
Correspondingly, the approach of $\langle r \rangle$-ratio to the thermal value is much clearer. %
Strikingly, the regime of $W$ where the $\langle r \rangle$-ratio starts to become localized is in quantitative agreement with where the mean-level spacing starts to exceed the minimum coupling.

We are led to the conclusion that even for significant interaction strength ($V=0.3$), numerical results may be strongly biased toward apparent localization for large $W$ due to these strong finite size effects.

\subsection{Supplementary Analysis for the SPT and DTC Models}

In order to demonstrate that the intervening ergodic region generically emerges between different MBL phases, we analyzed two additional models distinct from the symmetry breaking model of Eqn.~1: a model with an SPT transtion and a model with an DTC transition (Eqs.~2~and~3; data shown in Fig.~2e,f).
%
%We have shown that an ergodic region emerges between distinct MBL phases by analyzing two different models in the main text: a model with an SPT transtion and a model with an DTC transition (Eqs.~2~and~3; data shown in Fig.~2e,f). 
%\fm{Confused about this sentence... we also talk about the SG... Don't we just want to say: "In the main text we briefly introduced the SPT and DTC model. Here we complement the analysis by considering other metrics as well as the phase diagram"}
%
In this section, we provide additional numerical data for these models and construct the associated phase diagrams (analogous to the diagram in Fig.~1a). 

\begin{figure}
    \centering
    \includegraphics[width = \textwidth]{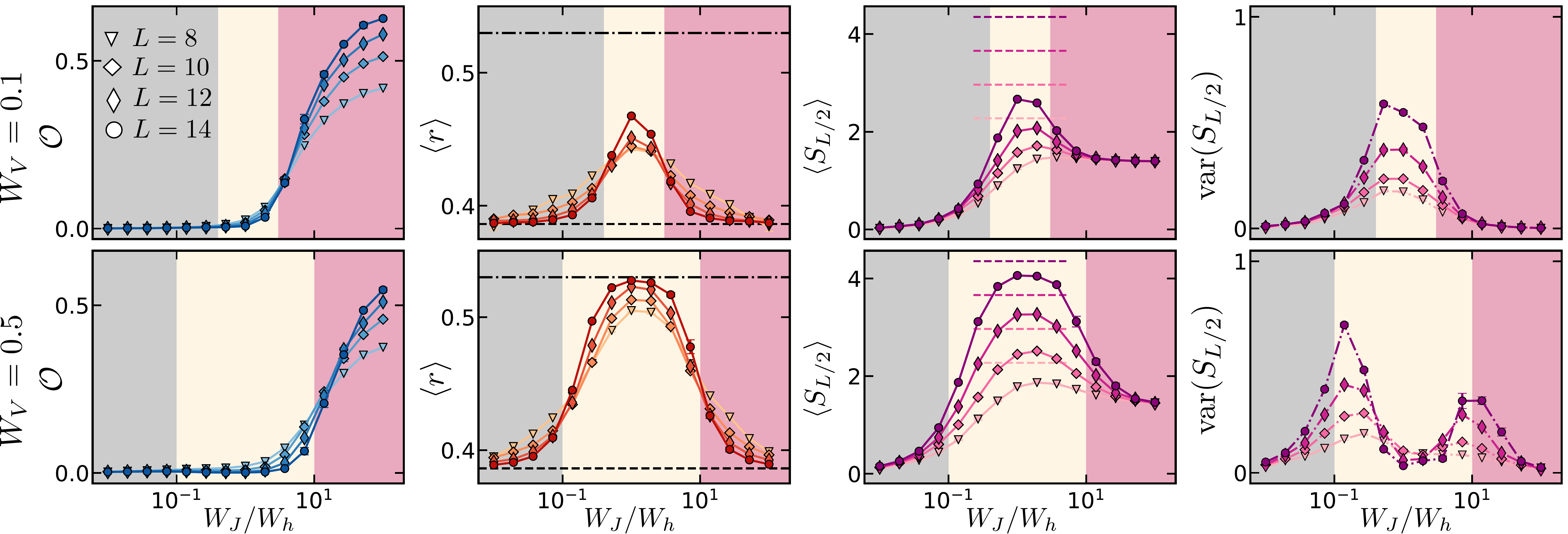}
    \caption{We present the order parameter $\mathcal{O}$, the $\langle r \rangle$-ratio, the half-chain entanglement $S_{L/2}$, and the variance of the entanglement $\mathrm{var}(S_{L/2})$ for interaction strengths $W_V\in \{0.1, 0.5\}$ for the SPT model (Eqn.~2) of the main text. For panels depicting the $\langle r \rangle$-value, the dash-dotted [dashed] line corresponds to the GOE [Poisson] value. At least $9 \cdot 10^2$ disorder averages were performed for each system size.}
    \label{fig:3}
\end{figure}

We begin with the SPT model in Eqn.~2.
In order to diagnose the different phases of the SPT model, we compute a string order parameter \cite{Bahri_SPTOP}:
\begin{equation}
\mathcal{O} = \left\llangle  \frac{1}{L}\sum_{i, j; i+2 \leq j-2} \bra{n} \sigma_i^z \sigma_{i+1}^y \left( \prod_{k = i +2}^{j-2} \sigma_k^x \right) \sigma_{j-1}^y \sigma_j^z \ket{n}^2\right\rrangle
\end{equation}
where $\ket{n}$ is the eigenstate at the center of the many-body spectrum and $\llangle \cdots \rrangle$ indicates averaging over disorder realizations. 
This order parameter is the non-local analogue of the Edwards-Anderson order parameter for the symmetry-breaking model and as such scales extensively in the MBL SPT phase and saturates to an $\mathcal{O}(1)$ constant in either the thermal or trivial MBL phases. 
The data for this order parameter, along with $\langle r \rangle$-ratio, half-chain entanglement $S_{L/2}$ and the variance of the half-chain entanglement $\mathrm{var}(S_{L/2})$ are shown in Fig.~\ref{fig:3} for interaction strength $W_V = 0.1, 0.5$.

\begin{figure}
    \centering
    \includegraphics[width = \textwidth]{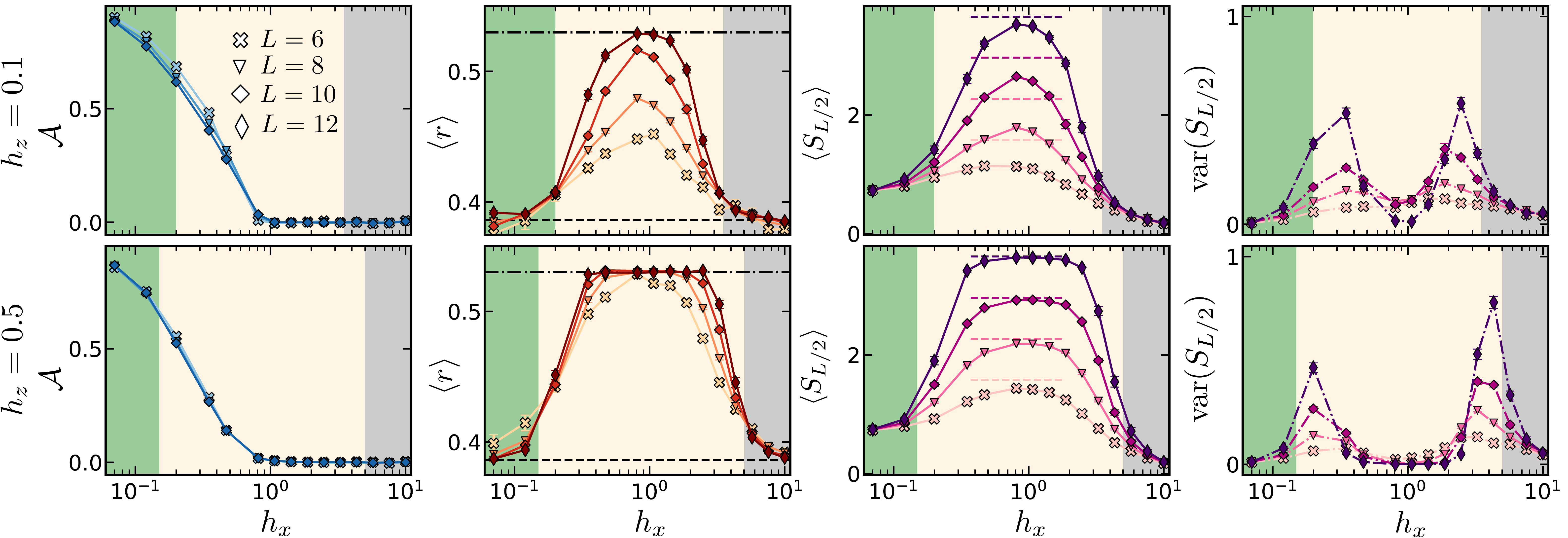}
    \caption{The order parameter $\mathcal{A}$, the $\langle r \rangle$-ratio, the half-chain entanglement $S_{L/2}$, and the variance of the entanglement $\mathrm{var}(S_{L/2})$ for interaction strengths $h_z\in \{0.1, 0.5\}$ for the DTC model (Eqn.~2) of the main text. For panels depicting the $\langle r \rangle$-value, the dash-dotted [dashed] line corresponds to the GOE [Poisson] value. At least $3 \cdot 10^2$ disorder averages were performed for each system size.
    }
    %Numerics for $\mathcal{A}$ are only presented up to system size $10$ due to computation cost.}
    \label{fig:4}
\end{figure}

We now turn to the DTC model in Eqn.~3, where the order parameter is given by:
\begin{equation}
\mathcal{A} = \frac{1}{N_T L}\sum_{t = 1}^{N_T} \sum_{i = 1}^{L} (-1)^t \text{Tr}\{\sigma_i^z(t T) \sigma_i^z (0) \}
\end{equation}
where $T$ represents the period of one Floquet cycle.
This order parameter is non-zero in the DTC phase and is zero in the trivial Floquet MBL phase \cite{Else_2016}.
In our numerics, $N_T$ is set to $20$. 
From our observations, this choice does not affect the numerical value of the order parameter. 
Data for this order parameter, along with $\langle r \rangle$-ratio and half-chain entanglement $S_{L/2}$ are shown in Fig.~\ref{fig:4} for interaction strengths $h_z = 0.1, 0.5$.

In both cases, we see evidence of an intervening thermal phase persistent to low interactions. 
The phase diagram for both the SPT and DTC models can be extracted using finite-size scaling (see the next subsection) of the $\langle r \rangle$-ratio numerics.  Both phase diagrams exhibit the intervening thermal phase the two MBL phases, as in Fig.~1, highlighting the generality of this feature.

\begin{figure}
    \centering
    \includegraphics[width = 0.7\textwidth]{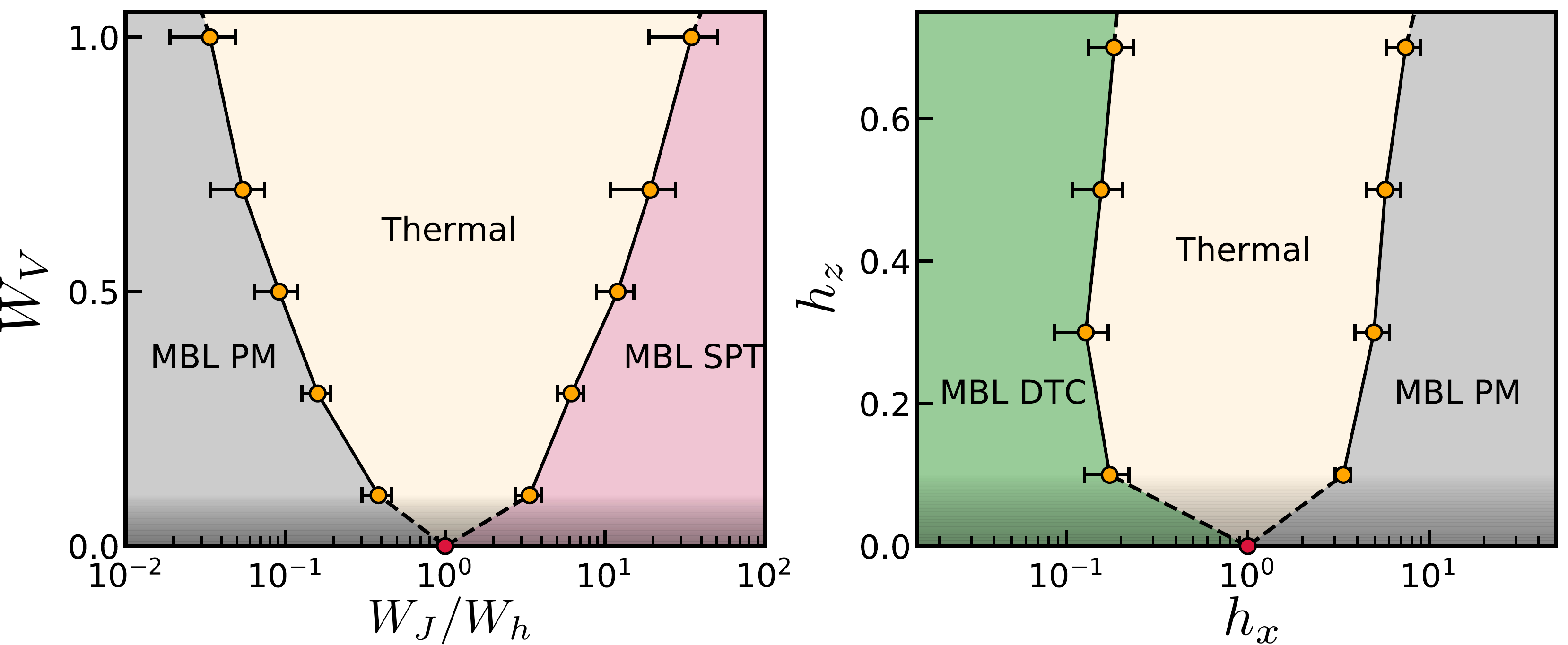}
    \caption{Phase diagrams of the SPT (left panel) and DTC (right panel) models of the main text. For all interaction strengths, we find an intervening thermal phase separating the two MBL phases. The red point on the $x$-axis of either diagram indicates the location of the non-interacting critical point. }
    \label{fig:5}
\end{figure}

\subsection{Finite-Size Scaling Analysis}
\label{sec:scaling}
\begin{figure}
    \centering
    \includegraphics[width = 0.8\textwidth]{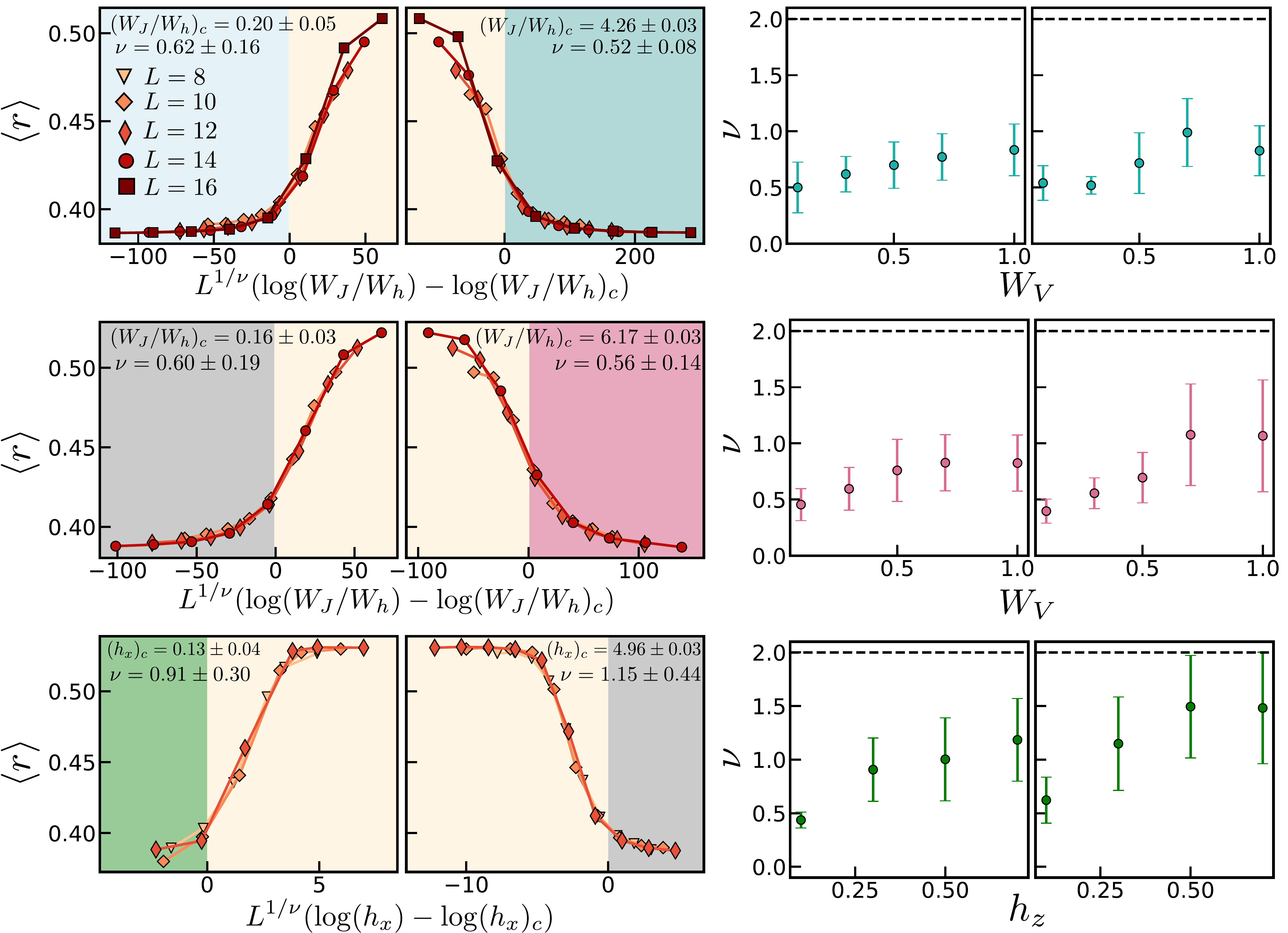}
    \caption{Left two columns: representative collapses for the $\langle r \rangle$-ratio for the models of Eqn.~1,~2,and~3 of the main text at interaction strengths $W_V = 0.3$, $W_V = 0.3$, and $h_z = 0.3$ respectively (top to bottom). The critical exponent extracted from these collapses are shown in each panel. Right two columns: the extracted critical exponents as a function of interaction strength. We note that each extracted exponent is well below the dashed line at $\nu = 2$, which is known analytically as the lower bound for critical exponent in generic disordered systems.  }
    \label{fig:6}
\end{figure}

\begin{figure}
    \centering
    \includegraphics[width = 0.8\textwidth]{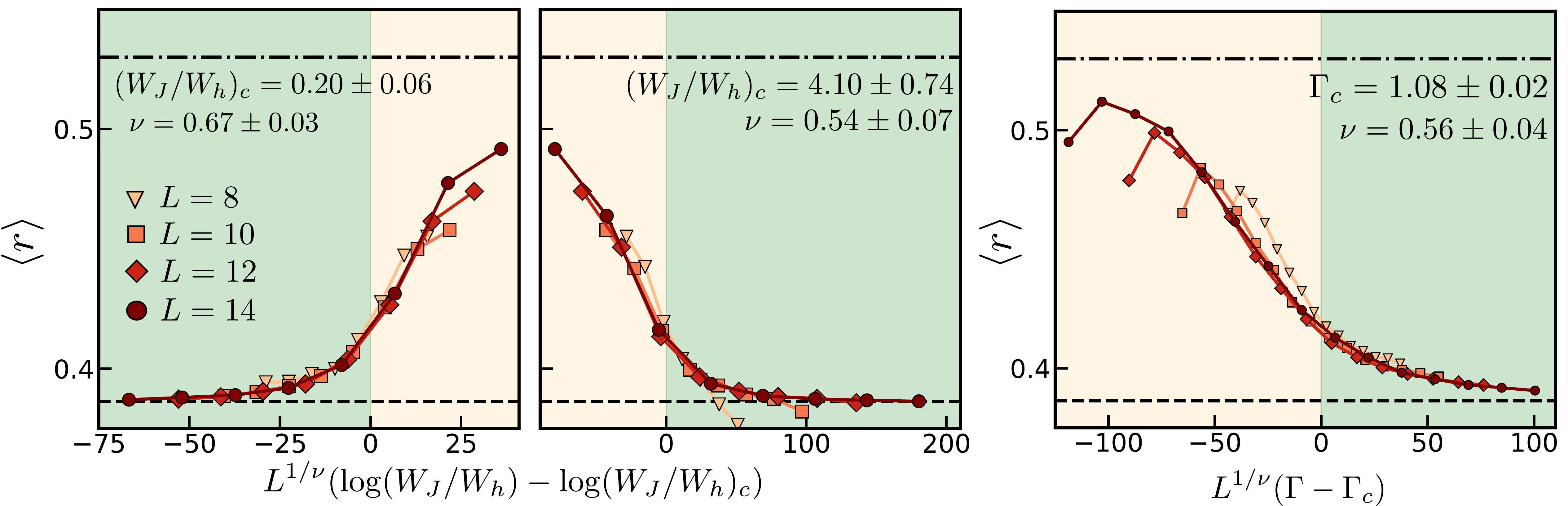}
    \caption{The left two panels depict representative collapses for the $\langle r\rangle$-ratio as a function of $W_J/W_h$ for $\Gamma = 0.5$ and $W_V = 0.3$. The dash-dotted [dashed] line corresponds to GOE [Poisson] value. Such collapses extract critical points for the two thermal MBL transitions along $W_J/W_h$ at fixed $\Gamma, W_V$. The right most panel depicts a collapse of the $\langle r \rangle$-ratio as a function of $\Gamma$ at $W_J/W_h = 1, W_V = 0.3$. This collapse extracts a critical point for the thermal MBL transition as we increase the symmetry-breaking field $\Gamma$.}
    \label{fig:6.5}
\end{figure}

In this section, we detail the finite-size scaling analysis performed to extract the phase boundaries of different models considered in the main text (Fig.~1 and Fig.~\ref{fig:5}) .
In particular, all phase boundaries are extracted via scaling collapse around a crossing point of the $\langle r \rangle$-ratio as a function some tuning parameter $\Delta$ ($\log(W_J/W_h)$ for the symmetry-breaking and SPT model, $\log(h_x)$ for the DTC model, and $\Gamma$ in the explicit symmetry-breaking analysis).
To perform the finite-size scaling analysis, we assume a standard $\langle r \rangle$-ratio scaling ansatz for the thermal-MBL transition \cite{Luitz}:
\begin{equation}
r_L(\Delta) = \tilde{f}(L^{1/\nu} (\Delta - \Delta_c))     
\end{equation}
where $r_L(\Delta)$ is the disorder averaged $\langle r \rangle$-ratio at $\Delta$ and length $L$. To extract the critical point, we first linearly interpolate the simulated data points $\{r_L(\Delta_i)\}$ as a function of $\Delta$ yielding an interpolation $\tilde{r}_L(\Delta)$ which we can then scale and sample from.
The critical point $\Delta_c$ and critical exponent $\nu$ are extracted by numerically minimizing the square residual quality function: 
\begin{equation}
\mathcal{Q}(\Delta_c, \nu) = \sum_{i, j} \sum_{k} (\tilde{r}_{L_i} ((\Delta_k - \Delta_c)L^{1/\nu}) - \tilde{r}_{L_j}((\Delta_k - \Delta_c)L^{1/\nu})^2     
\end{equation}
where $i,j$ index the system sizes that we simulate (e.g. $\{8, 10, 12, 14, 16\}$ for the model of Eqn.~1) and $k$ indexes the set of $\{(\Delta_k - \Delta)L^{1/\nu}\}$. The resulting scaling collapses for each model can be found in Fig.~\ref{fig:6} and the scaling collapses for the symmetry-breaking analysis can be found in Figure~\ref{fig:6.5}.
The errors in the critical point $\delta \Delta$ and the critical exponent $\delta \nu$ are estimated by adding a perturbation to each data point of the $\langle r\rangle$-ratio which is sampled from a uniform distribution with a width matching the standard error of that data point. The error reported is the standard deviation of the extracted distribution of $\Delta_c$ and $\nu$.
%
%For any given interaction strength, we use the $\langle r \rangle$-ratio as a function of the tuning parameter $\Delta$ ($W_J/W_h$ for the symmetry-breaking and SPT model and $h_x$ for the DTC model) to extract two critical points via finite-size scaling collapse.
%
%Namely, we use the standard $\langle r \rangle$-ratio scaling ansatz for the thermal-MBL transition \cite{Luitz}:
%$$r_L(\Delta) = \tilde{f}(L^{1/\nu} (\Delta - \Delta_c)) $$
%for $\Delta$ surrounding the crossing point associated with the transition of interest.
%
%To extract the critical point, we first linearly interpolate the simulated data points $\{r_L(\Delta_i)\}$ as a function of $\Delta$ yielding $\tilde{r}_L(\Delta)$.
%
%Subsequently, we finely sample $(\Delta - \Delta_c) L^{1/\nu}$ and compute the quality function 
%$$\mathcal{Q}(\Delta_c, \nu) = \sum_{i, j} \sum_k (\tilde{r}_{ik} - y_{jk})^2 $$
%

While all extracted critical exponents violate known analytic bounds \cite{Harris_1974, Chandran2015_Harris}, these results are consistent with previous numerical studies \cite{Luitz, Kjall2014_MBLOrder}.

\section{Additional Numerical Data for the Explicit Symmetry Breaking Analysis}

\begin{figure}
    \centering
    \includegraphics[width = \textwidth]{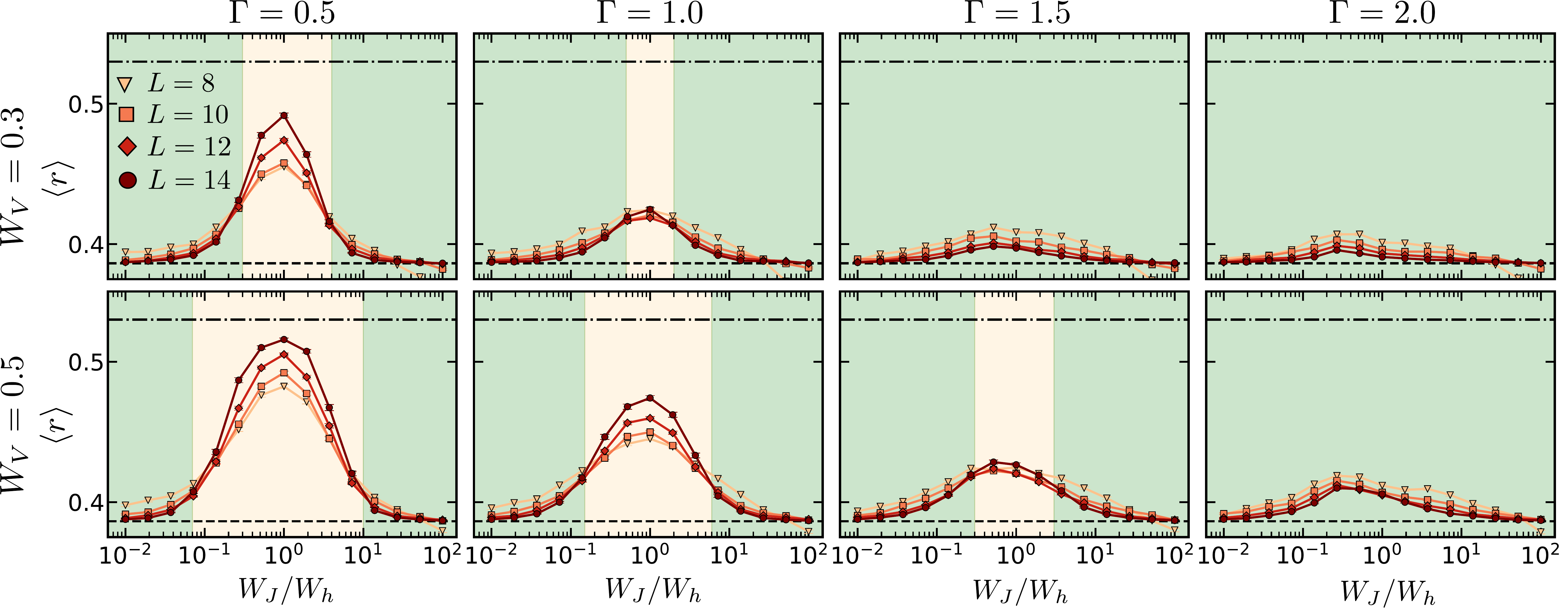}
    \caption{The two rows exhibit the $\langle r \rangle$-ratio as a function of $W_J/W_h$ for interaction strengths $W_V\in \{ 0.3, 0.5\}$ with $\Gamma \in \{0.5, 1.0, 1.5, 2.0\}$ in different columns. The dash-dotted [dashed] line corresponds to the GOE [Poisson] expectation of the $\langle r \rangle$-ratio. Each data point corresponds to averaging over at least $3 \cdot 10^2$ disorder realizations.}
    \label{fig:7}
\end{figure}

In the main text, we explored the effect of explicitly breaking the $\mathbb{Z}_2$-symmetry of Eqn.~1 by introducing a disorder-less longitudinal field $\sim \Gamma \sum_i \sigma_i^z$. Interestingly, we observed that, upon increasing $\Gamma$, the intervening thermal phase disappeared (Fig.~1b and Fig.~3). In this section, we provide additional numerical data highlighting the effects of this symmetry-breaking field on the intervening thermal phase. First, we consider the $\langle r \rangle$-ratio as a function of $W_J/W_h$ for $\Gamma \in \{0.5, 1, 1.5, 2\}$ and $V \in \{0.3, 0.5\}$ in Fig.~\ref{fig:7}. 

\begin{figure}
    \centering
    \includegraphics[width = 0.4\textwidth]{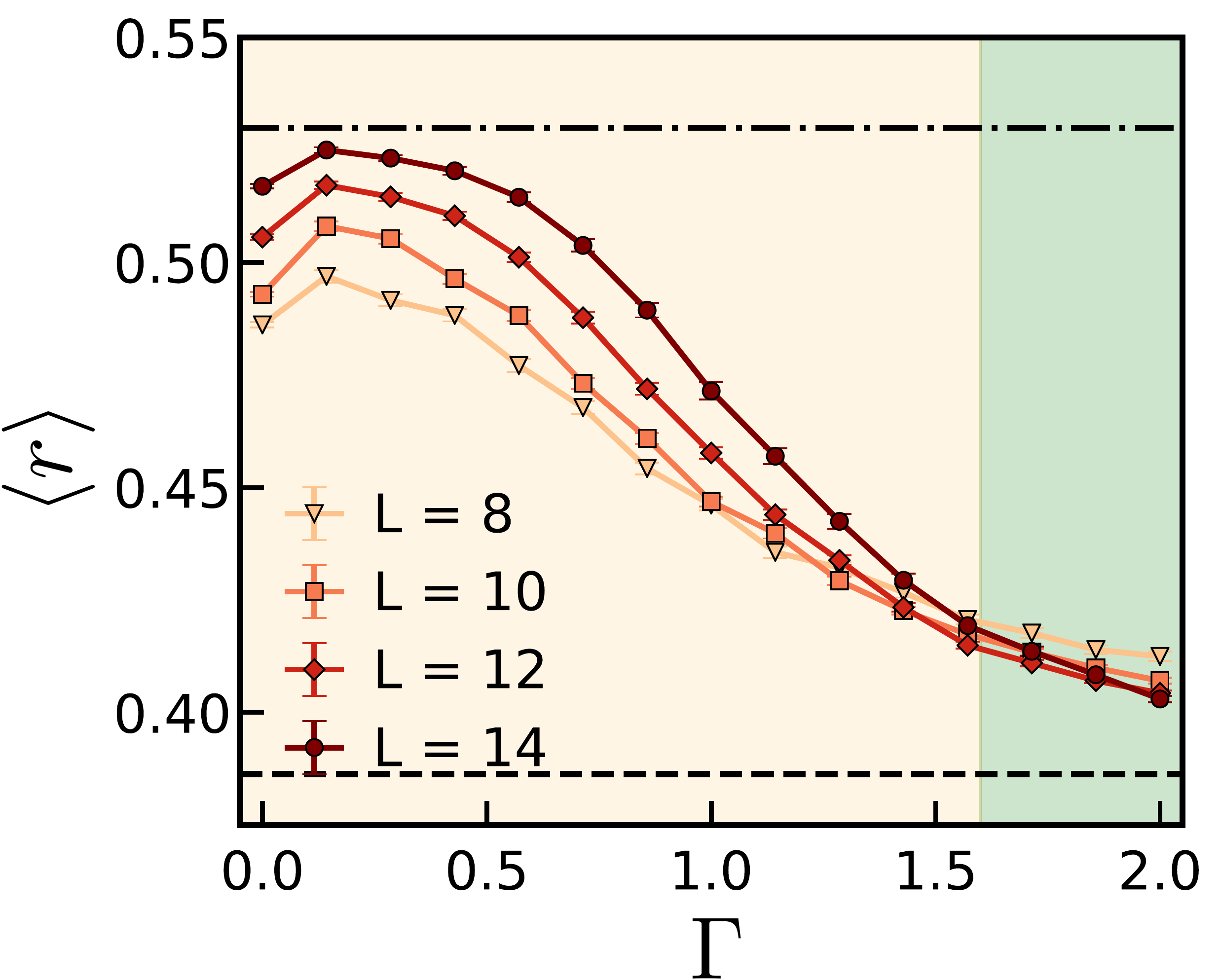}
    \caption{The $\langle r \rangle$-ratio as a function of the explicit symmetry breaking strength $\Gamma$ for $W_J/W_h = 1$. The dash-dotted [dashed] line corresponds to the GOE [Poisson] value.  Each data point corresponds to averaging over at least $3 \cdot 10^2$ disorder realizations. }
    \label{fig:8}
\end{figure}

For both interaction strengths, we find that, upon increasing $\Gamma$, the intervening thermal phase is suppressed and eventually disappears. Additionally, we show the $\langle r \rangle$-ratio as a function of $\Gamma$ for $W_J/W_h = 1$ and $W_V = 0.5$ in Fig.~\ref{fig:8}.
The crossing of the curves with different system sizes in Fig.~\ref{fig:8} indicates a transition between thermal and localized behavior. This is qualitatively the same as the results presented in Fig.~3b of the main text.

\section{Additional Details for Experimental Proposal}

%In the main text, we proposed an experimental protocol to observe an intervening thermal phase, wherein Rydberg dressing in an optical lattice was used to simulate the Hamiltonian in Eqn.~4.

%Want to say
%Proposed experimental protocol on neutral atom simulators with Rydberg dressing
%Our protocol hinged on the fact that the intervening thermal phase could be distinguished via the dynamics of local observables in experimentally accessible decoherence time scales.

\subsection{Dynamics of Local Observables in ``Zig-Zag'' and ``Linear'' Chain Geometries}

In the main text, we proposed an experimental protocol which can be naturally implemented in one-dimensional chains of neutral atoms trapped in an optical lattice. 
Here, optical dressing and Floquet engineering techniques are used in order to simulate time-evolution under the Hamiltonian of Eqn.~4.
%
%which could be naturally implemented in one-dimensional chains of neutral atoms trapped in an optical lattice.
%
%This protocol makes use of optical dressing along with Floquet engineering techniques in order to simulate time-evolution under the Hamiltonian of Eqn.~4.
%
In Fig.~4, we demonstrated that, by examining the dynamics of local observables, one could diagnose the phase diagram of the aforementioned Hamiltonian and see evidence of an intervening thermal phase. 
The numerics presented were all in the ``strong interaction'' regime ($\bar{J}_{i, i+2} = 0.6 \bar{J}_{i, i+1}$) which was motivated by the ``zig-zag'' geometry of Fig.~4a.
In this section, we provide additional numerics at ``strong interaction'' and further provide numerics in a ``weak interaction'' regime ($\bar{J}_{i, i+2} = 0.2 \bar{J}_{i, i+1}$) which can be realized via atoms arranged in a ``linear'' geometry (say a single row or column of a square optical lattice).

\begin{figure}
    \centering
    \includegraphics[width = 0.7\textwidth]{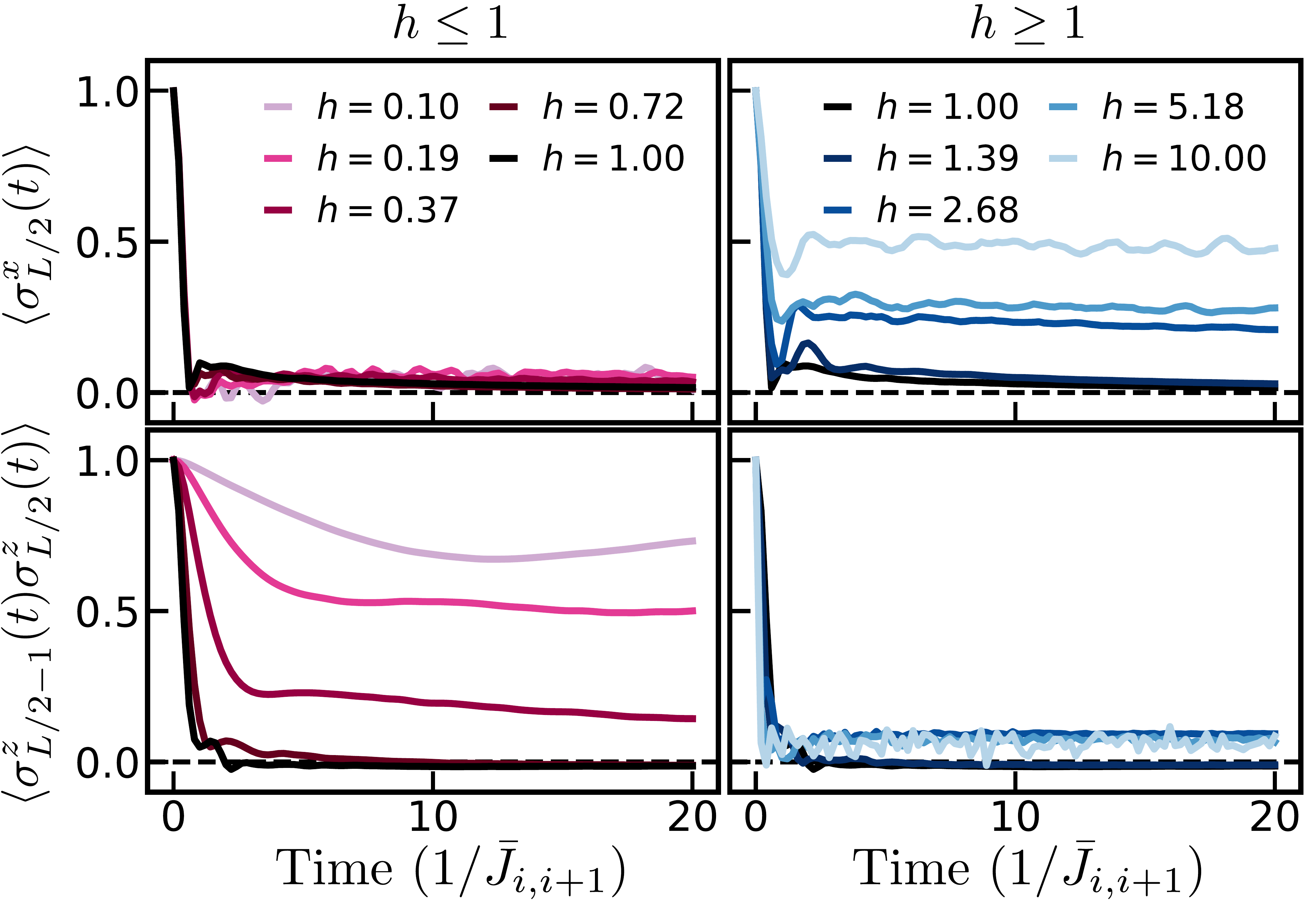}
    \caption{The two rows depict $\langle \sigma^x_{L/2}(t) \rangle$ and $\langle \sigma^z_{L/2-1}(t) \sigma^z_{L/2}(t) \rangle$ for a variety of $h$'s. The columns correspond to  $\bar{h}\leq 1$ and $\bar{h}\geq 1$ and highlight the contrast of these observables in the MBL PM and MBL SG phase. We note that curves around $\bar{h} = 1$ (e.g. $\bar{h} \in [0.72, 1.39]$) all observables plateau at the thermal expectation implying a finite-width intervening thermal phase. Each data point corresponds to averaging over at least $10^2$ disorder realizations.}
    \label{fig:9}
\end{figure}

In Fig.~\ref{fig:9}, we present the decay of $\bra{\psi_x} \sigma_{L/2}^{x} \ket{\psi_x} \equiv \langle \sigma_{L/2}^x \rangle$ and $\bra{\psi_z} \sigma_{L/2-1}^z \sigma_{L/2}^z \ket{\psi_z} \equiv \langle\sigma_{L/2-1}^z \sigma_{L/2}^z  \rangle $ for the ``strong interaction'' regime (See \cite{ft1} for the definition of $\ket{\psi_x}$ and $\ket{\psi_{zz}}$).
%e
Within $20/J$, all of the curves near $h = 1$ approximately saturate to zero, indicating the presence of the intervening thermal phase. We note that late-time plateaus for these curves are extracted by averaging each curve between $t \in [19/J, 20/J]$. These late-time plateaus are shown as a function of $h$ in Fig.~4d. 
We remark that the non-zero late-time plateau for $\langle \sigma_{L/2}^x (t)\rangle$  [$\langle\sigma_{L/2-1}^z \sigma_{L/2}^z  \rangle$] in the MBL SG [PM] phase is not a finite-time effect; the $l$-bit in these regimes has a finite overlap with the corresponding observable.

\begin{figure}
    \centering
    \includegraphics[width = 0.5\textwidth]{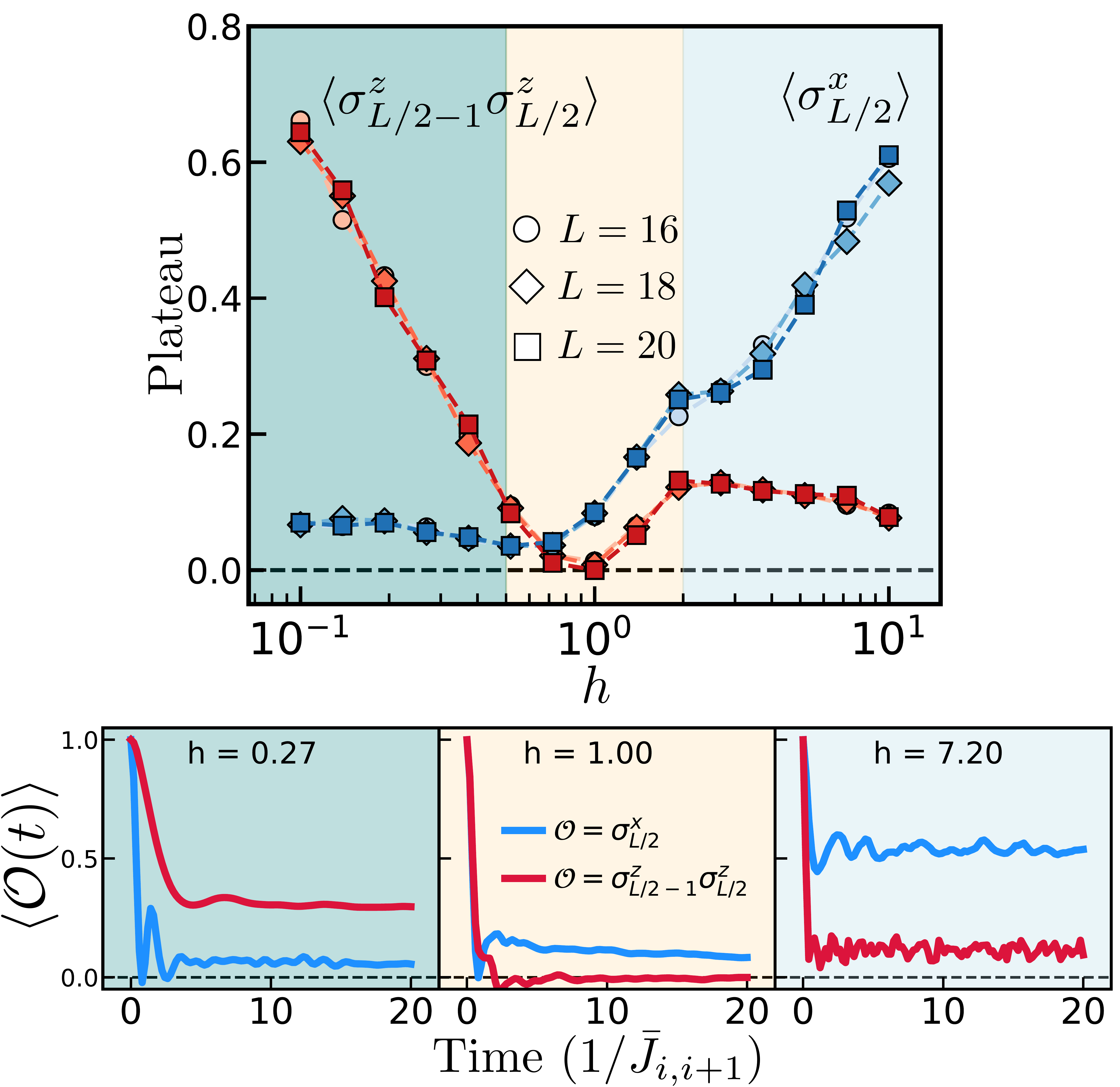}
    \caption{The top plot  depicts the late-time plateau (as extracted by averaging $\langle \mathcal{O}(t) \rangle$ over $t \in [19/J, 20/J]$ as a function of $h$. Representative decays for $h \in \{0.27, 1.00, 7.20\}$ are shown in the bottom three panels. Each data point corresponds to averaging over at least $10^2$ disorder realizations. }
    \label{fig:10}
\end{figure}

We conclude this section by presenting numerics in the ``weak interaction'' regime (with $\bar{J}_{i,i+2} = 0.2\bar{J}_{i, i+1}$). In the top panel of Fig.~\ref{fig:10}, we show the late-time plateaus of  $\langle \sigma_{L/2}^x (t)\rangle$  and $\langle\sigma_{L/2-1}^z \sigma_{L/2}^z  \rangle$ as a function of $h$ (analogous to the ``strong interaction'' plateaus that we showed in Fig.~4d of the main text). Observe that in the predicted thermal regime (the yellow region), it appears that the plateau of $\langle \sigma_{L/2}^x(t) \rangle$ is finite, contrary to the thermal prediction. This apparent contradiction is explained by examining the decays in the bottom panel of Fig.~\ref{fig:10}. Here, for $h = 1$, we see that, at $20/\bar{J}_{i, i+1}$,  $\langle \sigma_{L/2}^x(t) \rangle$ is far from saturating and is clearly downwards sloped. As such, in order to see clear evidence of the intervening thermal phase in the ``weak-interaction'' regime, it is necessary to go to later times. 

%We note that the presence of an intervening thermal phase is unclear from the extracted plateaus because, as one can see by the middle decay panel of Fig.~\ref{fig:10}, $\langle \sigma^x_{L/2}(t)\rangle$ is unable to reach it's late-time value in $20/J$. 
%\by{What do you want to say by this last sentence? It's not clear what do you mean by 'because' here... You may either modify this sentence to convey some clear/meaningful idea, or just remove it and merge this small paragraph with the previous one. }

%In Fig.~\ref{fig:10}, we find that, within the aforementioned time-scale, our curves do not saturate (Fig.~\ref{fig:10}). This is due to the fact that, at low interactions, the intervening thermal phase is thin and hence throughout the phase, one is close to a thermal-MBL transition. Due to the proximity to the transition, local observable equilibrate slowly and hence are difficult to observe plateau in experimentally accessible time scales.

\subsection{Trotterization Analysis}

Our experimental proposal hinged on implementing the Hamiltonian of Eqn.~4 via a two-stage Floquet drive (Fig.~4b in the main text). 
In this section, we present an analysis of the conditions required to ensure faithful engineering Floquet.

\begin{figure}
    \centering
    \includegraphics[width = 0.8\textwidth]{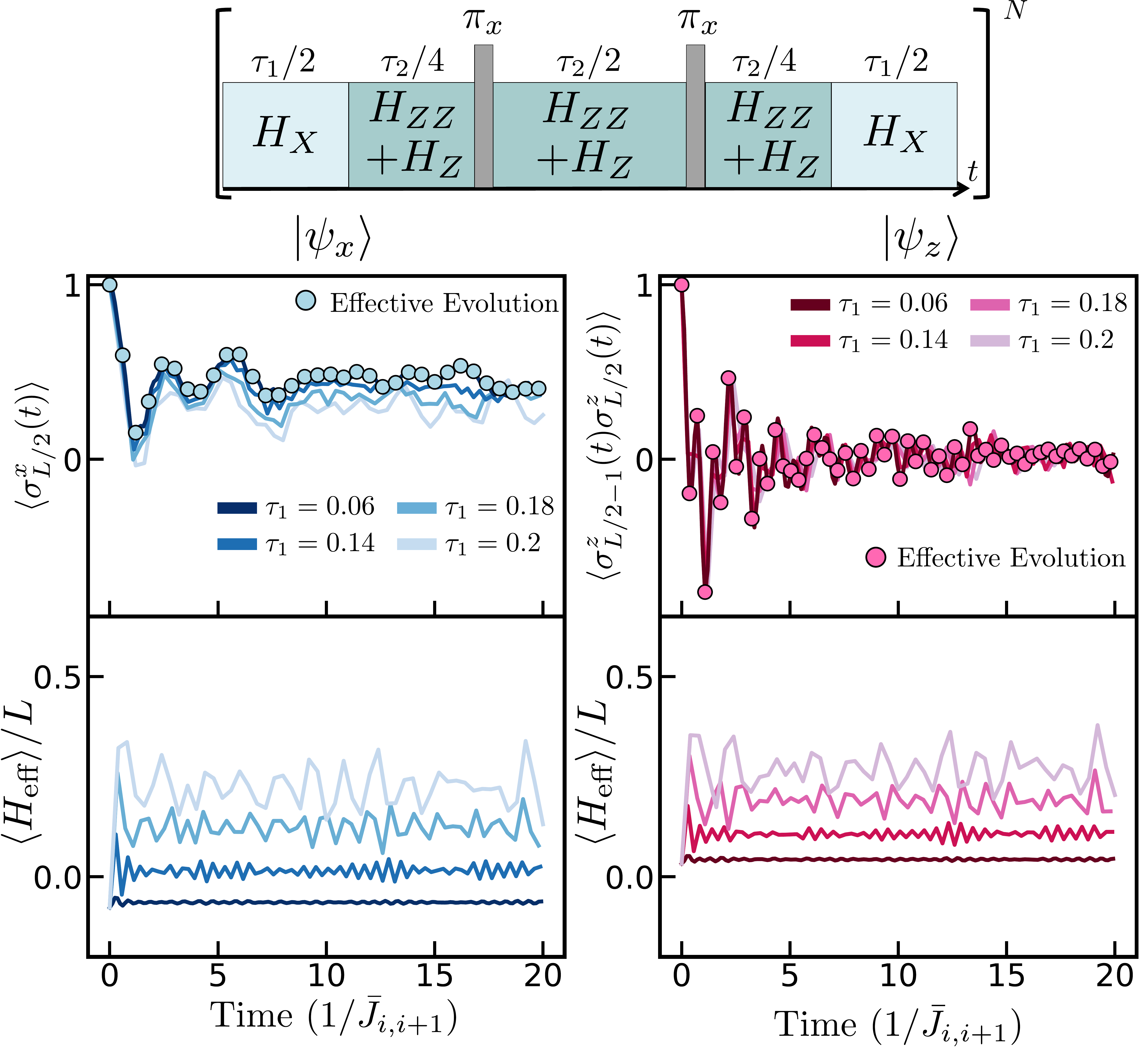}
    \caption{Trotterized evolution in the MBL PM phase ($\bar{h}_i= 5$, $\bar{J}_{i, i+1} = 1$, $\bar{J}_{i,i+2} = 0.6 \bar{J}_{i, i+1} $ and $\tau_1 = \tau_2$)---In the top row, we show a single disorder realization of $\langle \sigma^x_{L/2}(t) \rangle$ and $\langle \sigma^z_{L/2-1}(t) \sigma^z_{L/2}(t) \rangle$ under evolution by the effective Hamiltonian (points) and the trotterized evolution (line). The bottom row displays the expectation of the effective Hamiltonian per lattice site.}
    \label{fig:11}
\end{figure}

\begin{figure}
    \centering
    \includegraphics[width = 0.8\textwidth]{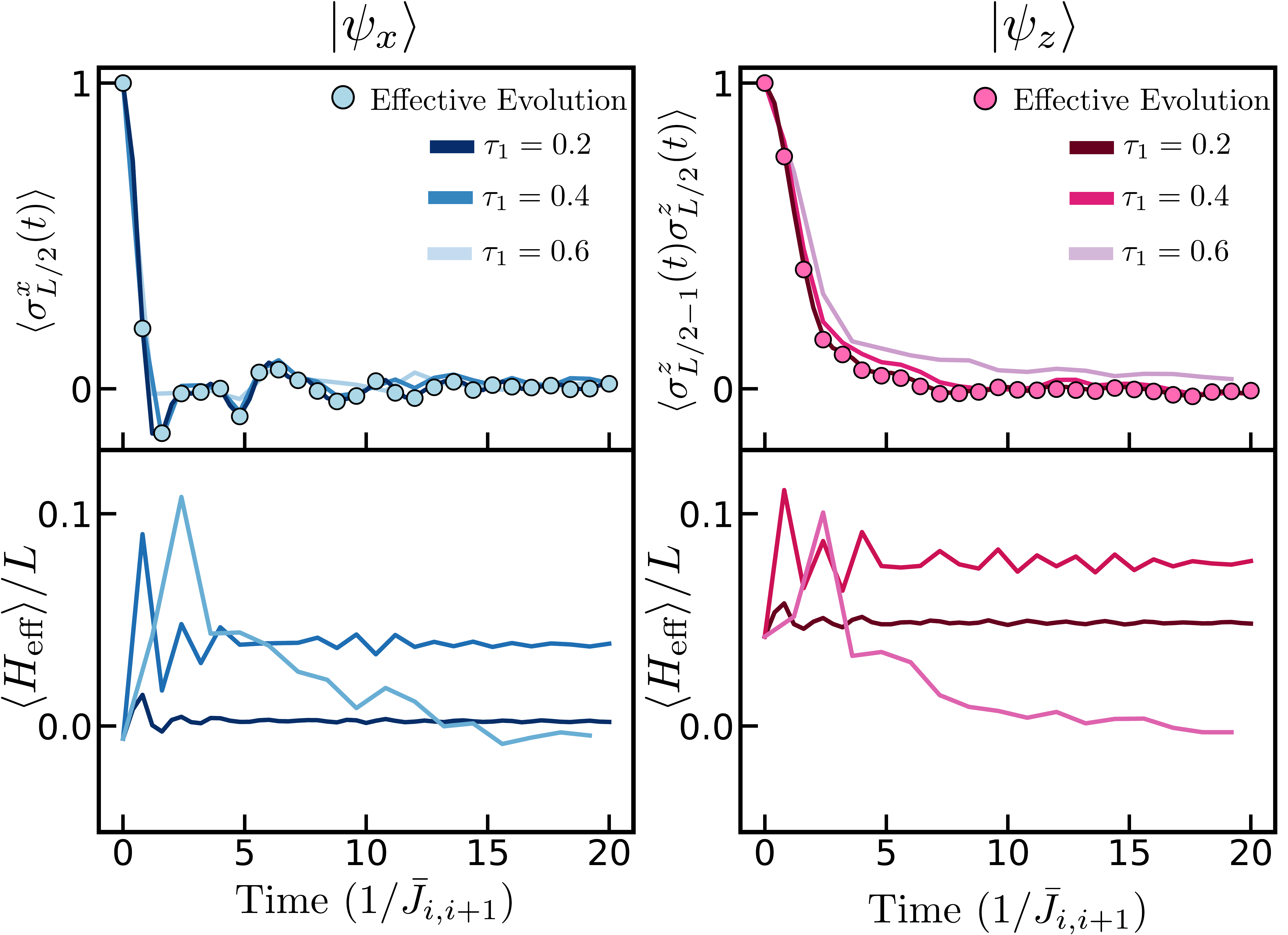}
    \caption{Trotterized evolution in the thermal phase ($\bar{h}_i= 1$, $\bar{J}_{i, i+1} = 1$, $\bar{J}_{i,i+2} = 0.6 \bar{J}_{i, i+1} $ and $\tau_1 = \tau_2$)---In the top row, a single disorder realization of $\langle \sigma^x_{L/2}(t) \rangle$ and $\langle \sigma^z_{L/2-1}(t) \sigma^z_{L/2}(t) \rangle$ is shown under both evolution by the effective Hamiltonian (points) and the true trotterized evolution (line). The bottom row displays the expectation of the effective Hamiltonian per lattice site.}
    \label{fig:12}
\end{figure}

In the main text, our driving protocol approximated the evolution under the time-independent Hamiltonian of Eqn.~4 as
\begin{equation}\label{eq-Approx1}
\exp\left(- iH_{\text{eff}}(\tau_1 + \tau_2)\right) \approx \exp\left(-i H_{X} \tau_1\right) \exp\left(-i H_{ZZ}\tau_2\right).
\end{equation}
While in the infinite-frequency limit, this sequence yields the correct effective Hamiltonian (Eqn.~4 of the main text), the leading order finite frequency corrections is linear in the inverse frequency of the drive.
% However, such an approximation and its associated pulse sequence (shown in Fig.~4b) is not ideal because, for experimentally accessible local energy scales and pulse timings, $\tau_1$ and $\tau_2$,  the first order term in Magnus expansion is non-negligible and faithful evolution of $H_{\eff}$ is not possible. 
Fortunately, such linear term can be cancelled by symmetrizing the Floquet sequence \cite{Soonwon_Pulse} (without requiring an additional $\pi$-pulse) as shown in the top of Fig.~\ref{fig:11}:
\begin{equation}\label{eq-Approx2}
\exp\left(-iH_{\text{eff}}(\tau_1 + \tau_2)\right) \approx \exp\left(-i  H_{X} \tau_1/2\right) \exp\left(-i H_{ZZ}\tau_2\right) \exp\left(-i H_{X} \tau_1/2\right)~.
\end{equation}
This enables us to consider larger values of $\tau_1$ and $\tau_2$ while keeping the simulation error small.
In particular, to leading order, the trotterization error is controlled by the small parameter
\begin{equation}\label{eq-Condition}
    \bar{h}_i \bar{J}_{i,i+1}\tau_1 \tau_2\cdot\mathrm{max}\{\bar{h}_i\tau_1, \bar{J}_{i,i+1}\tau_2\} \ll 1.
\end{equation}
With this Floquet sequence in hand, we seek to demonstrate that faithful Floquet evolution can be realized in experimentally relevant parameter regimes. 
To this end, we choose two typical sets of parameter values, ($\bar{h}_i= 5$, $\bar{J}_{i, i+1} = 1$, $\bar{J}_{i,i+2} = 0.6 \bar{J}_{i, i+1} $ and $\tau_1 = \tau_2$) and ($\bar{h}_i= 1$, $\bar{J}_{i, i+1} = 1$, $\bar{J}_{i,i+2} = 0.6 \bar{J}_{i, i+1} $ and $\tau_1 = \tau_2$ ), corresponding to the MBL PM and thermal phase respectively.
Using such parameters, we compare trotterized and effective Hamiltonian evolution for a single disorder realization (Figs.~\ref{fig:11}~and~\ref{fig:12}).
In particular, for both parameter sets, we compute the energy density with respect to the effective Hamiltonian $\langle H_{\mathrm{eff}} \rangle /L$, and the two local observables used to diagnose different phases, $\langle \sigma_{L/2}^x \rangle$ and $\langle\sigma_{L/2-1}^z \sigma_{L/2}^z \rangle $.
As we decrease the Floquet period $\tau_1$, both the Floquet heating effects and the discrepancy in local observables are quickly suppressed. 
For the MBL PM phase, when $\bar{J}_{i, i+1}\tau_1 \lesssim 0.14$ , the trotterized evolution very well approximate evolution under the time-independent Hamiltonian. 
In contrast, in the thermal regime, faithful evolution can be achieved with much longer pulse timings with $\bar{J}_{i, i+1}\tau_1 \lesssim 0.4$.
Both $\tau_1$'s are accessible in current experimental setups for an appropriate choice of local energy scale $\bar{J}_{i, i+1}$. For the MBL PM case ($\tau_1 = \tau_2 = 0.14/\bar{J}_{i, i+1}$), the decay of local observables start to saturate around $3/\bar{J}_{i,i+1}$, which corresponds approximately $20$ Floquet cycles. Similarly, for the thermal case  ($\tau_1 = \tau_2 = 0.4/\bar{J}_{i, i+1}$), local observables saturate in about $5/\bar{J}_{i,i+1}$ which is less than 20 Floquet cycles.

We end by noting that, even though $H_\eff$ is exactly conserved from the start of the evolution, for most choices of $\tau_1=\tau_2$ it remains flat.
This suggests that observed dynamics arise from higher order corrections to $H_\eff$ that are not included, rather than Floquet heating.

\section{Two-Body Resonance Counting at Infinite-Randomness}

In this section, we expand on the resonance counting criterion for the stability of localization of a non-interacting chain at infinite randomness against perturbative interactions. 
We consider a non-interacting Anderson-localized chain characterized by its density of single-particle states (DOS) $D(\varepsilon)$ and the localization length $\xi(\varepsilon)$ of single-particle orbitals (with energy $\varepsilon$).
At the infinite randomness fixed point, the localization length and DOS both diverge as $|\epsilon| \to 0$ and it is the interplay of such overlapping orbitals that could lead to two-body resonance proliferation.
We assume each orbital has a ``center'' at position $\alpha$ and an exponentially scaling envelope $\psi^*_\alpha \sim \frac{1}{\sqrt{\xi(\varepsilon_\alpha)}}e^{-|x - \alpha|/\xi(\varepsilon_\alpha)}$ determined by its energy $\varepsilon_\alpha$.
The presence of multiple centers (e.g. two in a typical state produced by the strong disorder renormalization group treatment of the Ising model) does not parametrically modify the estimates below.
Similarly, the presence of `pairing' terms in the fermionization of the Ising model is not parameterically important.

We consider a generic local interaction, which we schematically model by a density-density operator $\sim V \int dx\, \hat{n}(x) \hat{n}(x)$. 
Writing it in terms of the non-interacting orbitals, we have
\begin{equation}\label{eq-Vmatrix}
V_{\alpha \beta \gamma \delta} = V\int dx~\psi_{\alpha}(x) \psi_{\beta}(x) \psi^*_{\gamma}(x) \psi^*_{\delta}(x) \sim \frac{V}{\sqrt{\xi_{\alpha} \xi_{\beta} \xi_{\gamma} \xi_{\delta}}}\int~dx~e^{-(|x - \alpha|/\xi_{\alpha} +|x - \beta|/\xi_{\beta} + |x - \gamma|/\xi_{\gamma} + |x - \delta|/\xi_{\delta})}~.
\end{equation}
Two-body resonances occur when $V_{\alpha \beta \gamma \delta} > |(\varepsilon_{\alpha} - \varepsilon_{\delta}) - (\varepsilon_{\gamma} - \varepsilon_{\beta})| $. 
In general, any small finite strength of interactions produces some density of resonances, but this need not modify the ergodic properties of the system; instead it can ``dress'' the local conserved quantities to be many-particle operators---this is at the heart of MBL.
However, if the number of resonances in a localization volume becomes sufficiently large, then the local character of the conserved quantity is lost and we expect delocalization.
Counting the number of perturbative resonances, induced by interactions, can then identify instabilities to thermalization.

Owing to the localized nature of the single-particle orbitals, the matrix element will only be large whenever all four orbitals overlap.
Without loss of generality, we can take $\alpha$ to be the orbital with \emph{smallest} localization length $\xi_\alpha < \xi_\beta, \xi_\gamma, \xi_\delta$. For ease of notation let $\varepsilon = \varepsilon_\alpha$. 
This suggests the following organization of our counting: 
given such an orbital, first we compute how many other orbitals (labeled orbital $\delta$) exist within a block of size $\ell = \xi_\alpha$ around $\alpha$ and with energy $\delta\varepsilon$ around $\varepsilon$; second, given the energy difference between orbital $\alpha$ and $\delta$, what is the number of pairs of orbitals $\beta$ and $\gamma$ that have an energy difference within $V_{\alpha \beta \gamma \delta}$ of the initial pair.
Under this organization, one must have that both estimates diverge: the first ensures that there is always an initial pair that can transition, while the second ensures that, given a particular pair of orbitals, additional pairs can resonantly transition.
While the former can be simply estimated as $\delta \varepsilon D(\varepsilon) \ell$, the latter requires a more careful analysis.
Fixing the pair of resonances $\alpha$ and $\delta$, we must find the number of pairs of orbitals $\beta$ and $\gamma$ that satisfy three conditions:
(1) the within $\ell$ distance from orbital $\alpha$,
(2) their localization length is larger than $\ell$, and
(3) their energy difference close to the energy difference between $\alpha$ and $\gamma$ (where close is given by the strength of the matrix element).
The number $R$ of such pairs can be estimated as follows: given an orbital $\gamma$ within the block $\ell$, we need to find another orbital $\beta$ whose energy is in a window of size $V_{\alpha \beta \gamma \delta}$ around $\varepsilon_\gamma - \delta \varepsilon$.
At some energy $\varepsilon_\gamma < \varepsilon$, the number of such orbitals $\gamma$ is $\sim \ell D(\varepsilon_{\gamma}) d\varepsilon_\gamma$ and the number of corresponding orbitals $\beta$ is $\sim \ell D(\varepsilon_\beta) V_{\alpha \beta \gamma \delta}$, where $\varepsilon_\beta = \varepsilon_{\gamma} -  (\delta \varepsilon)$.
Integrating yields the total number of resonances:
\begin{equation}\label{purple-box}
R = \int_0^\varepsilon d\varepsilon_{\gamma}  \ell D(\varepsilon_\gamma) \ell D(\varepsilon_{\beta}) V_{\alpha \beta \gamma \delta} \sim \int_0^\varepsilon d\varepsilon_{\gamma}  \ell D(\varepsilon_\gamma) \ell D(\varepsilon_{\beta}) \frac{V \ell}{\sqrt{\xi_\alpha \xi_\beta \xi_{\gamma} \xi_{\delta}}}~.
\end{equation}
We make progress under the following approximation: take $\delta \varepsilon = C \varepsilon$ with a small $C$. 
Physically, this means that the initial orbitals have similar energies, and thus similar localization lengths, $\xi_\delta \approx \xi_\alpha = \ell$.

We can check that this counting argument reproduces previous work on interaction instabilities of localized systems in Ref.~\cite{Potter_2014}.
There, $D(\varepsilon)$ remains a constant, while the localization length diverges as a power-law, $\xi(\varepsilon) \sim \varepsilon^{-\nu}$.
The two conditions are then:
\begin{align}
    &C\varepsilon D(\varepsilon) \xi(\varepsilon) \sim \varepsilon^{1-\nu}\\
    &R \sim V \ell^2 \int_0^{\varepsilon}~d\varepsilon'~|\varepsilon' + C \varepsilon|^{\nu/2} |\varepsilon'|^{\nu/2} \sim \ell^2 |\varepsilon|^{\nu/2} |\varepsilon|^{1 + \nu/2} \sim \varepsilon^{1 -\nu}
\end{align}
Both quantities diverge when $1-\nu<0$, which agrees with previous estimates, $\nu > 1/d$ where $d=1$, using a diagramatic approach.

We can turn to the infinite randomness fixed point, which is characterized by a Dyson singularity with $D(\varepsilon) \sim [\varepsilon \log^3 \varepsilon]^{-1}$ and $\xi(\varepsilon) \sim \log\varepsilon$. 
We note that the $\xi(\varepsilon)$ corresponds to the \emph{typical} localization length.
Owing to the bi-locality of the free fermion wave functions \cite{Fisher_1995, Evers_2005}, the average localization length captures the distance between the two localization centers while the typical localization length captures the spread around each center---the latter is responsible for the mixing between orbitals and thus controls the matrix element.

In such systems we have:
\begin{equation}
    C \varepsilon \frac{1}{\varepsilon \log^3 \varepsilon} \log \varepsilon \sim \frac{1}{\log^2\varepsilon} \to 0
\end{equation}

\begin{align}
    R ~\sim~ & V \ell^2 \int_0^\varepsilon~d\varepsilon'~\frac{|\log(\varepsilon' + C \varepsilon)|^{-3 - 1/2} |\log(\varepsilon')|^{-3 -1/2} }{|\varepsilon' + C\varepsilon|||\varepsilon'|}    \notag \\
    \gtrsim~ & 2 V\ell^2 \frac{|\log(\varepsilon)|^{-3-1/2}}{|\varepsilon|} \int_0^{\varepsilon}~d \varepsilon'~\frac{|\log(\varepsilon')|^{-3 - 1/2}}{|\varepsilon'|} \notag \\ 
    \sim ~ & V \ell^2 \frac{|\log(\varepsilon)|^{-3 - 1/2}}{|\varepsilon|} |\log( \varepsilon)|^{-2-1/2} \sim 
    \frac{1}{\varepsilon} |\log(\varepsilon)|^{-4} 
\end{align}

While the latter condition diverges as $\varepsilon \to 0$, the former does not. This means that, within a block of size $\ell$ we are not guaranteed to find an appropriate orbital to start the resonance process.

\bibliographystyle{apsrev4-1} % Tell bibtex which bibliography style to use
\bibliography{refs} % Tell bibtex which .bib file to use (this one is some example file in TexLive's file tree)